\documentclass[11pt,a4paper]{article}
\pdfoutput=1
\usepackage{jheppub}
\usepackage{mathrsfs}

\DeclareUnicodeCharacter{2013}{--}
\DeclareUnicodeCharacter{2014}{---}
\DeclareUnicodeCharacter{2019}{'}

\allowdisplaybreaks[0]
\DeclareMathOperator{\sDet}{sDet}
\DeclareMathOperator{\Det}{Det}
\DeclareMathOperator{\Tr}{Tr}

\newcommand*\Laplace{\mathop{}\!\mathbin\bigtriangleup}

\newcommand{\VarPrimeSup}[3]{#1^{\prime #2 #3}}

\makeatletter\gdef\@fpheader{}\makeatother
\title{\boldmath Generating functional of correlators of twist-$2$ operators in
$\mathcal{N}=1$ SUSY Yang--Mills theory, II}

\author[a,1]{Marco Bochicchio,}
\author[a,b]{Mauro Papinutto,}
\author[a,b]{and Francesco Scardino}

\affiliation[a]{Physics Department, INFN Roma1,\\
Piazzale A. Moro 2, I-00185, Italy}
\affiliation[b]{Physics Department, Sapienza University,\\
Piazzale A. Moro 2, I-00185, Italy}

\emailAdd{marco.bochicchio@roma1.infn.it}
\emailAdd{mauro.papinutto@roma1.infn.it}
\emailAdd{francesco.scardino@uniroma1.it}

\abstract{This article represents the second installment, in which we calculate the generating functional of correlators of collinear twist-2 operators that are components of unbalanced superfields—i.e., superfields possessing an unequal number of dotted and undotted indices in their spinor representation—in $\mathcal{N} = 1$ SUSY SU($N$) YM theory. The analysis is performed in both Minkowskian and Euclidean space-time, in the conformal limit and its renormalization-group (RG) improved form to the leading and next-to-leading order of the large-$N$ expansion. The corresponding generating functional for correlators of balanced superfields was addressed in the first installment of this work. Finally, we compare our asymptotic RG-improved generating functional at the next-to-leading large-$N$ order with the corresponding nonperturbative object that originates from the glueball/gluinoball one-loop effective action, to which it should be asymptotic at short distances due to asymptotic freedom. It is noteworthy that we discover both objects share the structure of the logarithm of a functional superdeterminant. Consequently, our large-$N$ calculation imposes stringent ultraviolet asymptotic constraints on the nonperturbative solution of large-$N$ $\mathcal{N} = 1$ SUSY YM theory, which could serve as a crucial guide in the quest for such a solution.}

\begin{document}
\maketitle
\flushbottom
	
	\section{Introduction and physics motivations} \label{s1}

	This article, the second of a two-part series, deals with the generating functional for correlators of collinear twist-2 operators that are components of unbalanced superfields in $\mathcal{N} = 1$ supersymmetric (SUSY) SU($N$) Yang-Mills (YM) theory. In contrast, the first installment \cite{BPS41} focused on the analogous object for balanced superfields.\par
	Operators, and by extension superfields, are referred to as either balanced or unbalanced \cite{BPSpaper2,BPS1} depending on whether their spinor representation contains an equal or an unequal number of dotted and undotted indices, respectively.\par
	In essence, we calculate the short-distance asymptotic, renormalization-group (RG) improved generating functional of correlators of twist-2 operators to the leading and next-to-leading—i.e., leading nonplanar—order in the large-$N$ expansion \cite{tHooft:1973alw}. \par
	Significantly, we discover that our asymptotic generating functional at the leading nonplanar order possesses the structure of the logarithm of a functional superdeterminant, which mirrors the structure of the corresponding nonperturbative entity derived from the glueball/gluinoball one-loop effective action. \par
	Thus, our calculation establishes strict ultraviolet (UV) constraints on the prospective nonperturbative solution of large-$N$ $\mathcal{N} = 1$ SUSY YM theory, which could act as a crucial guide in finding that solution. \par
	Similar to the first installment \cite{BPS41}, our calculation consists of several stages, which are also of inherent interest on their own. \par
	First, we calculate at the lowest perturbative order the generating functional of conformal correlators of twist-2 operators having  maximal-spin component along the light-cone direction $p_+$ in Minkowskian space-time—which we term collinear twist-2 operators \cite{BPSpaper2,BPS1}—and its analytic continuation to Euclidean space-time. \par
	This conformal generating functional takes the form of the logarithm of a superdeterminant because, in the light-cone gauge, collinear twist-2 operators are quadratic in the fundamental fields. Consequently, the lowest-order generating functional of their correlators stems from a Gaussian integral \cite{BPSpaper2}, which in the SUSY case also includes anticommuting variables \cite{BPS41}. \par
	Second, we calculate—within a specific renormalization scheme \cite{MB1} that reduces the operator mixing of collinear twist-2 operators to the multiplicatively renormalizable case \cite{MB1,BPSpaper2}—the generating functional of the RG-improved UV asymptotic correlators. This functional inherits—as per our previous demonstration \cite{MB1,BPSpaper2,BPS41}—the aforementioned structure of the logarithm of a superdeterminant \cite{BPS41}, rather than a determinant as in pure YM theory \cite{BPSpaper2}.\par
	Third, we expand the above logarithm of a superdeterminant—while preserving its structure—to the leading and next-to-leading order in the large-$N$ expansion.\par
	It is noteworthy—as we review in this paper, following the initial prediction for pure YM theory \cite{Bochicchio:2016toi}, which was subsequently verified asymptotically in \cite{BPSpaper2,BPSL,QCD24}—that nonperturbatively the logarithm of a superdeterminant is expected to emerge at the next-to-leading order of the large-$N$ expansion from the Legendre transform \cite{BPSL,QCD24} of the glueball/gluinoball one-loop effective action, to which our leading-nonplanar RG-improved generating functional must be asymptotic due to asymptotic freedom (AF).\par
	Therefore, the structural correspondence between our 
	asymptotic result and the nonperturbative one provides a highly detailed quantitative constraint on the solution of the large-$N$ theory that could be an essential guide in seeking it.

	\section{Plan of the paper} \label{qq}
	
	In Sec. \ref{NP}—as a nonperturbative detour to motivate our perturbative and RG-improved computations—we determine the general structure of the generating  
	functional of large-$N$ correlators of unbalanced twist-2 operators that arises from the corresponding nonperturbative glueball/gluinoball one-loop effective action in $      
	\mathcal{N}=1$ SUSY YM theory. \par
	Secs. \ref{defSYM}-\ref{RGF}, in contrast, are dedicated to our perturbative and RG-improved calculations involving gluons and gluinos:\par
	In Sec. \ref{defSYM} we review the definition of $\mathcal{N}=1$ SUSY YM theory in the light-cone gauge.\par
	In Sec. \ref{twist2op} we introduce the twist-2 operators that form the components of unbalanced superfields.\par
	In Sec. \ref{functionalInt} we compute their Minkowskian conformal generating functional to the lowest perturbative order as a superdeterminant of a quadratic form.
	It is also expressed as a Fredholm superdeterminant using the correspondence from \cite{BPSpaper2}. \par
	In Sec. \ref{euclgen} we perform the analytic continuation of the Minkowskian conformal generating functional to Euclidean space-time.\par
	In Sec. \ref{s0} we determine the asymptotic RG-improved correlators by explicitly constructing the previously mentioned renormalization 
	scheme to all orders in perturbation theory. \par
	In Sec. \ref{RGF} we derive the corresponding generating functional of Euclidean asymptotic RG-improved correlators, which inherits the exact same structure of the 
	logarithm of a functional superdeterminant.\par
	Sec. \ref{Conc} presents our primary conclusions concerning the structural match of the logarithm of a superdeterminant emerging from the 
	nonperturbative large-$N$ one-loop effective action—in terms of glueballs and gluinoballs in Sec. \ref{NP}—with our asymptotic calculation—in terms 
	of gluons and gluinos in Sec. \ref{RGF}—at the leading nonplanar order.\par
	Finally, in Sec. \ref{sec:11} we report further developments and outlook. \par
	The Appendices contain fundamental definitions and supplementary calculations.

	\section{Nonperturbative effective action in large-$N$ $\mathcal{N}=1$ SUSY YM theory}\label{NP}

	For five decades, it has been established that $\mathcal{N}=1$ SUSY SU($N$) YM theory permits 't Hooft's large-$N$ topological expansion \cite{tHooft:1973alw} for the $n$-point connected correlators of gauge-invariant single-trace operators. In 't Hooft's double-line representation, the associated Feynman diagrams—following an appropriate gluing of oppositely oriented lines—are topologically categorized \cite{tHooft:1973alw,Veneziano:1976wm} by the sum over the genus $g$ of $n$-punctured closed Riemann surfaces, with each topology being weighted by a factor of $N^{\chi}$, where $\chi=2-2g-n$ is the Euler characteristic of the Riemann surface.\par
	As a result, the associated nonperturbative large-$N$ effective theory comprises an infinite spectrum of weakly interacting glueballs and gluinoballs, with a coupling of order $\frac{1}{N}$ \cite{tHooft:1973alw,Migdal:1977nu,Witten:1979kh} and masses proportional to the RG-invariant scale $\Lambda_{SYM}$. \par
	By assuming 't Hooft's topological expansion in $\mathcal{N}=1$ SUSY SU($N$) YM theory, the generating functional $\mathcal{W}^E[J_{S},J_T]=\log \mathcal{Z}^E[J_{S},J_T]$ for Euclidean connected correlators of single-trace bosonic and fermionic unbalanced operators $S$ and $T$, respectively, with
	\begin{align}
		&\mathcal{Z}^E[J_{S},J_T]=\frac{1}{\mathcal{Z}^E} \int \mathcal{D}A\mathcal{D}\chi\, e^{-  S_{SYM}+ \sum_s\int J_{S_s}S_s+  J_{T_s}T_s}
	\end{align}
	takes the form
	\begin{equation}
		\label{thooft}
		\mathcal{W}^E[J_{S},J_T]=\mathcal{W}^E_{\text{sphere}}[J_{S},J_T]+\mathcal{W}^E_{\text{torus}}[J_{S},J_T]+ \cdots \,.
	\end{equation}
	On a nonperturbative level, $\mathcal{W}^E_{\text{sphere}}[J_{S},J_T]$, which corresponds to the ('t Hooft-)planar contribution \cite{tHooft:1973alw} in perturbation theory, is given by a sum of tree diagrams that include glueball/gluinoball propagators and vertices, whereas $\mathcal{W}^E_{\text{torus}}[J_{S},J_T]$, which represents the leading-non('t Hooft-)planar contribution, is a sum of one-loop diagrams involving glueballs/gluinoballs. \par
	Nonperturbatively, $\mathcal{W}^E_{\text{torus}}[J_{S},J_T]$ is expected to possess the structure of a logarithm of a functional (super)determinant, a feature that was first predicted for the pure YM case \cite{Bochicchio:2016toi} based on fundamental principles and was later verified asymptotically \cite{BPSpaper2,BPSL,QCD24}. \par 
	Indeed, drawing an analogy to the pure YM case \cite{Bochicchio:2016toi,BPSL,QCD24}, in the yet-to-be-found nonperturbative solution of large-$N$ $\mathcal{N}=1$ SUSY YM theory, these same correlators should be calculable from the correlators of the glueball $\Phi$ and gluinoball $\Psi$ fields, which have an infinite number of components labelled by the spin $s$ and an internal quantum number $n$. The corresponding generating functional is schematically \cite{Bochicchio:2016toi,BPSL,QCD24}
	\begin{align}
		\mathcal{Z}^E_{\text{glueball/gluinoball}}[J_{\Phi},J_{\Psi}]= \frac{1}{\mathcal{Z}^{E}_{\text{glueball/gluinoball}}} \int\mathcal{D}\Phi \mathcal{D}\Psi\,e^{-S_{\text{glueball/gluinoball}}(\Phi,\Psi)+\int \Phi \ast_1 J_{\Phi}+\Psi \ast'_1 J_{\Psi}} \,,
	\end{align}
	with \cite{Bochicchio:2016toi,BPSL,QCD24}
	\begin{align}
		S_{\text{glueball/gluinoball}}(\Phi,\Psi) =&\frac{1}{2}\int\,\Phi\ast_2(-\Delta+M^2)\Phi + \Psi\ast'_2(-\Delta+M^2)\Psi\nonumber\\ &+\frac{1}{N}(\frac{1}{3}\Phi \ast_3\Phi\ast_3\Phi+\Psi \ast'_3\Phi\ast'_3\Psi)+\cdots
		\label{eq:action}
	\end{align}
	where $\ast_2$, $\ast_1$ and $\ast'_2$, $\ast'_1$ will be determined later; the ellipsis and the symbols $\ast_3$, $\ast'_3$ respectively represent $n$-glueball/gluinoball vertices with $n>3$ and some currently unknown operation on the glueball/gluinoball fields, which, by assuming locality and Euclidean invariance, might contain derivatives. Hence, nonperturbatively, the connected generating functional $\mathcal{W}^E_{\text{glueball/gluinoball}}[J_{\Phi},J_{\Psi}] = \log\mathcal{Z}^E_{\text{glueball/gluinoball}}[J_{\Phi},J_{\Psi}]$ is, up to one loop of glueballs/gluinoballs \cite{Bochicchio:2016toi,BPSL,QCD24}
	\begin{align}
		\label{glueballW}
		\mathcal{W}^E_{\text{glueball/gluinoball}}[J_{\Phi},J_{\Psi}] =& -S_{\text{glueball/gluinoball}}(\Phi_J,\Psi_J)+\int \Phi_J \ast_1 J_{\Phi}+\int \Psi_J \ast'_1 J_{\Psi} + \cdots \nonumber\\
		&+\frac{1}{2}\log\text{sDet}
		\begin{pmatrix}\ast'_2(-\Delta+M^2)+\frac{1}{N}\ast'_3\Phi_J\ast'_3& \frac{1}{N}\ast'_3\ast'_3\Psi_J\\ 
			\frac{1}{N}\ast'_3\ast'_3\Psi_J&\ast_2(-\Delta+M^2)+\frac{1}{N}\ast_3\Phi_J\ast_3 \end{pmatrix} \,,
	\end{align}
	where $\Phi_J,\Psi_J$ are found from
	\begin{equation}
		\frac{\delta S_{\text{glueball/gluinoball}}}{\delta\Phi}\Big\rvert_{\Phi_J}=\ast_1 J_{\Phi} 
	\end{equation}
	and
	\begin{equation}
		\frac{\delta S_{\text{glueball/gluinoball}}}{\delta\Psi}\Big\rvert_{\Psi_J}=\ast'_1 J_{\Psi}\,,
	\end{equation}
	with the superdeterminant being defined as \cite{zinn1993quantum}
	\begin{align}\label{eq:sdet}
		\text{sDet}\begin{pmatrix}A & B \\ C & D\end{pmatrix}&=\Det(A-BD^{-1}C)\Det(D)^{-1}\nonumber\\
		&= \Det(A)\Det(D-CA^{-1}B)^{-1}\,,
	\end{align}
	where $A$ and $D$ are bosonic entries and $B$ and $C$ are fermionic ones.\par
	The correspondence between $\mathcal{W}^E[J_{S},J_T]$ and $\mathcal{W}^E_{\text{glueball/gluinoball}}[J_{\Phi},J_{\Psi}]$ is established by matching their respective
	spectral representations—as a sum of free propagators with residues $R^{(s)}_{n},R'^{(s)}_{n}$—for the $2$-point correlators at $N =+\infty$ \cite{Migdal:1977nu} of $S_s$ and $T_s$. For example, for the $2$-point correlators of a renormalized twist-$2$ operator $\mathcal{O}_s$ of spin $s$, we explicitly obtain for the pure spin $s$ contribution\footnote{In general, the renormalized twist-$2$ operators $\mathcal{O}_s(p)$ mix with the derivatives of lower-spin twist-$2$ operators, so that the spectral representation in Eq. \eqref{eq:kallen} contains in general lower-spin contributions in the ellipsis \cite{Bochicchio:2023ols}.}\cite{Bochicchio:2023ols} 
	\begin{equation}\label{eq:kallen}
		\langle	\mathcal{O}_s(p)\mathcal{O}_s(-p)\rangle = \sum_{n=0}^\infty P^E_s (m^{(s)}_n)\frac{R^{(s)}_{n}}{p^2+{m^{(s)}}^2_n}+\cdots\,,
	\end{equation}
	with $s$ integer and \cite{Bochicchio:2023ols}
	\begin{align}
		\nonumber\label{eq:proj}
		P_{\mu_1\ldots\mu_s}^{E \nu_1\ldots\nu_s}(m)=&\sum_{j = 0}^{\left[\frac{s}{2}\right]}(-1)^j 2^{-j}\frac{s!}{j!(s-2j)!}\frac{(2s-2j-1)!!}{(2s-1)!!}\\
		&{\Pi^E}_{(\mu_1\mu_2}\ldots{\Pi^E}_{\mu_{2j-1}\mu_{2j}}{\Pi^E}^{(\nu_1\nu_2}\ldots{\Pi^E}^{\nu_{2j-1}\nu_{2j}}{\Pi^E}^{\nu_{2j+1}}_{\mu_{2j+1}}\ldots{\Pi^E}^{\nu_{s})}_{\mu_{s})} \,,
	\end{align}
	where the parentheses denote symmetrization of the indices in between, with
	\begin{equation} \label{PE}
		\Pi^E_{\mu\nu}(m)= \delta_{\mu\nu}+\frac{p_\mu p_\nu}{m^2}
	\end{equation}
	and
	\begin{equation}
		(2k-1)!! = \frac{(2k)!}{2^{k}k!} \,.
	\end{equation}
	This uniquely determines the coupling of $J_{\Phi},J_{\Psi}$ to the tower of glueball/gluinoball fields 
	\begin{align}
		&\Phi \ast_1 J_{\Phi} = \sum_{sn} \Phi_{sn} \sqrt{R^{(s)}_{n}} J_{\Phi_s}\nonumber\\
		&\Psi \ast'_1 J_{\Psi} = \sum_{sn} \Psi_{sn} \sqrt{R'^{(s)}_{n}} J_{\Psi_s}\,,
	\end{align}
	respectively, by setting $\ast_2,\ast'_2$ in accordance with the canonical normalization of the glueball/gluinoball kinetic term.\par
	Specifically, the canonical normalization that determines $\ast_2$ can be read from the canonical normalized propagator for maximal-spin component along the $p_z$ direction  \cite{Bochicchio:2023ols}
	\begin{equation}
		\langle	\Phi_{sn}(p)\Phi_{sn}(-p)\rangle = (-1)^s2^{s}\frac{(s!)^2}{(2s)!} \frac{p_z^{2s}}{p^2+{m^{(s)}}^2_n}\,.
	\end{equation}
	Remarkably, the large-momentum asymptotics of Eq. \eqref{eq:kallen} has been known by means of AF
	\begin{align}\label{eq:2asym}
		\sum_{n=1}^{\infty}  P_s^{E} (p) \frac{ R^{(s)}_n}{p^2+m^{(s)2}_n  }  
		&\sim - (-p^2)^{s} P_s^{E} (p) \frac{g^{-2}(p)}{\beta_0(\gamma'+1)} \biggl( \frac{g(\mu)}{g(p)} \biggr)^{2 \gamma' } 
		+ \cdots \nonumber \\
		&= - (-p^2)^{s} P_s^{E}(p) \frac{g^{-2}(\mu)}{\beta_0(\gamma'+1)} \biggl( \frac{g(\mu)}{g(p)} \biggr)^{2 \gamma' +2} 
		+ \cdots
	\end{align}
	up to possibly divergent contact terms in the ellipses \cite{Bochicchio:2023ols}, where $P_s^{E} (p)$ is obtained from $P_s^E(m)$ by the substitution $m^2\to -p^2$ in  Eqs. \eqref{eq:proj} and \eqref{PE} \cite{Bochicchio:2023ols}; $\gamma' = \frac{\gamma_0}{\beta_0}$, where $\gamma_0$ and $\beta_0$ are the one-loop anomalous dimension and beta function coefficients, respectively, and $g(p)$ is the running coupling with UV asymptotics 
	\begin{equation}
		g^2(p)\sim\frac{1}{\beta_0\log\left(\frac{p^2}{\Lambda^2_{SYM}}\right)}\left(1-\frac{\beta_1}{\beta_0^2}\frac{\log\log\left(\frac{p^2}{\Lambda^2_{SYM}}\right)}{\log\left(\frac{p^2}{\Lambda^2_{SYM}}\right)}\right)\,.
	\end{equation}
	\par
	Prior to this work, apart from the  nonperturbative structure of the planar $2$-point correlators in Eq. \eqref{eq:kallen} and their corresponding asymptotics in Eq. \eqref{eq:2asym}, no quantitative information was available for $\mathcal{W}^E[J_{S},J_T]$ and $\mathcal{W}^E_{\text{glueball/gluinoball}}[J_{\Phi},J_{\Psi}]$.\par 
	To fill this gap, in the following sections we will calculate the UV asymptotics of $\mathcal{W}^E[J_{S},J_T]$ and $\mathcal{W}^E_{\text{glueball/gluinoball}}[J_{\Phi},J_{\Psi}]$ for unbalanced twist-2 operators at the planar and leading-nonplanar order of the large-$N$ expansion, in order to put constraints on the structure of the glueball/gluinoball one-loop effective action above.

	\section{$\mathcal{N} = 1$ SUSY YM theory in the light-cone gauge \label{defSYM}} 
	
	The action for $\mathcal{N} = 1$ SUSY SU($N$) YM theory in Minkowskian space-time is given by \cite{Brink:1976bc,Belitsky:2004sc}
	\begin{equation}\label{eq:fundamentaltheory}
		S= \int -\frac{1}{4} F^a_{\mu\nu}F^{a\,\mu\nu} +\frac{i}{2}\bar{\chi}^a\gamma^\mu (D_\mu\chi)^a \,d^4x \,,
	\end{equation}
	with $a=1, \ldots, N^2-1$, where $\chi$ is a Majorana spinor satisfying $\chi=C\bar{\chi}^T$ in the adjoint representation of the SU($N$) Lie algebra, $C$ denotes the charge conjugation, and
	\begin{align}
		&F_{\mu\nu} = \partial_\mu A_\nu-\partial_\nu A_\mu +i\frac{g}{\sqrt{N}}\left[A_\mu,A_\nu\right]\nonumber\\
		&D_\mu \chi= \partial_\mu\chi+i\frac{g}{\sqrt{N}}\left[A_\mu,\chi\right]
	\end{align}  
	in matrix form, with $g^2= g^2_{YM}N$ being the 't Hooft coupling \cite{tHooft:1973alw}. This action is invariant under the following SUSY transformations \cite{Brink:1976bc,Belitsky:2003sh}
	\begin{align}
		&\delta A^a_\mu =-i \bar{\zeta}\gamma_\mu \chi^a\nonumber\\
		&\delta\chi^a=\frac{i}{2}\sigma^{\mu\nu}F_{\mu\nu}^a\zeta \,,
	\end{align}
	where $\sigma^{\mu\nu} = \frac{i}{2}\left[\gamma^\mu,\gamma^\nu\right]$ and $\zeta$ is a Majorana spinor. A Majorana spinor can be decomposed into a pair of complex-conjugate Weyl spinors, $\lambda_\alpha$ and $\bar{\lambda}_{\dot{\alpha}}$
	\begin{equation}
		\chi = \begin{pmatrix}
			\lambda_\alpha \\
			\bar{\lambda}^{\dot{\alpha}}
		\end{pmatrix}\hspace{1cm}\bar{\chi} = \chi^\dagger \gamma^0 =  \begin{pmatrix}
			\lambda^\alpha & \hspace{-0.1cm}
			\bar{\lambda}_{\dot{\alpha}}
		\end{pmatrix}\,.
	\end{equation}
	Accordingly, the action can be written as
	\begin{equation}
		S= - \int \frac{1}{4} F^a_{\mu\nu}F^{a\,\mu\nu} +i\bar{\lambda}^{a\,\dot{\alpha}}\sigma^\mu_{\alpha \dot{\alpha}} (D_\mu\lambda^\alpha)^a \,d^4x\,.
	\end{equation}
	The fundamental fields, $A_\mu$ and $\chi$, serve as interpolating fields for the gluon and the gluino, respectively.
	In the light-cone gauge $A_+ = 0$, $A_-$ and  
	\begin{equation}
		\chi_-=\Pi_{-} \chi=\begin{pmatrix}
			0 \\
			\lambda_2 \\
			\bar{\lambda}_{\dot{2}} \\
			0
		\end{pmatrix}
	\end{equation}
	can be integrated out, with $\Pi_{\pm} = \frac{1}{2} \gamma^{\mp}\gamma^{\pm}$. The resulting action is expressed solely in terms of physical fields—the transverse components of the gluon, $A$ and $\bar{A}$,
	\begin{align}
		& A = \frac{1}{\sqrt{2}} (A_1 + i A_2)\nonumber \\
		& \bar{A} = \frac{1}{\sqrt{2}} (A_1 - i A_2) 
	\end{align}
	and the plus component of the gluino, $\chi_+$,
	\begin{equation}
		\chi_+ = \Pi_+ \chi= \begin{pmatrix}
			\lambda_1 \\
			0 \\
			0 \\
			-\bar{\lambda}_{\dot{1}}
		\end{pmatrix}\,.
	\end{equation}
	For convenience, we define two anticommuting fields, $\lambda$ and $\bar \lambda$, such that their normalization
	\begin{align}
		\lambda = \frac{1}{2^{\frac{1}{4}}}\lambda_1 \nonumber \\
		\bar{\lambda} = \frac{1}{2^{\frac{1}{4}}}\bar{\lambda}_{\dot{1}}
	\end{align}
	is consistent with the canonical normalization of the kinetic term of the action in the light-cone gauge \cite{Belitsky:2004sc}
	\begin{align}
		&S(A, \bar A,\lambda,\bar{\lambda}) = \int -\bar{A}^a \square A^a-i\bar{\lambda}^a\square\partial_+^{-1}\lambda^a
		-2\frac{g}{\sqrt{N}} f^{abc}(A^a\partial_+\bar{A}^b\bar{\partial}\partial_+^{-1}A^c+\bar{A}^a\partial_+A^b\partial\partial_+^{-1}\bar{A}^c)\nonumber\\ &-2\frac{g^2}{N}f^{abc}f^{ade}\partial_+^{-1}(A^b\partial_+ \bar{A}^c)\partial_+^{-1}(\bar{A}^d\partial_+A^e) -2i\frac{g}{\sqrt{N}}f^{abc}\bar{\lambda}^a\lambda^b(\bar{\partial}A^c+\partial\bar{A}^c)\nonumber\\
		&-2i\frac{g^2}{N}f^{abe}f^{cde}\partial_+^{-1}(\bar{A}^a\partial_+ A^b+A^a\partial_+ \bar{A}^b)\partial_{+}^{-1}(\bar{\lambda}^c\lambda^d)+2\frac{g^2}{N}f^{abe}f^{cde}\partial_{+}^{-1}(\bar{\lambda}^a\lambda^b)\partial_{+}^{-1}(\bar{\lambda}^c\lambda^d)\nonumber\\
		&+2i\frac{g}{\sqrt{N}}f^{abc}\bar{A}^a\bar{\lambda}^b\partial_+^{-1}\partial\lambda^c+2i\frac{g}{\sqrt{N}}f^{abc}A^a\lambda^b\partial_+^{-1}\bar{\partial}\bar{\lambda}^c-2i\frac{g^2}{N}f^{abe}f^{cde}\bar{A}^a\bar{\lambda}^b\partial_{+}^{-1}(A^c\lambda^d) \,\,d^4x \,,
	\end{align} 
	with
	\begin{equation}
		\square= g^{\mu\nu}\partial_{\mu}\partial_{\nu}=\partial_0^2-\sum_{i=1}^{3}\partial_i^2\,,
	\end{equation}
	where we use the mostly minus metric $g_{\mu\nu}$ in Minkowskian space-time (Appendix A in \cite{BPS1}). \par
	The vacuum expectation value (v.e.v.) of a product of local gauge-invariant operators $\mathcal{O}_i$ that are independent of $A_{-}$ and $\chi_-$ is
	\begin{align}
		\langle\mathcal{O}_1(x_1)\ldots \mathcal{O}_n(x_n)\rangle =\frac{1}{Z}\int   \mathcal{D} A \mathcal{D} \bar{A} \mathcal{D} \lambda \mathcal{D} \bar{\lambda}\, e^{iS(A,\bar A,\lambda,\bar{\lambda})} \mathcal{O}_1(x_1)\ldots \mathcal{O}_n(x_n)\,.
	\end{align}
	At the leading order of perturbation theory, the v.e.v. simplifies to
	\begin{align}
		\langle \mathcal{O}_1(x_1)\ldots \mathcal{O}_n(x_n)\rangle =\frac{1}{Z}\int \mathcal{D} A \mathcal{D} \bar{A} \mathcal{D} \lambda \mathcal{D} \bar{\lambda}\,  e^{\int -i \bar{A}^a \square A^a+\bar{\lambda}^a\Box \partial_+^{-1} \lambda^a  \, d^4x}\mathcal{O}_1(x_1)\ldots \mathcal{O}_n(x_n)\,.
	\end{align}
	For the corresponding leading-order conformal generating functional, we have
	\begin{align}
		\mathcal{Z}_{\text{conf}}[J_{\mathcal{O}}]
		= \frac{1}{Z}\int \mathcal{D} A \mathcal{D} \bar{A} \mathcal{D} \lambda \mathcal{D} \bar{\lambda}\,  e^{\int -i\bar{A}^a \square A^a+\bar{\lambda}^a\Box \partial_+^{-1} \lambda^a \,  d^4x}\,\exp\left(\int d^4x\sum_{i}\, J_{\mathcal{O}_i}\mathcal{O}_i\right) ,
	\end{align}
	which is the supersymmetric counterpart to the conformal generating functional in YM theory \cite{BPSpaper2}.

	\section{Twist-2 unbalanced superfields in $\mathcal{N} = 1$ SUSY YM theory \label{twist2op}}
	\label{twist2}
	
	\subsection{Collinear twist-2 operators}
	
	Here we present the list of gauge-invariant collinear twist-2 operators that form the components of unbalanced superfields, as defined below, within $\mathcal{N} = 1$ SUSY YM theory \cite{Belitsky:2004sc,Belitsky:2003sh}
	\begin{itemize}
		\item \underline{gluon-gluon operators} 
		\subitem 
		\begin{align}
			S^A_s &= \frac{1}{2\sqrt{2}} \partial_+ \bar{A}^a(i\overrightarrow{\partial}_+ + i\overleftarrow{\partial}_+)^{s-2}C^{\frac{5}{2}}_{s-2}\Bigg(\frac{\overrightarrow{\partial}_+ - \overleftarrow{\partial}_+}{\overrightarrow{\partial}_++\overleftarrow{\partial}_+}\Bigg)\partial_+ \bar{A}^a \hspace{1cm} \text{ $s =2,4,6,\ldots$}\nonumber\\
			&=\frac{1}{2\sqrt{2}}\bar{A}^a\mathcal{Y}^{\frac{5}{2}}_{s-2}\left(\overrightarrow{\partial}_+,\overleftarrow{\partial}_+\right) \bar{A}^a
		\end{align}
		\subitem 
		\begin{align}
			\bar{S}^A_s &= \frac{1}{2\sqrt{2}} \partial_+ A^a(i\overrightarrow{\partial}_+ + i\overleftarrow{\partial}_+)^{s-2}C^{\frac{5}{2}}_{s-2}\Bigg(\frac{\overrightarrow{\partial}_+ - \overleftarrow{\partial}_+}{\overrightarrow{\partial}_++\overleftarrow{\partial}_+}\Bigg)\partial_+ A^a \hspace{1cm} \text{ $s =2,4,6,\ldots$}\nonumber\\
			&=\frac{1}{2\sqrt{2}}A^a\mathcal{Y}^{\frac{5}{2}}_{s-2}\left(\overrightarrow{\partial}_+,\overleftarrow{\partial}_+\right) A^a
		\end{align}
		with $PT=(-1)^s$ and $C=+1$, where $P$ denotes parity, $T$ is time reversal, and $C$ represents charge conjugation.
		\item \underline{gluino-gluino operators}
		\subitem 
		\begin{align}
			S^\lambda_s &=  \frac{1}{2\sqrt{2}} \bar{\lambda}^a(i\overrightarrow{\partial}_+ + i\overleftarrow{\partial}_+)^{s-1}C^{\frac{3}{2}}_{s-1}\Bigg(\frac{\overrightarrow{\partial}_+ - \overleftarrow{\partial}_+}{\overrightarrow{\partial}_++\overleftarrow{\partial}_+}\Bigg) \bar{\lambda}^a \hspace{1cm} \text{ $s = 2,4,6,\ldots$}\nonumber\\
			&=\frac{1}{2\sqrt{2}}\bar{\lambda}^a\mathcal{Y}^{\frac{3}{2}}_{s-1}\left(\overrightarrow{\partial}_+,\overleftarrow{\partial}_+\right) \bar{\lambda}^a
		\end{align}
		\subitem 
		\begin{align}
			\bar{S}^\lambda_s &=  \frac{1}{2\sqrt{2}} \lambda^a(i\overrightarrow{\partial}_+ + i\overleftarrow{\partial}_+)^{s-1}C^{\frac{3}{2}}_{s-1}\Bigg(\frac{\overrightarrow{\partial}_+ - \overleftarrow{\partial}_+}{\overrightarrow{\partial}_++\overleftarrow{\partial}_+}\Bigg) \lambda^a\hspace{1cm}\text{ $s = 2,4,6,\ldots$}\nonumber\\
			&=\frac{1}{2\sqrt{2}}\lambda^a\mathcal{Y}^{\frac{3}{2}}_{s-1}\left(\overrightarrow{\partial}_+,\overleftarrow{\partial}_+\right) \lambda^a
		\end{align}
		with $PT=(-1)^{s}$ and  $C=+1$.
		\item \underline{gluon-gluino operators}   
		\subitem
		\begin{align}
			T_s &=  \frac{1}{2}\lambda^a(i\overrightarrow{\partial}_+ + i\overleftarrow{\partial}_+)^{s-1}P^{(1,2)}_{s-1}\Bigg(\frac{\overrightarrow{\partial}_+ - \overleftarrow{\partial}_+}{\overrightarrow{\partial}_++\overleftarrow{\partial}_+}\Bigg) \partial_+\bar{A}^a \hspace{1cm} \text{ $s = 1,2,3,\ldots$}\nonumber\\
			\nonumber\\
			& = \frac{1}{2}\lambda^a  \mathcal{G}_{s-1}^{(1,2)}\bar{A}^a=\frac{1}{2} (-1)^{s-1}\bar{A}_a\mathcal{G}_{s-1}^{(2,1)}\lambda^a 
		\end{align}
		\subitem
		\begin{align}
			\bar{T}_s &=
			\frac{1}{2} \partial_+A^a(i\overrightarrow{\partial}_+ + i\overleftarrow{\partial}_+)^{s-1}P^{(2,1)}_{s-1}\Bigg(\frac{\overrightarrow{\partial}_+ - \overleftarrow{\partial}_+}{\overrightarrow{\partial}_++\overleftarrow{\partial}_+}\Bigg) \bar{\lambda}^a \hspace{1cm} \text{ $s = 1,2,3,\ldots$}\nonumber\\
			\nonumber\\
			& = \frac{1}{2}A^a  \mathcal{G}_{s-1}^{(2,1)}\bar{\lambda}^a=\frac{1}{2} (-1)^{s-1}\bar{\lambda}^a\mathcal{G}_{s-1}^{(1,2)}A^a\,, 
		\end{align}
	\end{itemize}
	with $C^{\alpha'}_n$ being Gegenbauer polynomials and $P^{(\alpha,\beta)}_n$ Jacobi polynomials (\ref{appB}),
	and
	\begin{align}
		\label{defY}
		\mathcal{Y}_{s-2}^{\frac{5}{2}}(\overrightarrow{\partial}_+,\overleftarrow{\partial}_+)
		&= \overleftarrow{\partial}_+ (i\overrightarrow{\partial}_++i\overleftarrow{\partial}_+)^{s-2}C^{\frac{5}{2}}_{s-2}\left(\frac{\overrightarrow{\partial}_+-\overleftarrow{\partial}_+}{\overrightarrow{\partial}_++\overleftarrow{\partial}_+}\right)\overrightarrow{\partial}_+\nonumber\\
		&=\frac{\Gamma(3)\Gamma(s+3)}{\Gamma(5)\Gamma(s+1)}i^{s-2}\nonumber\\
		&\quad\sum_{k=0}^{s-2} {s\choose k}{s\choose k+2}(-1)^{s-k} \overleftarrow{\partial}_{+}^{s-k-1} \overrightarrow{\partial}_{+}^{k+1} ,\nonumber\\
	\end{align}
	\begin{align}
		\mathcal{Y}_{s-1}^{\frac{3}{2}}(\overrightarrow{\partial}_+,\overleftarrow{\partial}_+) 
		&=  (i\overrightarrow{\partial}_++i\overleftarrow{\partial}_+)^{s-2}C^{\frac{3}{2}}_{s-1}\left(\frac{\overrightarrow{\partial}_+-\overleftarrow{\partial}_+}{\overrightarrow{\partial}_++\overleftarrow{\partial}_+}\right)\nonumber\\
		&=\frac{(s+1)}{2}i^{s-1} \sum_{k = 0}^{s-1}{s\choose k}{s\choose k+1}(-1)^{s-k-1} \overleftarrow{\partial_{+}}^{s-k-1}\overrightarrow{\partial_{+}}^{k} 
	\end{align}
	for even $s$, and
	\begin{align}
		\label{gdef}
		\mathcal{G}_{s-1}^{(1,2)} 
		&=(i\overrightarrow{\partial}_++ i\overleftarrow{\partial}_+)^{s-1}P^{(2,1)}_{s-1}\Bigg(\frac{\overrightarrow{\partial}_+- \overleftarrow{\partial}_+}{\overrightarrow{\partial_+}+\overleftarrow{\partial}_+}\Bigg) \overrightarrow{\partial}_+ \nonumber\\
		&=  i^{s-1} \sum_{k = 0}^{s-1}{s\choose k}{s+1\choose k+2} (-1)^{s-k-1} \overleftarrow{\partial_{+}}^{s-k-1} \overrightarrow{\partial_{+}}^{k+1} ,\nonumber\\
	\end{align}
	\begin{align}
		\mathcal{G}_{s-1}^{(2,1)} 
		&=\overleftarrow{\partial}_+(i\overrightarrow{\partial}_++ i\overleftarrow{\partial}_+)^{s-1}P^{(1,2)}_{s-1}\Bigg(\frac{\overrightarrow{\partial}_+- \overleftarrow{\partial}_+}{\overrightarrow{\partial_+}+\overleftarrow{\partial}_+}\Bigg)\nonumber\\
		&=   i^{s-1} \sum_{k = 0}^{s-1}{s+1\choose k}{s\choose k+1} (-1)^{s-k-1} \overleftarrow{\partial_{+}}^{s-k} \overrightarrow{\partial_{+}}^{k}\nonumber\\
	\end{align}
	for odd $s$. The spin projection on the $p_+$ direction is $s$ for $S^A_s$, $S^\lambda_s$, and $s + \frac{1}{2}$ for $T_s$, $\bar{T}_s$. The bosonic operators $S^A_s, \bar{S}^A_s, S^\lambda_s, \bar{S}^\lambda_s$ and fermionic operators $T_s, \bar{T}_s$ are Hermitian conjugates of one another, respectively.
	
	\subsection{Unbalanced superfields}
	
	The operators listed above constitute an irreducible representation of the superalgebra of SUSY transformations when restricted to the light-cone \cite{Belitsky:2003sh,Belitsky:1998gu}
	\begin{align}
		&\delta A = -2i\bar{\lambda}\zeta\nonumber\\
		&\delta \bar{A} = 2i\bar{\zeta}\lambda\nonumber\\
		&\delta\lambda=2\zeta\partial_+\bar{A}\nonumber\\
		&\delta\bar{\lambda} = 2\bar{\zeta}\partial_+A
	\end{align}
	with $\zeta$ being a Majorana spinor that satisfies $\Pi_+\zeta = 0$ \cite{BPS41}. These operators also diagonalize the matrix of anomalous dimensions to order $g^2$—where $\mathcal{N} = 1$ SUSY YM theory is indeed conformally invariant in the conformal scheme (\ref{B})—and appear as components (Sec. 6.3 in \cite{SS1}\footnote{In v2 of \cite{SS1}, the definitions of $\bar{S}^\lambda$ and $S^\lambda$ are swapped.}) of the unbalanced superfields
	\begin{align}
		\mathbb{W}^+_s(x,\theta,\bar{\theta}) \sim T_{s-1}+\theta S^A_s+\bar{\theta}\bar{S}^\lambda_{s}+\theta\bar{\theta}T_{s}+\theta\bar{\theta}\,\partial_+T_{s-1}
	\end{align}
	and
	\begin{align}
		\mathbb{W}^-_s(x,\theta,\bar{\theta}) \sim \bar{T}_{s-1}+\theta \bar{S}^A_s+\bar{\theta}S^\lambda_{s}+\theta\bar{\theta}\bar{T}_{s}+\theta\bar{\theta}\,\partial_+\bar{T}_{s-1}\,.
	\end{align}

	\section{Generating functional of Minkowskian conformal correlators \label{functionalInt}}

	The corresponding conformal generating functional is
	\begin{align}\label{eq:generatingfunctionalfirst}
		\mathcal{Z}_{\text{conf}}\left[J_{\bar{S}^A},\bar{J}_{S^A},J_{\bar{S}^\lambda},\bar{J}_{S^\lambda}, J_{\bar{T}},\bar{J}_T\right]=&\frac{1}{Z} \int \mathcal{D} A \mathcal{D} \bar{A} \mathcal{D} \lambda \mathcal{D} \bar{\lambda} e^{-\int d^4x\, i\bar{A}^a \square A^a-\bar{\lambda}^a\Box \partial_+^{-1} \lambda^a}\nonumber\\
		& \exp \Big( \int d^4 x\sum_s S^A_s \bar{J}_{S^A_s}+\bar{S}^A_s J_{\bar{S}^A_s}+S^\lambda_s \bar{J}_{S^\lambda_s}+\bar{S}^\lambda_s J_{\bar{S}^\lambda_s}+ \bar{J}_{T_s} T_s + \bar{T}_s J_{\bar{T}_s} \Big) \,.
	\end{align}
	The sources $J_{\bar{S}^A_s}$, $\bar{J}_{S^A_s}$, $J_{\bar{S}^\lambda_s}$,$\bar{J}_{S^\lambda_s}$ are bosonic, while $\bar{J}_{T_s}$,$ J_{\bar{T}_s}$ are fermionic. Thus, we can make the formal identification
	\begin{align}
		&\frac{\delta}{\delta J_{\bar{S}^A_s}(x)} \longleftrightarrow \bar{S}^A_{s}(x)\qquad  \frac{\delta}{\delta J_{\bar{S}^\lambda_s}(x)} \longleftrightarrow \bar{S}^\lambda_{s}(x)\nonumber \\
		&\frac{\delta}{\delta \bar{J}_{S^A_{s}}(x)} \longleftrightarrow S^A_{s}(x)\qquad  \frac{\delta}{\delta \bar{J}_{S^\lambda_{s}}(x)} \longleftrightarrow S^\lambda_{s}(x)\nonumber \\
		&\frac{\delta}{\delta \bar{J}_{T_{s}}(x)} \longleftrightarrow T_{s}(x) \qquad -\frac{\delta}{\delta  J_{\bar{T}_s}(x)} \longleftrightarrow \bar{T}_{s}(x)\,.
	\end{align}
	The generating functional of connected correlators, $\mathcal{W}_{\text{conf}} = \log \mathcal{Z}_{\text{conf}}$, follows from this.
	For instance,
	\begin{equation}
		\langle S^A_{s_1}(x)S^A_{s_2}(y) \rangle = \, \frac{\delta}{\delta \bar{J}_{S^A_{s_1}}(x)} \frac{\delta}{\delta \bar{J}_{S^A_{s_2}}(y)}\mathcal{W}_{\text{conf}}\,\Bigg|_{J = 0}
	\end{equation}
	and
	\begin{equation}
		\langle T_{s_1}(x)\bar{T}_{s_2}(y) \rangle = \, \frac{\delta}{\delta \bar{J}_{T_{s_1}}(x)} \Big(-\frac{\delta}{\delta  J_{\bar{T}_{s_2}}(y)}\Big) \mathcal{W}_{\text{conf}}\,\Bigg|_{J = 0}\,.
	\end{equation}
	Henceforth, we will omit the specification $\big|_{J = 0}$. \par

	\subsection{Connected generating functional $\mathcal{W}_{\text{conf}}$ as the $\log$ of a superdeterminant of a quadratic form}

	The functional integral presented above is quadratic in the elementary fields and can therefore be computed exactly \cite{BPSpaper2}. By utilizing the symmetry properties \cite{BPSpaper2} of the Gegenbauer polynomials (\ref{appB}), we find
	\begin{align}
		&\mathcal{Z}_{\text{conf}}\left[J_{\bar{S}^A},\bar{J}_{S^A},J_{\bar{S}^\lambda},\bar{J}_{S^\lambda}, J_{\bar{T}},\bar{J}_T\right]\nonumber\\
		 &= \frac{1}{Z} \int \mathcal{D}A\mathcal{D}\bar{A} \mathcal{D} \lambda \mathcal{D} \bar{\lambda} \exp\Bigg(-\frac{1}{2}\int d^4x\, \begin{pmatrix}
			\bar{A}^a(x) &A^a(x)
		\end{pmatrix} \mathcal{M}_{AA}^{ab}  \begin{pmatrix}
			&A^b(x) &\\
			&\bar{A}^b(x) & 
		\end{pmatrix}\nonumber\\
		&\quad-\frac{1}{2}\int d^4x\, \begin{pmatrix}
			\bar{\lambda}^a(x)&\lambda^a(x)
		\end{pmatrix} \mathcal{M}_{\lambda\lambda}^{ab} \begin{pmatrix}
			&\lambda^b(x) &\\
			&\bar{\lambda}^b(x) & 
		\end{pmatrix}\nonumber\\
		&\quad+\frac{1}{4}\int d^4x\,\begin{pmatrix}
			&\bar{\lambda}^a(x) &\lambda^a(x)
		\end{pmatrix} \mathcal{M}^{ab}_{\lambda A} \begin{pmatrix}
			&A^b(x)  &\\
			&\bar{A}^b(x) & 
		\end{pmatrix}\nonumber\\
		&\quad+\frac{1}{4}\int d^4x\,\begin{pmatrix}
			&\bar{A}^a(x)&A^a(x) 
		\end{pmatrix} \mathcal{M}_{A\lambda}^{ab} \begin{pmatrix}
			&\lambda^b(x)  &\\
			&\bar{\lambda}^b(x) & 
		\end{pmatrix}\Bigg)\,,
	\end{align}
	with
	\begin{align}
		\label{matrixM}
		& \mathcal{M}_{AA}^{ab}=\delta^{ab}\scalebox{0.78}{$
			\begin{pmatrix}i\square & -\frac{1}{\sqrt{2}}\sum_s\bar{J}_{S^A_{s}}\otimes\mathcal{Y}_{s-2}^{\frac{5}{2}}\\ -\frac{1}{\sqrt{2}}\sum_s J_{\bar{S}^A_s}\otimes\mathcal{Y}_{s-2}^{\frac{5}{2}}&i\square \end{pmatrix}$}\,,\qquad\qquad \mathcal{M}_{\lambda\lambda}^{ab}=\delta^{ab}\scalebox{0.75}{$
			\begin{pmatrix}-\square\partial_+^{-1}& -\frac{1}{\sqrt{2}}\sum_s\bar{J}_{S^\lambda_{s}}\otimes\mathcal{Y}_{s-1}^{\frac{3}{2}}\\ -\frac{1}{\sqrt{2}} \sum_sJ_{\bar{S}^\lambda_s}\otimes\mathcal{Y}_{s-1}^{\frac{3}{2}}&-\square\partial_+^{-1}  \end{pmatrix}$}\,,\nonumber\\\nonumber\\
		& \mathcal{M}_{A\lambda}^{ab}=\frac{\delta^{ab}}{2}\scalebox{0.8}{$
			\begin{pmatrix}\sum_s\bar{J}_{T_s}\otimes\mathcal{G}_{s-1}^{(2,1)}(-1)^{s-1} & 0\\ 0&- \sum_sJ_{\bar{T}_s}\otimes\mathcal{G}_{s-1}^{(2,1)}\end{pmatrix}$}\,,\qquad\qquad\mathcal{M}_{\lambda A}^{ab}=\frac{\delta^{ab}}{2}\scalebox{0.75}{$
			\begin{pmatrix} \sum_sJ_{\bar{T}_s} \otimes\mathcal{G}_{s-1}^{(1,2)}(-1)^{s-1}& 0\\ 
				0&-\sum_s\bar{J}_{T_s}\otimes  \mathcal{G}_{s-1}^{(1,2)} \end{pmatrix}$}\,,
	\end{align}
	where the symbol $\otimes$ indicates that the right and left derivatives do not act on the sources $J$.
	We can group these matrices into a single supermatrix
	\begin{align}
		\mathcal{X}^{ab} = 	\begin{pmatrix}\mathcal{M}^{ab}_{\lambda\lambda} & \mathcal{M}^{ab}_{\lambda A}\\ \mathcal{M}^{ab}_{A\lambda}&\mathcal{M}^{ab}_{AA} \end{pmatrix}\,,
	\end{align}
	so that the generating functional takes the form
	\begin{align}
		&\mathcal{Z}_{\text{conf}}\left[J_{\bar{S}^A},\bar{J}_{S^A},J_{\bar{S}^\lambda},\bar{J}_{S^\lambda}, J_{\bar{T}},\bar{J}_T\right]\nonumber\\& = \frac{1}{Z} \int \mathcal{D}A\mathcal{D}\bar{A} \mathcal{D} \lambda \mathcal{D} \bar{\lambda} 
		\exp\Bigg(-\frac{1}{2}\int d^4x\, \begin{pmatrix}
			\bar{\lambda}^a(x)&\lambda^a(x)&	\bar{A}^a(x) &A^a(x)
		\end{pmatrix} \mathcal{X}^{ab}\begin{pmatrix}
			&\lambda^b(x) &\\
			&\bar{\lambda}^b(x) & \\
			&A^b(x) &\\
			&\bar{A}^b(x) & 
		\end{pmatrix}\Bigg)
	\end{align}
	that reduces to the superdeterminant \cite{zinn1993quantum}
	\begin{align}
		\mathcal{Z}_{\text{conf}}\left[J_{\bar{S}^A},\bar{J}_{S^A},J_{\bar{S}^\lambda},\bar{J}_{S^\lambda}, J_{\bar{T}},\bar{J}_T\right] = \sDet^{\frac{1}{2}}\left(\mathcal{X}\right)\,.
	\end{align}
	More explicitly, this gives
	\begin{align}
		\label{superdet}
		\mathcal{Z}_{\text{conf}}\left[J_{\bar{S}^A},\bar{J}_{S^A},J_{\bar{S}^\lambda},\bar{J}_{S^\lambda}, J_{\bar{T}},\bar{J}_T\right]
		&= \Det^{\frac{1}{2}}(\mathcal{M}_{\lambda\lambda} )\Det^{-\frac{1}{2}}\left(\mathcal{M}_{AA}-\mathcal{M}_{A\lambda}\mathcal{M}_{\lambda\lambda}^{-1}\mathcal{M}_{\lambda A}\right)\nonumber\\
		&=\Det^{\frac{1}{2}}(\mathcal{M}_{\lambda\lambda} )\Det^{-\frac{1}{2}}(\mathcal{M}_{AA})\Det^{-\frac{1}{2}}\left(\mathcal{I}-\mathcal{M}_{AA}^{-1}\mathcal{M}_{A\lambda}\mathcal{M}_{\lambda\lambda}^{-1}\mathcal{M}_{\lambda A}\right)\nonumber\\
		&= \Det^{-\frac{1}{2}}(\mathcal{M}_{AA})\Det^{\frac{1}{2}}\left(\mathcal{M}_{\lambda\lambda}-\mathcal{M}_{\lambda A }\mathcal{M}_{AA}^{-1}\mathcal{M}_{A\lambda }\right)\nonumber\\
		&= \Det^{-\frac{1}{2}}(\mathcal{M}_{AA})\Det^{\frac{1}{2}}(\mathcal{M}_{\lambda\lambda})\Det^{\frac{1}{2}}\left(\mathcal{I}-\mathcal{M}_{\lambda\lambda}^{-1}\mathcal{M}_{\lambda A }\mathcal{M}_{AA}^{-1}\mathcal{M}_{A\lambda }\right)\,,
	\end{align}
	where the two different results depend on whether one first integrates over the fermionic or the bosonic variables. \par
	For the first factors in the first and third equalities above, we find the relatively straightforward expressions
	\begin{align}
		&\Det^{\frac{1}{2}}(\mathcal{M}_{\lambda\lambda}) =\Det^{\frac{1}{2}}\Bigg(\mathcal{I}-\frac{1}{2} \partial_+\square^{-1}\bar{J}_{S^\lambda_{s_1}} \otimes\mathcal{Y}_{s_1-1}^{\frac{3}{2}} \partial_+\square^{-1}J_{\bar{S}^\lambda_{s_2}} \otimes\mathcal{Y}_{s_2-1}^{\frac{3}{2}} \Bigg)\nonumber\\
	\end{align}
	and
	\begin{align}
		&\Det^{-\frac{1}{2}}(\mathcal{M}_{AA}) =\Det^{-\frac{1}{2}}\Bigg(\mathcal{I}-\frac{1}{2} i\square^{-1}\bar{J}_{S^A_{s_1}} \otimes\mathcal{Y}_{s_1-2}^{\frac{5}{2}} i\square^{-1}J_{\bar{S}^A_{s_2}} \otimes\mathcal{Y}_{s_2-2}^{\frac{5}{2}} \Bigg)\,,
	\end{align}
	respectively, where $\mathcal{I}$ is the identity in both color and space-time, and summation over repeated spin indices is implied.\par
	Calculating the remaining two factors is significantly more complex. In this paper, we will explicitly compute only $\Det^{-\frac{1}{2}}\left(\mathcal{I}-\mathcal{M}_{AA}^{-1}\mathcal{M}_{A\lambda}\mathcal{M}_{\lambda\lambda}^{-1}\mathcal{M}_{\lambda A}\right)$ from the second line of Eq. \eqref{superdet}, as the other result in the fourth line is equivalent. We define, by convention,
	\begin{align}
		\mathcal{I}-\mathcal{M}_{AA}^{-1}\mathcal{M}_{A\lambda}\mathcal{M}_{\lambda\lambda}^{-1}\mathcal{M}_{\lambda A} = \mathcal{Q}=	\begin{pmatrix}\mathcal{Q}_{11}& \mathcal{Q}_{12}\\ \mathcal{Q}_{21}&\mathcal{Q}_{22} \end{pmatrix}\,.
	\end{align}
	The diagonal entries are given by
	\begin{align}
		\mathcal{Q}_{11} &=\mathcal{I}-\frac{1}{4}\Big(\mathcal{I}-\frac{1}{2} i\square^{-1}\bar{J}_{S^A_{s_1}} \otimes\mathcal{Y}_{s_1-2}^{\frac{5}{2}} i\square^{-1}J_{\bar{S}^A_{s_2}} \otimes\mathcal{Y}_{s_2-2}^{\frac{5}{2}} \Big)^{-1}i\square^{-1}\bar{J}_{T_{s_3}}\otimes\mathcal{G}_{s_3-1}^{(2,1)}(-1)^{s_3-1}\nonumber\\
		&\quad\Big(\mathcal{I}-\frac{1}{2} \partial_+\square^{-1}\bar{J}_{S^\lambda_{s_4}} \otimes\mathcal{Y}_{s_4-1}^{\frac{3}{2}} \partial_+\square^{-1}J_{\bar{S}^\lambda_{s_5}} \otimes\mathcal{Y}_{s_5-1}^{\frac{3}{2}} \Big)^{-1}\partial_+\square^{-1}J_{\bar{T}_{s_6}} \otimes\mathcal{G}_{s_6-1}^{(1,2)}(-1)^{s_6-1}\nonumber\\
		&+\frac{1}{8}\Big(\mathcal{I}-\frac{1}{2} i\square^{-1}\bar{J}_{S^A_{s_1}} \otimes\mathcal{Y}_{s_1-2}^{\frac{5}{2}} i\square^{-1}J_{\bar{S}^A_{s_2}} \otimes\mathcal{Y}_{s_2-2}^{\frac{5}{2}} \Big)^{-1}i\square^{-1}\bar{J}_{S^A_{s_3}} \otimes\mathcal{Y}_{s_3-2}^{\frac{5}{2}}i\square^{-1}J_{\bar{T}_{s_4}}\otimes\mathcal{G}_{s_4-1}^{(2,1)}\nonumber\\
		&\quad\Big(\mathcal{I}-\frac{1}{2} \partial_+\square^{-1}\bar{J}_{S^\lambda_{s_5}} \otimes\mathcal{Y}_{s_5-1}^{\frac{3}{2}} \partial_+\square^{-1}J_{\bar{S}^\lambda_{s_6}} \otimes\mathcal{Y}_{s_6-1}^{\frac{3}{2}} \Big)^{-1}\partial_+\square^{-1}J_{\bar{S}^\lambda_{s_7}} \otimes\mathcal{Y}_{s_7-1}^{\frac{3}{2}}\nonumber\\
		&\quad\partial_+\square^{-1}J_{\bar{T}_{s_8}} \otimes\mathcal{G}_{s_8-1}^{(1,2)}(-1)^{s_8-1}
	\end{align}
	and
	\begin{align}
		\mathcal{Q}_{22} =
		&\mathcal{I}-\frac{1}{4}\Big(\mathcal{I}-\frac{1}{2} i\square^{-1}J_{\bar{S}^A_{s_1}} \otimes\mathcal{Y}_{s_1-2}^{\frac{5}{2}}i\square^{-1}\bar{J}_{S^A_{s_2}} \otimes\mathcal{Y}_{s_2-2}^{\frac{5}{2}}  \Big)^{-1}i\square^{-1}J_{\bar{T}_{s_3}}\otimes\mathcal{G}_{s_3-1}^{(2,1)}\nonumber\\
		&\quad\Big(\mathcal{I}-\frac{1}{2} \partial_+\square^{-1}J_{\bar{S}^\lambda_{s_4}} \otimes\mathcal{Y}_{s_4-1}^{\frac{3}{2}} \partial_+\square^{-1}\bar{J}_{S^\lambda_{s_5}} \otimes\mathcal{Y}_{s_5-1}^{\frac{3}{2}} \Big)^{-1}\partial_+\square^{-1}\bar{J}_{T_{s_6}}\otimes  \mathcal{G}_{s_6-1}^{(1,2)} \nonumber\\
		&+\frac{1}{8}\Big(\mathcal{I}-\frac{1}{2} i\square^{-1}J_{\bar{S}^A_{s_1}} \otimes\mathcal{Y}_{s_1-2}^{\frac{5}{2}}i\square^{-1}\bar{J}_{S^A_{s_2}} \otimes\mathcal{Y}_{s_2-2}^{\frac{5}{2}}  \Big)^{-1}i\square^{-1}J_{\bar{S}^A_{s_3}} \otimes\mathcal{Y}_{s_3-2}^{\frac{5}{2}}\nonumber\\
		&\quad i\square^{-1}\bar{J}_{T_{s_4}}\otimes\mathcal{G}_{s_4-1}^{(2,1)}(-1)^{s_4-1}\nonumber\\
		&\quad\Big(\mathcal{I}-\frac{1}{2} \partial_+\square^{-1}J_{\bar{S}^\lambda_{s_5}} \otimes\mathcal{Y}_{s_5-1}^{\frac{3}{2}} \partial_+\square^{-1}\bar{J}_{S^\lambda_{s_6}} \otimes\mathcal{Y}_{s_6-1}^{\frac{3}{2}} \Big)^{-1}\partial_+\square^{-1}J_{\bar{S}^\lambda_{s_7}} \otimes\mathcal{Y}_{s_7-1}^{\frac{3}{2}}\nonumber\\
		&\quad\partial_+\square^{-1}\bar{J}_{T_{s_8}}\otimes  \mathcal{G}_{s_8-1}^{(1,2)}\,.
	\end{align}
	The off-diagonal entries are
	\begin{align}
		\mathcal{Q}_{12}=
		&\frac{1}{4\sqrt{2}}\Big(\mathcal{I}-\frac{1}{2} i\square^{-1}J_{\bar{S}^A_{s_1}} \otimes\mathcal{Y}_{s_1-2}^{\frac{5}{2}}i\square^{-1}\bar{J}_{S^A_{s_2}} \otimes\mathcal{Y}_{s_2-2}^{\frac{5}{2}}  \Big)^{-1}i\square^{-1}J_{\bar{T}_{s_3}}\otimes\mathcal{G}_{s_3-1}^{(2,1)}\nonumber\\
		&\quad\Big(\mathcal{I}-\frac{1}{2} \partial_+\square^{-1}J_{\bar{S}^\lambda_{s_4}} \otimes\mathcal{Y}_{s_4-1}^{\frac{3}{2}} \partial_+\square^{-1}\bar{J}_{S^\lambda_{s_5}} \otimes\mathcal{Y}_{s_5-1}^{\frac{3}{2}} \Big)^{-1}\partial_+\square^{-1}J_{\bar{S}^\lambda_{s_6}} \otimes\mathcal{Y}_{s_6-1}^{\frac{3}{2}}\nonumber\\
		&\quad\partial_+\square^{-1}J_{\bar{T}_{s_7}} \otimes\mathcal{G}_{s_7-1}^{(1,2)}(-1)^{s_7-1}\nonumber\\
		&-\frac{1}{4\sqrt{2}}\Big(\mathcal{I}-\frac{1}{2} i\square^{-1}J_{\bar{S}^A_{s_1}} \otimes\mathcal{Y}_{s_1-2}^{\frac{5}{2}}i\square^{-1}\bar{J}_{S^A_{s_2}} \otimes\mathcal{Y}_{s_2-2}^{\frac{5}{2}}  \Big)^{-1}i\square^{-1}J_{\bar{S}^A_{s_3}} \otimes\mathcal{Y}_{s_3-2}^{\frac{5}{2}}\nonumber\\
		&\quad  i\square^{-1}\bar{J}_{T_{s_4}}\otimes\mathcal{G}_{s_4-1}^{(2,1)}(-1)^{s_4-1}\nonumber\\
		&\quad\Big(\mathcal{I}-\frac{1}{2} \partial_+\square^{-1}J_{\bar{S}^\lambda_{s_5}} \otimes\mathcal{Y}_{s_5-1}^{\frac{3}{2}} \partial_+\square^{-1}\bar{J}_{S^\lambda_{s_6}} \otimes\mathcal{Y}_{s_6-1}^{\frac{3}{2}} \Big)^{-1}\partial_{+}\square^{-1}J_{\bar{T}_{s_7}} \otimes\mathcal{G}_{s_7-1}^{(1,2)}(-1)^{s_7-1}
	\end{align}
	and
	\begin{align}
		\mathcal{Q}_{21} &=\frac{1}{4\sqrt{2}}\Big(\mathcal{I}-\frac{1}{2} i\square^{-1}\bar{J}_{S^A_{s_1}} \otimes\mathcal{Y}_{s_1-2}^{\frac{5}{2}} i\square^{-1}J_{\bar{S}^A_{s_2}} \otimes\mathcal{Y}_{s_2-2}^{\frac{5}{2}} \Big)^{-1}i\square^{-1}\bar{J}_{T_{s_3}}\otimes\mathcal{G}_{s_3-1}^{(2,1)}(-1)^{s_3-1}\nonumber\\
		&\quad\Big(\mathcal{I}-\frac{1}{2} \partial_+\square^{-1}\bar{J}_{S^\lambda_{s_4}} \otimes\mathcal{Y}_{s_4-1}^{\frac{3}{2}} \partial_+\square^{-1}J_{\bar{S}^\lambda_{s_5}} \otimes\mathcal{Y}_{s_5-1}^{\frac{3}{2}} \Big)^{-1}\partial_+\square^{-1}\bar{J}_{S^\lambda_{s_6}} \otimes\mathcal{Y}_{s_6-1}^{\frac{3}{2}}\nonumber\\
		&\quad\partial_+\square^{-1}\bar{J}_{T_{s_7}}\otimes  \mathcal{G}_{s_7-1}^{(1,2)} \nonumber\\
		&-\frac{1}{4\sqrt{2}}\Big(\mathcal{I}-\frac{1}{2} i\square^{-1}\bar{J}_{S^A_{s_1}} \otimes\mathcal{Y}_{s_1-2}^{\frac{5}{2}} i\square^{-1}J_{\bar{S}^A_{s_2}} \otimes\mathcal{Y}_{s_2-2}^{\frac{5}{2}} \Big)^{-1}i\square^{-1}\bar{J}_{S^A_{s_3}} \otimes\mathcal{Y}_{s_3-2}^{\frac{5}{2}}\nonumber\\
		&\quad  i\square^{-1}J_{\bar{T}_{s_4}}\otimes\mathcal{G}_{s_4-1}^{(2,1)}\nonumber\\
		&\quad \Big(\mathcal{I}-\frac{1}{2} \partial_+\square^{-1}\bar{J}_{S^\lambda_{s_5}} \otimes\mathcal{Y}_{s_5-1}^{\frac{3}{2}} \partial_+\square^{-1}J_{\bar{S}^\lambda_{s_6}} \otimes\mathcal{Y}_{s_6-1}^{\frac{3}{2}} \Big)^{-1}\partial_+\square^{-1}\bar{J}_{T_{s_7}}\otimes  \mathcal{G}_{s_7-1}^{(1,2)}\,.
	\end{align}
	From the determinant of a block matrix \cite{BPSpaper2}, 
	\begin{equation}
		\Det\mathcal{Q}=
		\Det\left(\mathcal{Q}_{22}\right)
		\Det\left(\mathcal{Q}_{11}-\mathcal{Q}_{12}\mathcal{Q}_{22}^{-1}\mathcal{Q}_{21}\right)
	\end{equation}
	we find
	\begin{align}
		\Det^{-\frac{1}{2}}\mathcal{Q}&=\Det^{-\frac{1}{2}}\left(\mathcal{I}-\mathcal{M}_{AA}^{-1}\mathcal{M}_{A\lambda}\mathcal{M}_{\lambda\lambda}^{-1}\mathcal{M}_{\lambda A}\right)\nonumber\\
		&=
		\Det^{-\frac{1}{2}}\left(\mathcal{Q}_{11}\right)
		\Det^{-\frac{1}{2}}\left(\mathcal{Q}_{22}\right)\Det^{-\frac{1}{2}}\left(\mathcal{I}-\mathcal{Q}_{11}^{-1}\mathcal{Q}_{12}\mathcal{Q}_{22}^{-1}\mathcal{Q}_{21}\right)\,.
	\end{align}
	Consequently, the connected generating functional is
	\begin{align}
		\label{wgen1}
		\mathcal{W}_{\text{conf}}\left[\bar{J}_{S^A},J_{\bar{S}^A},\bar{J}_{S^\lambda},J_{\bar{S}^\lambda},\bar{J}_{T},J_{\bar T}\right] 
		&=-\frac{1}{2}\log\Det \Bigg(\mathcal{I}-\frac{1}{2} i\square^{-1}\bar{J}_{S^A_{s_1}} \otimes\mathcal{Y}_{s_1-2}^{\frac{5}{2}} i\square^{-1}J_{\bar{S}^A_{s_2}} \otimes\mathcal{Y}_{s_2-2}^{\frac{5}{2}} \Bigg) \nonumber\\
		&\quad+\frac{1}{2}\log\Det\Bigg(\mathcal{I}-\frac{1}{2} \partial_+\square^{-1}\bar{J}_{S^\lambda_{s_1}} \otimes\mathcal{Y}_{s_1-1}^{\frac{3}{2}} \partial_+\square^{-1}J_{\bar{S}^\lambda_{s_2}} \otimes\mathcal{Y}_{s_2-1}^{\frac{3}{2}} \Bigg) \nonumber\\
		&\quad-\frac{1}{2}\log\Det\left(\mathcal{Q}_{11}\right)-\frac{1}{2}\log\Det\left(\mathcal{Q}_{22}\right)\nonumber\\
		&\quad-\frac{1}{2}\log\Det\left(\mathcal{I}-\mathcal{Q}_{11}^{-1}\mathcal{Q}_{12}\mathcal{Q}_{22}^{-1}\mathcal{Q}_{21}\right)\,.
	\end{align}
	For clarity, we display the generating functional separately for the bosonic sector,
	\begin{align}
		\mathcal{W}_{\text{conf}}\left[\bar{J}_{S^A},J_{\bar{S}^A},\bar{J}_{S^\lambda},J_{\bar{S}^\lambda},0,0\right] 
		&=-\frac{1}{2}\log\Det\Bigg(\mathcal{I}-\frac{1}{2} i\square^{-1}\bar{J}_{S^A_{s_1}} \otimes\mathcal{Y}_{s_1-2}^{\frac{5}{2}} i\square^{-1}J_{\bar{S}^A_{s_2}} \otimes\mathcal{Y}_{s_2-2}^{\frac{5}{2}} \Bigg)\nonumber\\
		&\quad+\frac{1}{2}\log\Det\Bigg(\mathcal{I}-\frac{1}{2} \partial_+\square^{-1}\bar{J}_{S^\lambda_{s_1}} \otimes\mathcal{Y}_{s_1-1}^{\frac{3}{2}} \partial_+\square^{-1}J_{\bar{S}^\lambda_{s_2}} \otimes\mathcal{Y}_{s_2-1}^{\frac{3}{2}} \Bigg)
	\end{align}
	and for the fermionic sector, where $\mathcal{Q}_{12}$ and $\mathcal{Q}_{21}$ vanish, and $\mathcal{Q}_{11}$ and $\mathcal{Q}_{22}$ simplify considerably
	\begin{align}
		\mathcal{W}_{\text{conf}}\left[0,0,0,0,\bar{J}_{T},J_{\bar T}\right] 
		&=-\frac{1}{2}\log\Det \Bigg(\mathcal{I}-\frac{1}{4}i\square^{-1}\bar{J}_{T_{s_1}}\otimes\mathcal{G}_{s_1-1}^{(2,1)}(-1)^{s_1-1}\partial_+\square^{-1}J_{\bar{T}_{s_2}} \otimes\mathcal{G}_{s_2-1}^{(1,2)}(-1)^{s_2-1}\Bigg)\nonumber\\
		&\quad-\frac{1}{2}\log\Det\Bigg(\mathcal{I}-\frac{1}{4}i\square^{-1}J_{\bar{T}_{s_1}}\otimes\mathcal{G}_{s_1-1}^{(2,1)}\partial_+\square^{-1}\bar{J}_{T_{s_2}}\otimes  \mathcal{G}_{s_2-1}^{(1,2)} \Bigg)\,.
	\end{align}
	These two determinants are in fact identical \cite{BPS41}, leading to
	\begin{align}
		\label{wgenT1}
		&\mathcal{W}_{\text{conf}}\left[0,0,0,0,\bar{J}_{T},J_{\bar T}\right] =-\log\Det\Bigg(\mathcal{I}-\frac{1}{4}i\square^{-1}J_{\bar{T}_{s_1}}\otimes\mathcal{G}_{s_1-1}^{(2,1)}\partial_+\square^{-1}\bar{J}_{T_{s_2}}\otimes  \mathcal{G}_{s_2-1}^{(1,2)} \Bigg)\,.
	\end{align}
	If we instead use the fourth line of Eq. \eqref{superdet}, we equivalently get
	\begin{align}
		\label{wgenT2}
		&\mathcal{W}_{\text{conf}}\left[0,0,0,0,\bar{J}_{T},J_{\bar T}\right] =\log\Det\Bigg(\mathcal{I}-\frac{1}{4}\partial_+\square^{-1}\bar{J}_{T_{s_1}}\otimes  \mathcal{G}_{s_1-1}^{(1,2)} i\square^{-1}J_{\bar{T}_{s_2}}\otimes\mathcal{G}_{s_2-1}^{(2,1)}\Bigg)\,.
	\end{align}
	After rescaling the operators
	\begin{align}
		\label{rescale}
		&S^{'A}_s(x)=\frac{1}{N}\frac{2\Gamma(5)\Gamma(s+1)}{\Gamma(3)\Gamma(s+3)}S^A_s(x)\nonumber\\
		&\bar{S}^{'A}_s(x)=\frac{1}{N}\frac{2\Gamma(5)\Gamma(s+1)}{\Gamma(3)\Gamma(s+3)}	\bar{S}^A_s(x)\nonumber\\
		&S^{'\lambda}_s(x)=\frac{1}{N}\frac{4}{s+1}	S^\lambda_s(x)\nonumber\\
		&\bar{S}^{'\lambda}_s(x)=\frac{1}{N}\frac{4}{s+1}	\bar{S}^\lambda_s(x)\nonumber\\
		&T'_s = \frac{2}{N}T_s\,,
	\end{align}
	so that their $2$-point correlators are of order $1$ for large $N$, we obtain the more explicit forms
	\begin{align}
		\label{wexplicit}
		&{\mathcal{W}_{\text{conf}}}\left[\bar{J}_{S^{'A}},J_{\bar{S}^{'A}},\bar{J}_{S^{'\lambda}},J_{\bar{S}^{'\lambda}},0,0\right]=\nonumber\\
		&-\frac{N^2-1}{2}\log\Det\Bigg(I-\frac{2}{N^2}\sum_{k_1=0}^{s_1-2}{s_1\choose k_1}{s_1\choose k_1+2}(i\overrightarrow{\partial}_+)^{s_1-k_1-1}i \square^{-1}J_{\bar{S}^{'A}_{s_1}}(i\overrightarrow{\partial}_+)^{k_1+1}\nonumber\\ 
		&\quad\sum_{k_2=0}^{s_2-2}{s_2\choose k_2}{s_2\choose k_2+2}(i\overrightarrow{\partial}_+)^{s_2-k_2-1}i \square^{-1}\bar{J}_{S^{'A}_{s_2}}(i\overrightarrow{\partial}_+)^{k_2+1}\Bigg)\nonumber\\
		&+\frac{N^2-1}{2}\log\Det\Bigg(I-\frac{2}{N^2}\sum_{k_1=0}^{s_1-1}{s_1\choose k_1}{s_1\choose k_1+1}(i\overrightarrow{\partial}_+)^{s_1-k_1-1}i\partial_+ i\square^{-1}J_{\bar{S}^{'\lambda}_{s_1}}(i\overrightarrow{\partial}_+)^{k_1}\nonumber\\ 
		&\quad\sum_{k_2=0}^{s_2-1}{s_2\choose k_2}{s_2\choose k_2+1}(i\overrightarrow{\partial}_+)^{s_2-k_2-1}i\partial_+ i\square^{-1}\bar{J}_{S^{'\lambda}_{s_2}}(i\overrightarrow{\partial}_+)^{k_2}\Bigg)
	\end{align}
	and
	\begin{align}
		\label{wexplicitT1}
		&\mathcal{W}_{\text{conf}}\left[0,0,0,0,\bar{J}_{T'},J_{\bar T'}\right] =\nonumber\\
		&	-(N^2-1)\log\Det\Bigg(I+\frac{1}{N^2}
		\sum_{k_1 = 0}^{s_1-1}{s_1+1\choose k_1}{s_1\choose k_1+1} 
		(i\overrightarrow{\partial}_+)^{s_1-k_1}i\square^{-1}J_{\bar T'_{s_1}} (i\overrightarrow{\partial}_+)^{k_1}   \nonumber\\
		&\quad\sum_{k_2 = 0}^{s_2-1}{s_2\choose k_2}{s_2+1\choose k_2+2} 
		(i\overrightarrow{\partial}_+)^{s_2-k_2-1}i\partial_+i\square^{-1}\bar{J}_{T'_{s_2}} (i\overrightarrow{\partial}_+)^{k_2+1}\Bigg)\,,
	\end{align}
	where $I$ is the identity in space-time with the kernel
	\begin{align}
		I \rightarrow \delta^{(4)}(x-y)
	\end{align}
	and, after calculating the color trace, we have used the definitions in Eqs. \eqref{defY}, \eqref{gdef} and \cite{BPSpaper2}
	\begin{align}
		\label{arrowkey}
		i\square^{-1}\overleftarrow{\partial}_{+}^{s-k-1} = (-1)^{s-k-1}\overrightarrow{\partial}_{+}^{s-k-1} i\square^{-1}
	\end{align}
	which arises from (minus) the propagator in coordinate representation \cite{BPS1}
	\begin{equation}
		\label{propagator}
		i \square^{-1} \rightarrow \frac{1}{4\pi^2}\frac{1}{\rvert x-y \rvert^2-i\epsilon}\,.
	\end{equation}

	\subsection{Connected generating functional $\Gamma_{\text{conf}}$ as the $\log$ of a Fredholm superdeterminant \label{kernel}}

	Using the correspondence in \cite{BPSpaper2}, we can rewrite the generating functional $\mathcal{W}_{\text{conf}}$—which is the logarithm of a (super)determinant of a quadratic form—as the logarithm of a (super)determinant $\Gamma_{\text{conf}}$ of integral operators that are formally of the Fredholm type
	\begin{align}
		\label{genMg1}
		\Gamma_{\text{conf}}\left[\bar{j}_{S^{A}},j_{\bar{S}^{A}},\bar{j}_{S^{\lambda}},j_{\bar{S}^{\lambda}},0,0\right] = &
		-\frac{N^2-1}{2}\log\Det \left[\mathbb{I}-2\mathcal{D}^{-1}_Aj_{\bar{S}^A}\mathcal{D}^{-1}_A\bar{j}_{S^A}\right]\nonumber\\
		&+\frac{N^2-1}{2}\log\Det \left[\mathbb{I}-2\mathcal{D}^{-1}_\lambda j_{\bar{S}^\lambda}\mathcal{D}^{-1}_\lambda \bar{j}_{S^\lambda}\right]
	\end{align}
	and
	\begin{align}
		\label{genMg2}
		\Gamma_{\text{conf}}\left[0,0,0,0,\bar{j}_{T},j_{\bar{T}}\right]
		= 
		-(N^2-1)\log\Det\left[\mathbb{I}-\mathcal{D}_{\bar{T}}^{-1}j_{\bar T}\mathcal{D}^{-1}_T \bar j_{T} \right]\,,
	\end{align}
	where $\mathbb{I}$ represents the identity in both space-time and the discrete indices defined below. The corresponding kernels are:
	\begin{itemize}
		\item \underline{gluon-gluon  kernel}
		\subitem 
		\begin{align}
			\label{kernelG}
			&\mathcal{D}^{-1}_{A\,s_1k_1,s_2k_2}=\frac{1}{2}\frac{\Gamma(3)\Gamma(s_1+3)}{\Gamma(5)\Gamma(s_1+1)}{s_1\choose k_1}{s_2\choose k_2+2}(-i\partial_{+})^{s_1-k_1+k_2}i\square^{-1}\nonumber\\ &\rightarrow	\mathcal{D}^{-1}_{A\,s_1k_1,s_2k_2}(x-y) =\frac{1}{8\pi^2}\frac{\Gamma(3)\Gamma(s_1+3)}{\Gamma(5)\Gamma(s_1+1)}{s_1\choose k_1}{s_2\choose k_2+2}(-i\partial_{+})^{s_1-k_1+k_2}\frac{1}{\rvert x-y\rvert^2-i\epsilon}
		\end{align}
		\item \underline{gluino-gluino  kernel}
		\subitem
		\begin{align}
			\label{kernelL}
			&\mathcal{D}^{-1}_{\lambda\,s_1k_1,s_2k_2}=\frac{1}{2}\frac{s_1+1}{2}{s_1\choose k_1}{s_2\choose k_2+1}(-i\partial_{+})^{s_1-k_1+k_2-1}(-i\partial_+)i\square^{-1}\nonumber\\ &\rightarrow\mathcal{D}^{-1}_{\lambda\,s_1k_1,s_2k_2}(x-y) =\frac{1}{8\pi^2}\frac{s_1+1}{2}{s_1\choose k_1}{s_2\choose k_2+1}(-i\partial_{+})^{s_1-k_1+k_2-1}(-i\partial_+)\frac{1}{\rvert x-y\rvert^2-i\epsilon}
		\end{align}
		\item \underline{gluon-gluino kernels}
		\begin{align}
			&\mathcal{D}^{-1}_{T\,s_1k_1,s_2k_2}=-\frac{i}{2}{s_1+1\choose k_1+2}{s_2+1\choose k_2}(-i\partial_{+})^{s_1-k_1+k_2}(-i\partial_+)i\square^{-1}\nonumber\\ &\rightarrow	\mathcal{D}^{-1}_{T\,s_1k_1,s_2k_2}(x-y) =-\frac{i}{8\pi^2}{s_1+1\choose k_1+2}{s_2+1\choose k_2}(-i\partial_{+})^{s_1-k_1+k_2}(-i\partial_+)\frac{1}{\rvert x-y\rvert^2-i\epsilon}\\
			&\vspace{5cm}\nonumber\\
			&\mathcal{D}^{-1}_{\bar{T}\,s_1k_1,s_2k_2}=-\frac{i}{2}{s_1\choose k_1}{s_2\choose k_2+1}(-i\partial_{+})^{s_1-k_1+k_2}i\square^{-1}\nonumber\\ &\rightarrow \mathcal{D}^{-1}_{\bar{T}\,s_1k_1,s_2k_2}(x-y) =-\frac{i}{8\pi^2}{s_1\choose k_1}{s_2\choose k_2+1}(-i\partial_{+})^{s_1-k_1+k_2}\frac{1}{\rvert x-y\rvert^2-i\epsilon}\,.
		\end{align}
	\end{itemize}
	These kernels are coupled to the currents $j_{\mathcal{O}_{sk}}$, which are dual to the component operators $\mathcal{O}_{sk}$ used in constructing the conformal operators $\mathcal{O}_s$ \cite{BPS1}
	\begin{equation}
		\mathcal{O}_s = \sum_{k=0}^{l}\mathcal{O}_{sk}\,.
	\end{equation}
	Thus, the $n$-point correlators can be written as \cite{BPSpaper2}
	\begin{align}
		\label{gammacorr}
		\langle \mathcal{O}_{s_1}(x_1)\ldots\mathcal{O}_{s_n}(x_n)\rangle
		=\sum_{k_1 = 0}^{l_1} \frac{\delta}{\delta j_{\mathcal{O}_{s_1k_1}}(x_1)}\cdots\sum_{k_n = 0}^{l_n}\frac{\delta}{\delta j_{\mathcal{O}_{s_nk_n}}(x_n)} {\Gamma_{\text{conf}}}[j_{\mathcal{O}}]
	\end{align}
	and the following identity is satisfied
	\begin{align} \label{73}
		\sum_{k_1 = 0}^{l_1} \frac{\delta}{\delta j_{\mathcal{O}_{s_1k_1}}(x_1)}\cdots\sum_{k_n = 0}^{l_n}\frac{\delta}{\delta j_{\mathcal{O}_{s_nk_n}}(x_n)} {\Gamma_{\text{conf}}}[j_{\mathcal{O}}]
		= \frac{\delta}{\delta J_{\mathcal{O}_{s_1}}(x_1)}\cdots\frac{\delta}{\delta J_{\mathcal{O}_{s_n}}(x_n)} {\mathcal{W}_{\text{conf}}}[J_{\mathcal{O}}]
	\end{align}
	based on the correspondence in \cite{BPSpaper2}. Equivalently, as per the above equation, $\Gamma_{\text{conf}}[j_{\mathcal{O}_{sk}}]$ coincides with $\mathcal{W}_{\text{conf}}[J_{\mathcal{O}_{s}}]$ through the identification
	\begin{equation} \label{74}
		j_{\mathcal{O}_{sk}} = J_{\mathcal{O}_{s}}
	\end{equation}
	for every $k$.
	
	\section{Generating functional of Euclidean conformal correlators}\label{euclgen}

	\subsection{Analytic continuation to Euclidean space-time}

	The Minkowskian correlators can be analytically continued to Euclidean space-time \cite{BPS1,BPSpaper2,BPS41} through the substitutions
	\begin{equation} \label{ab}
		x_+\rightarrow -i x_{z}
	\end{equation}
	and
	\begin{equation} \label{bc}
		\frac{1}{\rvert x \rvert^2-i\epsilon}\rightarrow-\frac{1}{x^2}\,.
	\end{equation}
	The analytically continued operators are:
	\begin{itemize}
		\item \underline{gluon-gluon operators}
		\subitem 
		\begin{equation}
			S^A_s \rightarrow \frac{1}{2\sqrt{2}}(-1)^{s+1} \partial_z \bar{A}^{E\,a}(\overrightarrow{\partial_z} + \overleftarrow{\partial_z})^{s-2}C^{\frac{5}{2}}_{s-2}\Bigg(\frac{\overrightarrow{\partial_z} - \overleftarrow{\partial_z}}{\overrightarrow{\partial_z}+\overleftarrow{\partial_z}}\Bigg)\partial_z \bar{A}^{E\,a} = S^{A\,E}_s \hspace{1cm} 
		\end{equation}
		\item \underline{gluino-gluino operators}
		\subitem \begin{equation}
			O^\lambda_s \rightarrow  \frac{1}{2\sqrt{2}} (-1)^{s-1}\bar{\lambda}^{E\,a}(\overrightarrow{\partial_z} + \overleftarrow{\partial_z})^{s-1}C^{\frac{3}{2}}_{s-1}\Bigg(\frac{\overrightarrow{\partial_z} - \overleftarrow{\partial_z}}{\overrightarrow{\partial_z}+\overleftarrow{\partial_z}}\Bigg) \bar{\lambda}^{E\,a}=S^{\lambda\,E}_s\hspace{1cm} 
		\end{equation}
		\item \underline{gluon-gluino operators} 
		\begin{align}
			&T_s \rightarrow \frac{1}{2}i(-1)^{s-1}\hspace{0.08cm} \lambda^{E\,a} (\overrightarrow{\partial_z} + \overleftarrow{\partial_z})^{s-1}P^{(2,1)}_{s-1}\Bigg(\frac{\overrightarrow{\partial_z} - \overleftarrow{\partial_z}}{\overrightarrow{\partial_z}+\overleftarrow{\partial_z}}\Bigg) \partial_z\bar{A}^{E\,a} = T^E_s\hspace{1cm} \nonumber\\
			&\bar{T}_s \rightarrow  \hspace{0.08cm}\frac{1}{2}i(-1)^{s-1}\partial_zA^{E\,a}(\overrightarrow{\partial_z} + \overleftarrow{\partial_z})^{s-1}P^{(1,2)}_{s-1}\Bigg(\frac{\overrightarrow{\partial_z} - \overleftarrow{\partial_z}}{\overrightarrow{\partial_z}+\overleftarrow{\partial_z}}\Bigg) \bar{\lambda}^{E\,a}=\bar{T}^E_s   
		\end{align}
	\end{itemize}
	
	\subsection{Analytic continuation of $\mathcal{W}_{\text{conf}}$}
	
	Consequently, the Euclidean generating functionals are
	\begin{align}
		\label{wexplicitE}
		&{\mathcal{W}^E_{\text{conf}}}\left[\bar{J}_{S^{'A\,E}},J_{\bar{S}^{'A\,E}},\bar{J}_{S^{'\lambda\,E}},J_{\bar{S}^{'\lambda\,E}},0,0\right]\nonumber\\
		&=-\frac{N^2-1}{2}\log\Det\Bigg(I-\frac{2}{N^2}\sum_{k_1=0}^{s_1-2}{s_1\choose k_1}{s_1\choose k_1+2}(-\overrightarrow{\partial}_z)^{s_1-k_1-1}\Laplace^{-1}J_{\bar{S}^{'A\,E}_{s_1}}(-\overrightarrow{\partial}_z)^{k_1+1}\nonumber\\ 
		&\quad\sum_{k_2=0}^{s_2-2}{s_2\choose k_2}{s_2\choose k_2+2}(-\overrightarrow{\partial}_z)^{s_2-k_2-1}\Laplace^{-1}\bar{J}_{S^{'A\,E}_{s_2}}(-\overrightarrow{\partial}_z)^{k_2+1}\Bigg)\nonumber\\
		&+\frac{N^2-1}{2}\log\Det\Bigg(I-\frac{2}{N^2}\sum_{k_1=0}^{s_1-1}{s_1\choose k_1}{s_1\choose k_1+1}(-\overrightarrow{\partial}_z)^{s_1-k_1-1}\partial_z\Laplace^{-1}J_{\bar{S}^{'\lambda\,E}_{s_1}}(-\overrightarrow{\partial}_z)^{k_1}\nonumber\\ 
		&\quad\sum_{k_2=0}^{s_2-1}{s_2\choose k_2}{s_2\choose k_2+1}(-\overrightarrow{\partial}_z)^{s_2-k_2-1}\partial_z\Laplace^{-1}\bar{J}_{S^{'\lambda\,E}_{s_2}}(-\overrightarrow{\partial}_z)^{k_2}\Bigg)
	\end{align}
	and
	\begin{align}
		\label{wexplicitT1E}
		&\mathcal{W}^E_{\text{conf}}\left[0,0,0,0,\bar{J}_{T^{'E}},J_{\bar{T}^{'E}}\right] \nonumber\\
		&=-(N^2-1)\log\Det\Bigg(I+\frac{1}{N^2}
		\sum_{k_1 = 0}^{s_1-1}{s_1+1\choose k_1}{s_1\choose k_1+1} 
		(-\overrightarrow{\partial}_z)^{s_1-k_1}\Laplace^{-1}J_{\bar {T}^{'E}_{s_1}} (-\overrightarrow{\partial}_z)^{k_1}   \nonumber\\
		&\quad\sum_{k_2 = 0}^{s_2-1}{s_2\choose k_2}{s_2+1\choose k_2+2} 
		(-\overrightarrow{\partial}_z)^{s_2-k_2-1}(-\partial_z)\Laplace^{-1}\bar{J}_{{T}^{'E}_{s_2}} (-\overrightarrow{\partial}_z)^{k_2+1}\Bigg)\,,
	\end{align}
	with
	\begin{equation}
		\Laplace= \delta_{\mu\nu}\partial_{\mu}\partial_{\nu}=\partial_4^2+\sum_{i=1}^{3}\partial_i^2
	\end{equation}
	and
	\begin{equation}
		-\Laplace^{-1} \rightarrow \frac{1}{4\pi^2}\frac{1}{(x-y)^2} \,.
	\end{equation}

	\subsection{Analytic continuation of $\Gamma_{\text{conf}}$ \label{kernelE}}

	Correspondingly, by performing the analytic continuation of Eqs. \eqref{genMg1} and \eqref{genMg2}, we have
	\begin{align}
		\Gamma^E_{\text{conf}}\left[\bar{j}_{S^{A\,E}},j_{\bar{S}^{A\,E}},\bar{j}_{S^{\lambda\,E}},j_{\bar{S}^{\lambda\,E}},0,0\right] = 
		&-\frac{N^2-1}{2}\log\Det \left[\mathbb{I}-2\mathcal{D}^{-1}_{E\,A}j_{\bar{S}^{A\,E}}\mathcal{D}^{-1}_{E\,A}\bar{j}_{S^{A\,E}}\right]\nonumber\\
		&+\frac{N^2-1}{2}\log\Det \left[\mathbb{I}-2\mathcal{D}^{-1}_{E\,\lambda} j_{\bar{S}^{\lambda\,E}}\mathcal{D}^{-1}_{E\,\lambda}  \bar{j}_{S^{\lambda\,E}}\right]
	\end{align}
	and
	\begin{align}
		\Gamma^E_{\text{conf}}\left[0,0,0,0,\bar{j}_{T^E},j_{\bar{T}^E}\right]= 
		-(N^2-1)\log\Det\left[\mathbb{I}-\mathcal{D}_{E\,\bar{T}}^{-1} j_{\bar{T}^E}\mathcal{D}^{-1}_{E\,T} \bar{j}_{T^E}\right]\,,
	\end{align}
	with the following kernels: 
	\begin{itemize}
		\item \underline{gluon-gluon kernel}
		\subitem 
		\begin{align}
			\label{EkernelG}
			&\mathcal{D}^{-1}_{E\,A\,s_1k_1,s_2k_2}=\frac{1}{2}\frac{\Gamma(3)\Gamma(s_1+3)}{\Gamma(5)\Gamma(s_1+1)}{s_1\choose k_1}{s_2\choose k_2+2}\partial_{z}^{s_1-k_1+k_2}\Laplace^{-1}\nonumber\\ &\rightarrow 	\mathcal{D}^{-1}_{E\,A\,s_1k_1,s_2k_2}(x-y) =-\frac{1}{8\pi^2}\frac{\Gamma(3)\Gamma(s_1+3)}{\Gamma(5)\Gamma(s_1+1)}{s_1\choose k_1}{s_2\choose k_2+2}\partial_{z}^{s_1-k_1+k_2}\frac{1}{ (x-y)^2}
		\end{align}
		\item \underline{gluino-gluino kernel}
		\subitem
		\begin{align}
			\label{EkernelL}
			&\mathcal{D}^{-1}_{E\,\lambda\,s_1k_1,s_2k_2}=\frac{1}{2}\frac{s_1+1}{2}{s_1\choose k_1}{s_2\choose k_2+1}\partial_{z}^{s_1-k_1+k_2-1}\partial_z\Laplace^{-1}\nonumber\\ &\rightarrow \mathcal{D}^{-1}_{E\,\lambda\,s_1k_1,s_2k_2}(x-y)=-\frac{1}{8\pi^2}\frac{s_1+1}{2}{s_1\choose k_1}{s_2\choose k_2+1}\partial_{z}^{s_1-k_1+k_2-1}\partial_z\frac{1}{ (x-y)^2}
		\end{align}
		\item \underline{gluon-gluino kernels}
		\begin{align}
			&\mathcal{D}^{-1}_{E\,T\,s_1k_1,s_2k_2}=-\frac{i}{2}{s_1+1\choose k_1+2}{s_2+1\choose k_2}\partial_{z}^{s_1-k_1+k_2}\partial_z\Laplace^{-1}\nonumber\\ &\rightarrow \mathcal{D}^{-1}_{E\,T\,s_1k_1,s_2k_2}(x-y)=\frac{i}{8\pi^2}{s_1+1\choose k_1+2}{s_2+1\choose k_2}\partial_{z}^{s_1-k_1+k_2}\partial_z\frac{1}{ (x-y)^2}\\
			&\vspace{5cm}\nonumber\\
			&\mathcal{D}^{-1}_{E\,\bar{T}\,s_1k_1,s_2k_2}=-\frac{i}{2}{s_1\choose k_1}{s_2\choose k_2+1}\partial_{z}^{s_1-k_1+k_2}\Laplace^{-1}\nonumber\\ &\rightarrow \mathcal{D}^{-1}_{E\,\bar{T}\,s_1k_1,s_2k_2}(x-y)=\frac{i}{8\pi^2}{s_1\choose k_1}{s_2\choose k_2+1}\partial_{z}^{s_1-k_1+k_2}\frac{1}{ (x-y)^2}
		\end{align}
	\end{itemize}

	\section{RG-improvement of the Euclidean correlators}  \label{s0}
	
	\subsection{Operator mixing  \label{app:C1}}
	
	Here, we briefly outline our method for constructing RG-improved asymptotic correlators in a particular scheme, following \cite{BPSpaper2,MB1}.
	The renormalized Euclidean correlators
	\begin{equation}\label{key}
		\langle \mathcal{O}_{k_1}(x_1)\ldots \mathcal{O}_{k_n}(x_n) \rangle = G^{(n)}_{k_1 \ldots k_n}( x_1,\ldots,  x_n; \mu, g(\mu))
	\end{equation}
	obey the Callan-Symanzik equation
	\begin{align}\label{CallanSymanzik}
		& \Big(\sum_{\alpha = 1}^n x_\alpha \cdot \frac{\partial}{\partial x_\alpha} + \beta(g)\frac{\partial}{\partial g} +\sum_{\alpha = 1}^n D_{\mathcal{O}_\alpha}\Big)G^{(n)}_{k_1 \ldots k_n}  \nonumber\\
		&+ \sum_a \Big(\gamma_{k_1a}(g) G^{(n)}_{ak_2 \ldots k_n}+ \gamma_{k_2a}(g) G^{(n)}_{k_1 a k_3 \ldots k_n} \cdots +\gamma_{k_n a}(g) G^{(n)}_{k_1 \ldots a}\Big) = 0\,,
	\end{align}
	that  has the solution
	\begin{align}\label{csformula}
		G^{(n)}_{k_1 \ldots k_n}(\lambda x_1,\ldots, \lambda x_n; \mu, g(\mu)) = \sum_{j_1 \ldots j_n} Z_{k_1 j_1} (\lambda)\ldots Z_{k_n j_n}(\lambda)\hspace{0.1cm} \lambda^{-\sum_{i=1}^nD_{\mathcal{O}_{j_i}}} G^{(n)}_{j_1 \ldots j_n }( x_1, \ldots, x_n; \mu, g(\frac{\mu}{\lambda}))\,,
	\end{align}
	where $D_{\mathcal{O}_i}$ is the canonical dimension of $\mathcal{O}_i(x)$, and $\gamma(g)=\gamma_0 g^2+ \cdots$ is the matrix of anomalous dimensions, with
	\begin{equation}\label{eqZ-rg}
		\Bigg(\frac{\partial}{\partial g} + \frac{\gamma(g)}{\beta(g)}\Bigg)Z(\lambda) = 0
	\end{equation}
	in matrix form, and
	\begin{equation} \label{ZZ-rg}
		Z(\lambda) = P\exp \Big(\int_{g(\mu)}^{g(\frac{\mu}{\lambda})}\frac{\gamma(g')}{\beta(g')} dg'\Big)\,.
	\end{equation}
	Eq. \eqref{csformula} simplifies greatly if a renormalization scheme exists in which $Z(\lambda)$ can be diagonalized to all orders of perturbation theory \cite{MB1} and, more specifically, is one-loop exact with eigenvalues $Z_{\mathcal{O}_i}(\lambda)$ \cite{MB1}
	\begin{equation} \label{zz}
		Z_{\mathcal{O}_i}(\lambda) = \Bigg(\frac{g(\mu)}{g(\frac{\mu}{\lambda})}\Bigg)^{\frac{\gamma_{0\mathcal{O}_i}}{\beta_0}} \,,
	\end{equation}
	where $\gamma_{0\mathcal{O}_i}$ are the eigenvalues of $\gamma_0$. In this scheme, Eq. \eqref{csformula} reduces to a single term
	\begin{align}\label{csformuladiag}
		G^{(n)}_{j_1 \ldots j_n}(\lambda x_1,\ldots, \lambda x_n; \mu, g(\mu)) =Z_{\mathcal{O}_{j_1}}(\lambda) \ldots Z_{\mathcal{O}_{j_n}}(\lambda)\,\lambda^{-\sum_{i=1}^nD_{\mathcal{O}_{j_i}}}
		G^{(n)}_{j_1 \ldots j_n }( x_1, \ldots, x_n; \mu, g(\frac{\mu}{\lambda}))\,.
	\end{align}
	Then, as $\lambda \rightarrow 0$ in any renormalization scheme, the renormalized correlator on the right-hand side above allows for a perturbative asymptotic expansion in terms of the renormalized coupling $g(\frac{\mu}{\lambda})$ at the scale $\frac{\mu}{\lambda}$
	\begin{align} \label{eq:expansion}
		G^{(n)}_{j_1 \ldots j_n }( x_1, \ldots, x_n; \mu, g(\frac{\mu}{\lambda}))
		\sim\,&{\mathcal G}^{(n,0)}_{j_1 \ldots j_n }( x_1, \ldots, x_n; \mu)+g^2(\frac{\mu}{\lambda}) \, {\mathcal G}^{(n,2)}_{j_1 \ldots j_n }( x_1, \ldots, x_n; \mu)\nonumber\\
		&+ g^4(\frac{\mu}{\lambda}) \, {\mathcal G}^{(n,4)}_{j_1 \ldots j_n }( x_1, \ldots, x_n; \mu)
		+\cdots\,.
	\end{align}
	Naturally, the first term in this expansion, being independent of the coupling, represents the conformal contribution at zero coupling
	\begin{equation}
		{\mathcal G}^{(n,0)}_{j_1 \ldots j_n }( x_1, \ldots, x_n; \mu) = G^{(n)}_{\text{conf} \, j_1 \ldots j_n }( x_1, \ldots, x_n) \, ,
	\end{equation}
	since renormalized operators at zero coupling coincide with conformal ones. 
	The higher-order corrections in Eq. \eqref{eq:expansion} stem from nonconformal interactions due to the non-zero beta function. Consequently, the conformal contribution is corrected at higher orders in the renormalized coupling as shown, with the renormalized operators being non-conformal at higher orders in any renormalization scheme.
	Nevertheless, if the conformal contribution is non-zero, for fixed $x_1, \ldots, x_n$, all higher-order terms in Eq. \eqref{eq:expansion} are suppressed relative to the conformal one by powers of
	\begin{align} 
		g^2(\frac{\mu}{\lambda}) &\sim \dfrac{1}{\beta_0 \log(\frac{\mu^2}{\lambda^2\Lambda_{SYM}^2})}\left(1-\frac{\beta_1}{\beta_0^2}\frac{\log\log(\frac{\mu^2}{\lambda^2\Lambda_{SYM}^2})}{\log(\frac{\mu^2}{\lambda^2\Lambda_{SYM}^2})}\right)\nonumber\\
		&\sim \dfrac{1}{\beta_0 \log(\frac{1}{\lambda^2})}\left(1-\frac{\beta_1}{\beta_0^2}\frac{\log\log(\frac{1}{\lambda^2})}{\log(\frac{1}{\lambda^2})}\right) \nonumber \\
	\end{align}
	—i.e., asymptotically by inverse powers of $\log \frac{1}{\lambda}$—even though they are generally non-conformal.
	Hence, the corresponding UV asymptotics for fixed $x_1, \ldots, x_n$ is, as $\lambda \rightarrow 0$,
	\begin{align} \label{eqrg}
		\langle \mathcal{O}_{j_1}(\lambda x_1)\ldots\mathcal{O}_{j_n}(\lambda x_n)\rangle 
		\sim \,\frac{Z_{\mathcal{O}_{j_1}}(\lambda)\ldots Z_{\mathcal{O}_{j_n}}(\lambda)}{\lambda^{D_{\mathcal{O}_1}+\cdots+D_{\mathcal{O}_n}}} G^{(n)}_{\text{conf}\,j_1 \ldots j_n }( x_1, \ldots, x_n)\,.
	\end{align}
	We designate the scheme described above as the nonresonant diagonal scheme \cite{MB1}. Its existence is established via the Poincaré-Dulac theorem through the following differential-geometric interpretation of operator mixing \cite{MB1}.
	We view a finite change of basis of the renormalized operators,
	\begin{equation}\label{linearcomb}
		\mathcal{O}'(x) = S(g) \mathcal{O}(x)
	\end{equation}
	in matrix notation, as a real-analytic, invertible gauge transformation $S(g)$ that is dependent on $g \equiv g(\mu)$. Consequently, the matrix
	\begin{equation}\label{eq:connection}
		A(g) = -\frac{\gamma(g)}{\beta(g)} = \frac{1}{g} \Big(\frac{\gamma_0}{\beta_0} + \cdots\Big)
	\end{equation}
	that appears in the differential equation for $Z(\lambda)$,
	\begin{equation}
		\Big(\frac{\partial}{\partial g} - A(g)\Big) Z(\lambda) = 0
	\end{equation}
	defines a connection $A(g)$
	\begin{eqnarray} \label{sys2}
		A(g)= \frac{1}{g} \left(A_0 + \sum^{\infty}_ {n=1} A_{2n} g^{2n} \right)\,,
	\end{eqnarray}
	with a regular singularity at $g = 0$ that transforms as
	\begin{equation}
		A'(g) = S(g)A(g)S^{-1}(g) + \frac{\partial S(g)}{\partial g} S^{-1}(g)
	\end{equation}
	under the gauge transformation $S(g)$, where
	\begin{equation}
		\mathcal{D} = \frac{\partial }{\partial g} - A(g)
	\end{equation}
	is the corresponding covariant derivative.
	As a result, $Z(\lambda)$ acts as a Wilson line that transforms according to
	\begin{equation}
		Z'(\lambda) = S(g(\mu))Z(\lambda)S^{-1}(g(\frac{\mu}{\lambda}))\,.
	\end{equation}
	It is a consequence of the Poincaré-Dulac theorem~\cite{MB1} that if any two eigenvalues $\lambda_1, \lambda_2, \ldots$ of the matrix $\frac{\gamma_0}{\beta_0}$, arranged in non-increasing order $\lambda_1 \geq \lambda_2 \geq \ldots$, do not differ by a positive even integer $2k$,
	\begin{equation} \label{nr}
		\lambda_i - \lambda_j - 2k \neq 0
	\end{equation}
	for $i \leq j$ i.e., they are nonresonant then a gauge transformation exists that puts $A(g)$ into the canonical nonresonant form \cite{MB1}
	\begin{equation} \label{1loop}
		A'(g) = \frac{\gamma_0}{\beta_0}\frac{1}{g}
	\end{equation}
	that is one-loop exact to all orders of perturbation theory. Therefore, if $\frac{\gamma_0}{\beta_0}$ is also diagonalizable, Eq. \eqref{zz} follows. More precisely, Eq. \eqref{1loop} is obtained constructing a basis of renormalized operators, order by order in perturbation theory, such that in Eq. \eqref{sys2} only the one-loop contribution in Eq. \eqref{1loop} survives. The proof is reported in the following section.
	
	\subsection{Nonresonant diagonal renormalization scheme  \label{app:C2}}
	
	To make this paper self-contained, we present the order-by-order construction of the nonresonant diagonal scheme in perturbation theory, following \cite{MB1}.\par
	The construction is inductive on $k=1,2, \cdots$, demonstrating that once $A_0$ and the first $k-1$ matrix coefficients $A_2,\cdots,A_{2(k-1)}$ in Eq. \eqref{sys2} are set to the canonical nonresonant form of Eq. \eqref{1loop}—i.e., $A_0$ is diagonal and $A_2,\cdots,A_{2(k-1)}$ $=0$—a real-analytic gauge transformation can be found that leaves them unchanged while setting the $k$-th coefficient $A_{2k}$ to zero as well. \par
	The initial step of the induction ($k=0$) involves diagonalizing $A_0$—with eigenvalues in non-increasing order—by means of a constant gauge transformation. \par
	At the $k$-th step, we select the gauge transformation
	\begin{eqnarray}
		S_k(g)=1+ g^{2k} H_{2k}\,,
	\end{eqnarray}
	where $H_{2k}$ is a matrix to be determined. Its inverse is
	\begin{eqnarray}
		S^{-1}_k(g)= (1+ g^{2k} H_{2k})^{-1} = 1- g^{2k} H_{2k} + \cdots\,,
	\end{eqnarray}
	with the ellipsis representing terms of order higher than $g^{2k}$.
	The effect of $S_k(g)$ on the connection $A(g)$ provides
	\begin{align} \label{ind}
		A'(g) 
		&=  2k g^{2k-1} H_{2k} ( 1- g^{2k} H_{2k})^{-1}+  (1+ g^{2k} H_{2k}) A(g)( 1- g^{2k} H_{2k})^{-1} \nonumber \\
		&=  2k g^{2k-1} H_{2k} ( 1- g^{2k} H_{2k})^{-1} +  (1+ g^{2k} H_{2k})  \frac{1}{g} \left(A_0 + \sum^{\infty}_ {n=1} A_{2n} g^{2n} \right)  ( 1- g^{2k} H_{2k})^{-1} \nonumber \\
		&=  2k g^{2k-1} H_{2k} ( 1- \cdots) +  (1+ g^{2k} H_{2k})  \frac{1}{g} \left(A_0 + \sum^{\infty}_ {n=1} A_{2n} g^{2n} \right)  ( 1- g^{2k} H_{2k}+\cdots) \nonumber \\
		&=  2k g^{2k-1} H_{2k}  +    \frac{1}{g} \left(A_0 + \sum^k_ {n=1} A_{2n} g^{2n} \right) + g^{2k-1} (H_{2k}A_0-A_0H_{2k}) + \cdots\nonumber \\
		&=   g^{2k-1} (2k H_{2k} + H_{2k} A_0 - A_0 H_{2k})  + A_{2(k-1)}(g) + g^{2k-1} A_{2k}+ \cdots\,,\nonumber \\
	\end{align}
	where we have omitted all terms that contribute to orders higher than $g^{2k-1}$, and
	\begin{align}
		A_{2(k-1)}(g) =  \frac{1}{g} \left(A_0 + \sum^{k-1}_ {n=1} A_{2n} g^{2n} \right)
	\end{align}
	is the part of $A(g)$ unaffected by the gauge transformation $S_k(g)$, thus satisfying the induction hypotheses—i.e., that $A_2, \cdots, A_{2(k-1)}$ are zero. \par
	Thus, according to Eq. \eqref{ind}, the $k$-th matrix coefficient $A_{2k}$ can be removed from the expansion of $A'(g)$ up to order $g^{2k-1}$ if an $H_{2k}$ exists such that
	\begin{align}
		&A_{2k}+(2k H_{2k} + H_{2k} A_0 - A_0 H_{2k}\nonumber\\
		&= A_{2k}+ (2k-ad_{A_0}) H_{2k}=0\,,
	\end{align}
	with $ad_{A_0}Y=[A_0,Y]$.
	If the operator $ad_{A_0}-2k$ is invertible, the unique solution for $H_{2k}$ is
	\begin{eqnarray}
		H_{2k}=(ad_{A_0}-2k)^{-1} A_{2k}\,.
	\end{eqnarray}
	To complete the induction, we must therefore demonstrate that if the eigenvalues of $A_0$ are nonresonant, $ad_{A_0}-2k$ is invertible, meaning its kernel is trivial. 
	Now, $ad_{\Lambda}-2k$, acting as a linear operator on matrices, is diagonal with eigenvalues $\lambda_{i}-\lambda_{j}-2k$ and eigenvectors $E_{ij}$, whose only non-zero entries are $(E_{ij})_{ij}$. The eigenvectors $E_{ij}$, normalized so that $(E_{ij})_{ij}=1$, constitute an orthonormal basis for the space of matrices. Consequently, $E_{ij}$ is in the kernel of $ad_{\Lambda}-2k$ if and only if $\lambda_{i}-\lambda_{j}-2k=0$. Since, by assumption, $\lambda_{i}-\lambda_{j}-2k \neq 0$ for all $i,j$, the kernel of $ad_{\Lambda}-2k$ contains only the zero matrix, and the construction is complete.

	\subsection{Anomalous dimensions of unbalanced twist-$2$ operators in $\mathcal{N} = 1$ SUSY YM theory}
	
	We define the bare collinear twist-2 operators for $s \geq 1$ and $k \geq 0$ as \cite{Belitsky:1998gc}
	\begin{equation}\label{eq:125}
		\mathcal{O}^{(k)}_{Bs} = (i\partial_+)^{k}\mathcal{O}_{Bs}
	\end{equation}
	which, at the leading order of perturbation theory and for $k>0$, are conformal descendants \cite{Braun:2003rp} of the corresponding primary conformal operator 
	$\mathcal{O}^{(0)}_{Bs}=\mathcal{O}_{Bs}$.
	Due to operator mixing, we obtain for the renormalized operators \cite{Belitsky:1998gc,Braun:2003rp}
	\begin{equation}\label{m}
		\mathcal{O}^{(k)}_{s}= \sum^{s}_{i} Z_{si} \mathcal{O}^{(k+s-i)}_{B i} \,,
	\end{equation}
	where $Z$ is the bare mixing matrix and 
	\begin{equation}
		\gamma(g) =- \frac{\partial Z}{\partial \log \mu} Z^{-1}=\sum_{j=0}^{\infty}\gamma_j \, g^{2j+2}
	\end{equation}
	is the matrix of anomalous dimensions—generally lower triangular—with $\gamma_0$ being diagonal in the $\overline{MS}$ scheme \cite{Belitsky:1998gc,Braun:2003rp}.\par
	For $\mathcal{O} = S^{A},\bar{S}^{A},S^{\lambda},\bar{S}^{\lambda},T,\bar{T}$, $\gamma_0$ is diagonal with eigenvalues \cite{Belitsky:2004sc}
	\begin{equation}
		\gamma_{0\mathcal{O}_s} = \frac{1}{4 \pi^2} \Big( \tilde{\gamma}_{0\mathcal{O}_s} - \frac{3}{2}\Big)\,,
	\end{equation}
	where \cite{Belitsky:2004sc}
	\begin{align}
		&\tilde{\gamma}_{0S^{A}_s} = 2\psi(s + 1)- 2\psi(1) \nonumber\\
		&\tilde{\gamma}_{0S^{\lambda}_s} = 2\psi(s + 1)- 2\psi(1) 
	\end{align}
	and \cite{Belitsky:2004sc,Belitsky:2003sh}
	\begin{align}
		\tilde{\gamma}_{0T_s} 
		=\begin{cases}
			\psi(s+1)-\psi(1),&s=2, 4 , \ldots\\
			\psi(s+2)-\psi(1), & s=1,3, \ldots \,.
		\end{cases}
	\end{align}
	We have confirmed numerically that the first $10^4$ eigenvalues of $\frac{{\gamma}_{0\mathcal{O}_s}}{\beta_0}$ are nonresonant, with $\beta_0 = \frac{3}{(4\pi)^2}$. Furthermore, a number theoretic proof of the nonresonant condition for the anomalous dimensions of twist-$2$ operators in $\mathcal{N}=1$ SUSY YM has been given in \cite{S1,S2}.

	\section{RG-improved generating functionals} \label{RGF}
	
	\subsection{$\mathcal{W}$} \label{W}

	In the 't Hooft large-$N$ expansion, the generating functional of Euclidean $n$-point correlators separates into its planar contribution $\mathcal{W}^E_{\text{sphere}}$ and its leading-order nonplanar contribution $\mathcal{W}^E_{\text{torus}}$ \cite{BPSpaper2,BPSL,QCD24}. The UV asymptotics of $\mathcal{W}^E_{\text{sphere}}$ and $\mathcal{W}^E_{\text{torus}}$ as $\lambda \rightarrow 0$ can be determined from the RG-improved correlators explicitly calculated in \ref{npoint}, \ref{9}, and \ref{rgcorr} according to Eq. \eqref{eqrg}, and from the conformal generating functionals in Eqs. \eqref{wexplicitE} and \eqref{wexplicitT1E}, satisfying
	\begin{equation}
		\mathcal{W}^E_{\text{asym sphere}}[J_{\mathcal{O'}^E},\lambda]=-N^2 \mathcal{W}^E_{\text{asym torus}}[J_{\mathcal{O'}^E},\lambda]\,,
	\end{equation}
	where
	\begin{align}
		\label{wexplicitERG}
		&{\mathcal{W}^E_{\text{asym torus}}}\left[\bar{J}_{S^{'A\,E}},J_{\bar{S}^{'A\,E}},\bar{J}_{S^{'\lambda\,E}},J_{\bar{S}^{'\lambda\,E}},0,0,\lambda\right]=\nonumber\\
		&+\frac{1}{2}\log\Det\Bigg(I-\frac{2}{N^2}\sum_{k_1=0}^{s_1-2}{s_1\choose k_1}{s_1\choose k_1+2}(-\overrightarrow{\partial}_z)^{s_1-k_1-1}\Laplace^{-1}\frac{Z_{S_{s_1}^A}(\lambda)}{\lambda^{s_1+2}}J_{\bar{S}^{'A\,E}_{s_1}}(-\overrightarrow{\partial}_z)^{k_1+1}\nonumber\\ 
		&\quad\sum_{k_2=0}^{s_2-2}{s_2\choose k_2}{s_2\choose k_2+2}(-\overrightarrow{\partial}_z)^{s_2-k_2-1}\Laplace^{-1}\frac{Z_{S_{s_2}^A}(\lambda)}{\lambda^{s_2+2}}\bar{J}_{S^{'A\,E}_{s_2}}(-\overrightarrow{\partial}_z)^{k_2+1}\Bigg)\nonumber\\
		&-\frac{1}{2}\log\Det\Bigg(I-\frac{2}{N^2}\sum_{k_1=0}^{s_1-1}{s_1\choose k_1}{s_1\choose k_1+1}(-\overrightarrow{\partial}_z)^{s_1-k_1-1}\partial_z\Laplace^{-1}\frac{Z_{S_{s_1}^\lambda}(\lambda)}{\lambda^{s_1+2}}J_{\bar{S}^{'\lambda\,E}_{s_1}}(-\overrightarrow{\partial}_z)^{k_1}\nonumber\\ 
		&\quad\sum_{k_2=0}^{s_2-1}{s_2\choose k_2}{s_2\choose k_2+1}(-\overrightarrow{\partial}_z)^{s_2-k_2-1}\partial_z\Laplace^{-1}\frac{Z_{S_{s_2}^\lambda}(\lambda)}{\lambda^{s_2+2}}\bar{J}_{S^{'\lambda\,E}_{s_2}}(-\overrightarrow{\partial}_z)^{k_2}\Bigg)
	\end{align}
	and
	\begin{align}
		\label{wexplicitT1ERG}
		&\mathcal{W}^E_{\text{asym torus}}\left[0,0,0,0,\bar{J}_{T^{'E}},J_{\bar{T}^{'E}},\lambda\right] =\nonumber\\
		&	\log\Det\Bigg(I+\frac{1}{N^2}
		\sum_{k_1 = 0}^{s_1-1}{s_1+1\choose k_1}{s_1\choose k_1+1} 
		(-\overrightarrow{\partial}_z)^{s_1-k_1}\Laplace^{-1}\frac{Z_{T_{s_1}}(\lambda)}{\lambda^{s_1+2}}J_{\bar {T}^{'E}_{s_1}} (-\overrightarrow{\partial}_z)^{k_1}   \nonumber\\
		&\quad\sum_{k_2 = 0}^{s_2-1}{s_2\choose k_2}{s_2+1\choose k_2+2} 
		(-\overrightarrow{\partial}_z)^{s_2-k_2-1}(-\partial_z)\Laplace^{-1}\frac{Z_{T_{s_2}}(\lambda)}{\lambda^{s_2+2}}\bar{J}_{{T}^{'E}_{s_2}} (-\overrightarrow{\partial}_z)^{k_2+1}\Bigg)\,.
	\end{align}

	\subsection{$\Gamma$}\label{gamma}
	
	We also present a more concise formula for the asymptotic RG-improved generating functionals, expressed as Fredholm (super)determinants
	\begin{align}
		\label{eq:asymglue}
		\resizebox{0.88\textwidth}{!}{%
			$\begin{aligned}
				&\Gamma^E_{\text{asym}} \left[ \overline{j}_{\VarPrimeSup{S}{A}{E}},\; j_{\VarPrimeSup{\overline{S}}{A}{E}},\; \overline{j}_{\VarPrimeSup{S}{\lambda}{E}},\; j_{\VarPrimeSup{\overline{S}}{\lambda}{E}},\; 0,\; 0,\; \lambda \right] \nonumber \\
				&= 
				-\frac{N^2 - 1}{2} \log \Det \Bigg[ I\, \delta_{s_1 s_2} \delta_{k_1 k_2} 
				- \frac{2}{N^2} \binom{s_1}{k_1} \binom{s}{k+2} \partial_z^{s_1 - k_1 + k} \Laplace^{-1} 
				\frac{Z_{\VarPrimeSup{\overline{S}}{A}{E}}(\lambda)}{\lambda^{s+2}} j_{\VarPrimeSup{\overline{S}}{A}{E}_{sk}} 
				\binom{s}{k} \binom{s_2}{k_2 + 2} \partial_z^{s - k + k_2} 
				\Laplace^{-1} \frac{Z_{\VarPrimeSup{\overline{S}}{A}{E}_{s_2}}(\lambda)}{\lambda^{s_2 + 2}}
				\overline{j}_{\VarPrimeSup{S}{E}{A}_{s_2 k_2}} \Bigg]  \nonumber\\
				& + \frac{N^2 - 1}{2} \log \Det \Bigg[ I\, \delta_{s_1 s_2} \delta_{k_1 k_2} 
				- \frac{2}{N^2}\binom{s_1}{k_1} \binom{s}{k+1} \partial_z^{s_1 - k_1 + k-1}\partial_z \Laplace^{-1} 
				\frac{Z_{\VarPrimeSup{\overline{S}}{\lambda}{E}}(\lambda)}{\lambda^{s+2}} j_{\VarPrimeSup{\overline{S}}{\lambda}{E}_{sk}} 
				\binom{s}{k} \binom{s_2}{k_2 + 1} \partial_z^{s - k + k_2-1} 
				\partial_z\Laplace^{-1} \frac{Z_{\VarPrimeSup{\overline{S}}{\lambda}{E}_{s_2}}(\lambda)}{\lambda^{s_2 + 2}} 
				\overline{j}_{\VarPrimeSup{S}{E}{\lambda}_{s_2 k_2}} \Bigg]
			\end{aligned}$
		}\\
	\end{align}
	and
	\begin{align}\label{eq:asymgluino}
		\resizebox{0.88\textwidth}{!}{%
			$\begin{aligned}
				&\Gamma^E_{\text{asym}}\left[0,0,0,0,\bar{j}_{T^E},j_{\bar{T}^E},\lambda\right]\nonumber\\
				&= 
				-(N^2-1)\log\Det\left[I\, \delta_{s_1 s_2} \delta_{k_1 k_2} -\frac{1}{N^2}{s_1\choose k_1}{s\choose k+1}\partial_{z}^{s_1-k_1+k}\Laplace^{-1}\frac{Z_{T^E}(\lambda)}{\lambda^{s+2}}\bar {j}_{T_{sk}^E}{s+1\choose k+2}{s_2+1\choose k_2}\partial_{z}^{s-k+k_2}\partial_z\Laplace^{-1} \frac{Z_{\bar{T}^E}(\lambda)}{\lambda^{s_2+2}} j_{\bar{T}_{s_2k_2}^E}\right]\,
			\end{aligned}$
		}\\
	\end{align}
	that are derived from the corresponding conformal objects (Sec. \ref{kernelE}).
	
	\section{Conclusions}  \label{Conc}

	We are now prepared to present the main results of this article.\par
	Nonperturbatively, the generating functional for the glueball/gluinoball one-loop correlators is given by
	\begin{align}
		\label{glueballW1loop_tot}
		\resizebox{0.89\textwidth}{!}{%
			$\begin{aligned}
				&\mathcal{W}^E_{\text{glueball/gluinoball 1-loop }}[J_{\Phi},J_{\Psi}] \nonumber\\
				&=\frac{1}{2}\log\text{sDet}
				\begin{pmatrix}\ast'_2(-\Delta+M^2)+\frac{1}{N}\ast'_3\Phi_J\ast'_3& \frac{1}{N}\ast'_3\ast'_3\Psi_J\nonumber\\ 
					\frac{1}{N}\ast'_3\ast'_3\Psi_J&\ast_2(-\Delta+M^2)+\frac{1}{N}\ast_3\Phi_J\ast_3 \end{pmatrix} \nonumber\\
				&=+\frac{1}{2}\log\text{Det}
				\left(\ast'_2(-\Delta+M^2)+\frac{1}{N}\ast'_3\Phi_J\ast'_3\right)-\frac{1}{2}\log\text{Det}
				\left(\ast_2(-\Delta+M^2)+\frac{1}{N}\ast_3\Phi_J\ast_3\right)\nonumber\\
				&\quad+\frac{1}{2}\log\text{Det}
				\left[\mathcal{I}-\left(\ast'_2(-\Delta+M^2)+\frac{1}{N}\ast'_3\Phi_J\ast'_3\right)^{-1}\frac{1}{N}\ast'_3\ast'_3\Psi_J\left(\ast_2(-\Delta+M^2)+\frac{1}{N}\ast_3\Phi_J\ast_3\right)^{-1}\frac{1}{N}\ast'_3\ast'_3\Psi_J\right] ,
			\end{aligned}$
		}\\
	\end{align} 
	where we have used the first line in Eq. \eqref{eq:sdet}. By setting $J_{\Psi}=0$, we find, up to an inconsequential constant
	\begin{align}
		\label{glueballW1loop_glue}
		\mathcal{W}^E_{\text{glueball/gluinoball 1-loop }}[J_{\Phi},0] 
		=&+\frac{1}{2}\log\text{Det}
		\left(\mathcal{I}+(\ast'_2(-\Delta+M^2))^{-1}\frac{1}{N}\ast'_3\Phi_J\ast'_3\right)\nonumber\\
		&-\frac{1}{2}\log\text{Det}
		\left(\mathcal{I}+(\ast_2(-\Delta+M^2))^{-1}\frac{1}{N}\ast_3\Phi_J\ast_3\right)\,.
	\end{align}
	In a similar manner, setting $J_{\Phi}=0$, we obtain, up to an irrelevant constant
	\begin{align}
		\label{glueballW1loop_gluino}
		\resizebox{0.89\textwidth}{!}{%
			$\begin{aligned}
				\mathcal{W}^E_{\text{glueball/gluinoball 1-loop }}[0,J_{\Psi}] =+\frac{1}{2}\log\text{Det}
				\left[\mathcal{I}-\left(\ast'_2(-\Delta+M^2)\right)^{-1}\frac{1}{N}\ast'_3\ast'_3\Psi_J\left(\ast_2(-\Delta+M^2)\right)^{-1}\frac{1}{N}\ast'_3\ast'_3\Psi_J\right] .\nonumber
			\end{aligned}$
		}\\
	\end{align}
	The corresponding RG-improved objects, derived from Eqs. \eqref{eq:asymglue}, \eqref{eq:asymgluino}, \eqref{gammacorr}, \eqref{73}, and the identification in \eqref{74}, are
	\begin{align}\label{eq:asymgluetorus}
		\resizebox{0.89\textwidth}{!}{%
			$\begin{aligned}
				&\Gamma^E_{\text{asym torus}} \left[ \overline{j}_{\VarPrimeSup{S}{A}{E}},\; j_{\VarPrimeSup{\overline{S}}{A}{E}},\; \overline{j}_{\VarPrimeSup{S}{\lambda}{E}},\; j_{\VarPrimeSup{\overline{S}}{\lambda}{E}},\; 0,\; 0,\; \lambda \right] \nonumber \\
				&= +\frac{1}{2} \log \Det \Bigg[I\begin{pmatrix} \delta_{s_1 s_2} \delta_{k_1 k_2} &0 \nonumber\\ 
					0 & \delta_{s_1 s_2} \delta_{k_1 k_2}  \end{pmatrix} +\partial_z^{s_1 - k_1 + k_2} \Laplace^{-1}\begin{pmatrix}0&\sqrt{2} \binom{s_1}{k_1} \binom{s_2}{k_2+2}
					\frac{Z_{\VarPrimeSup{\overline{S}}{A}{E}}(\lambda)}{\lambda^{s_2+2}}\frac{1}{N} j_{\VarPrimeSup{\overline{S}}{A}{E}_{s_2k_2}}\nonumber \\ 
					\sqrt{2}\binom{s_1}{k_1} \binom{s_2}{k_2 + 2} 
					\frac{Z_{\VarPrimeSup{\overline{S}}{A}{E}_{s_2}}(\lambda)}{\lambda^{s_2 + 2}} 
					\frac{1}{N}\overline{j}_{\VarPrimeSup{S}{E}{A}_{s_2 k_2}} &0 \end{pmatrix} \Bigg]\nonumber \\
				&\quad -\frac{1}{2} \log \Det \Bigg[I\begin{pmatrix} \delta_{s_1 s_2} \delta_{k_1 k_2} &0\nonumber \\ 
					0 &\delta_{s_1 s_2} \delta_{k_1 k_2}  \end{pmatrix} +\partial_z^{s_1 - k_1 + k_2-1}\partial_z \Laplace^{-1}\begin{pmatrix}0&\sqrt{2}\binom{s_1}{k_1} \binom{s_2}{k_2+1}
					\frac{Z_{\VarPrimeSup{\overline{S}}{\lambda}{E}}(\lambda)}{\lambda^{s_2+2}}\frac{1}{N} j_{\VarPrimeSup{\overline{S}}{\lambda}{E}_{s_2k_2}}\nonumber \\ 
					\sqrt{2}	\binom{s_1}{k_1} \binom{s_2}{k_2 + 1} \frac{Z_{\VarPrimeSup{\overline{S}}{\lambda}{E}_{s_2}}(\lambda)}{\lambda^{s_2 + 2}} 
					\frac{1}{N}	\overline{j}_{\VarPrimeSup{S}{E}{\lambda}_{s_2 k_2}} &0 \end{pmatrix} \Bigg]
			\end{aligned}$
		}\\
	\end{align}
	and
	\begin{align}\label{eq:asymgluinotorus}
		\resizebox{0.89\textwidth}{!}{%
			$\begin{aligned}
				&\Gamma^E_{\text{asym torus}}\left[0,0,0,0,\bar{j}_{T^E},j_{\bar{T}^E},\lambda\right]\nonumber\\
				&= 
				+\log\Det\left[I\,\delta_{s_1 s_2} \delta_{k_1 k_2} -{s_1\choose k_1}{s\choose k+1}\partial_{z}^{s_1-k_1+k}\Laplace^{-1}\frac{Z_{T^E}(\lambda)}{\lambda^{s+2}}\frac{1}{N}\bar {j}_{T_{sk}^E}{s+1\choose k+2}{s_2+1\choose k_2}\partial_{z}^{s-k+k_2}\partial_z\Laplace^{-1} \frac{Z_{\bar{T}^E}(\lambda)}{\lambda^{s_2+2}} \frac{1}{N}j_{\bar{T}_{s_2k_2}^E}\right]\,,
			\end{aligned}$
		}\\
	\end{align}
	that are UV asymptotic to the nonperturbative ones as $\lambda \rightarrow 0$ due to AF. Respectively,
	\begin{align}
		\Gamma^E_{\text{asym torus}} \left[ \overline{j}_{\VarPrimeSup{S}{A}{E}},\; j_{\VarPrimeSup{\overline{S}}{A}{E}},\; \overline{j}_{\VarPrimeSup{S}{\lambda}{E}},\; j_{\VarPrimeSup{\overline{S}}{\lambda}{E}},\; 0,\; 0,\; \lambda \right]  \sim \mathcal{W}^E_{\text{glueball/gluinoball 1-loop }}[J_{\Phi},0] 
	\end{align}
	and
	\begin{align}
		\Gamma^E_{\text{asym torus}}\left[0,0,0,0,\bar{j}_{T^E},j_{\bar{T}^E},\lambda\right] 
		\sim \mathcal{W}^E_{\text{glueball/gluinoball 1-loop }}[0,J_{\Psi}] \,.
	\end{align}
	for an appropriate selection of the glueball $\Phi$ and gluinoball $\Psi$ interpolating fields (Sec. \ref{NP}), where, to maintain notational simplicity, the coordinate rescaling by the factor $\lambda$ in the nonperturbative generating functional is implicit. \par
	As previously noted in \cite{BPS41}, the aforementioned structure of the logarithm of a functional (super)determinant connects with the specific features of the topology of leading nonplanar diagrams in the large-$N$ expansion of the SU($N$) theory as opposed to the (bare) U($N$) one \cite{BPSL,QCD24}. This includes resolving an issue \footnote{This issue will be discussed in detail in the next section.} regarding the spin-statistics theorem in $\mathcal{N}=1$ SUSY YM theory \cite{BPSSUSY} that is analogous to the one in pure SU($N$) YM theory \cite{BPSL,QCD24}.\par
	Thus, the matching of the $\log \Det$ structure between the above nonperturbative and UV-asymptotic RG-improved generating functionals in the large-$N$ expansion to the leading-nonplanar order imposes strong qualitative and quantitative UV constraints on the prospective nonperturbative solution of large-$N$ $\mathcal{N} = 1$ SUSY YM theory. This matching could serve as an essential guide in the search for such a solution.

	\section{Further developments and outlook}\label{sec:11}
	 
	        The results reported in the previous sections nicely intertwine with a number of recent new ideas and possible developments that we briefly    
		summarize in the following, also illuminating the further physical meaning of the computations in the present paper \footnote{We would like to  thank the referee for raising these points.}:\par
	         The relation between the conformal renormalization scheme in \cite{Braun:2003rp} and the nonresonant one in the present paper. \par
		 The changes of UV asymptotics arising in the resonant case \cite{Becchetti:2021for}. \par
	         The proof that the mixing of twist-$2$ operators is nonresonant \cite{S1,S2}. \par
                  The existence of a new sector in the large-$N$ SU($N$) expansion, as opposed to the U($N$) one, of $\mathcal{N}=1$ SUSY YM  
                  theory -- involving pinched tori for twist-$2$ operators and the spin-statistics theorem -- extending to the supersymmetric case our previous 
                  treatment \cite{BPSL} in the pure SU($N$) YM theory. \par
                  Some remarks about the matching between the structure of the one-loop glueball/gluinoball effective action and the corresponding 
                  RG-improved perturbative object for twist-$2$ operators. \par
                  Stronger nonperturbative UV constraints for $2$-point correlators of twist-$2$ operators, involving the computation of the actual nonperturbative 
                  large-$N$ coupling in the aforementioned new sector of SU($N$) YM and $\mathcal{N}=1$ SUSY YM theory. \par

		\subsection{Conformal and nonresonant renormalization scheme}

	\subsubsection{Generalities about renormalization of twist-$2$ operators}
	
	In dimensional regularization, $d=4-2\epsilon$, the bare and renormalized gauge coupling are related by
	\begin{equation}
		g_0=\mu^{\epsilon} Z_g(g,\epsilon)\, g ,
		\label{eq:g0def}
	\end{equation}
	where $g_0$ is $\mu$ independent and $Z_g(g,\epsilon)$ is a formal power series in $g$ meromorphic in $\epsilon$.\par
	Generic operators only mix with other operators having the same canonical dimension and quantum numbers. Hence, twist-$2$ operators only mix with descendants of twist-$2$ operators of lower spin, since they have minimal twist, so that $ Z(g,\epsilon)$ in Eq. \eqref{eq:OpRen} is lower triangular.\par
	 We denote by $\mathcal O^{(k)}_{Bs}= (i\partial_+)^{k}\mathcal{O}_{Bs}$ the bare operators which, at the leading order of perturbation theory and for $k>0$, are conformal descendants \cite{Braun:2003rp} of the corresponding twist-$2$ primary conformal operators, and by $\mathcal O^{(k)}_{s}$ the renormalized ones,
	\begin{equation}
	\mathcal{O}^{(k)}_{s}= \sum^{s}_{i=0} Z_{si}(g,\epsilon) \mathcal{O}^{(k+s-i)}_{B i} \,,
		\label{eq:OpRen}
	\end{equation}
	where $Z(g, \epsilon)$ is the bare mixing matrix. In matrix notation Eq. \eqref{eq:OpRen} reads
\begin{equation}
	O^{(k)}_{s}=  Z(g,\epsilon) O^{(k)}_{Bs}\,,
	\label{eq:OpRen2}
\end{equation}
where we have introduced the vector 
\begin{align}\label{eq:vector}
	O^{(k)}_s &= \begin{pmatrix}
		\mathcal O_s^{(k)} \\
		\mathcal O_{s-1}^{(k+1)}\\
		\vdots \\
		\mathcal O_0^{(k+s)} 
	\end{pmatrix} .
\end{align}	
In a minimal-subtraction scheme, $ Z(g,\epsilon)$ has the general structure
	\begin{equation}
		 Z(g,\epsilon)
		=
	\mathbb{I}
		+\sum_{r=1}^{\infty} g^{2r}\, Z^{(r)}(\epsilon),
		\qquad
		 Z^{(r)}(\epsilon)
		=
		\sum_{k=1}^{r}\frac{ Z^{(r,k)}}{\epsilon^{k}} .
		\label{eq:Zloop}
	\end{equation}
	Equivalently, we may reorganize the same expansion as
	\begin{equation}
		 Z(g,\epsilon)
		=
		\mathbb{I}
		+\frac{ Z^{[1]}(g)}{\epsilon}
		+\frac{ Z^{[2]}(g)}{\epsilon^{2}}
		+\frac{ Z^{[3]}(g)}{\epsilon^{3}}
		+\cdots ,
		\label{eq:Zpole}
	\end{equation}
	where the matrix coefficient $ Z^{[k]}(g)$ starts at order $g^{2k}$. The matrix of anomalous dimensions
	\begin{equation}
		\gamma(g)
		=
		-\frac{\partial Z(g,\epsilon)}{\partial\log\mu} Z^{-1}(g,\epsilon)
		=
		-\beta(g,\epsilon)\,\frac{\partial Z(g,\epsilon)}{\partial g}\, Z^{-1}(g,\epsilon) 
		= \sum_{j=0}^{\infty}\gamma_j \, g^{2j+2}\
		\label{eq:gammadef}
	\end{equation}
is actually independent of $\epsilon$, where
	\begin{equation}
		\beta(g,\epsilon)
		\equiv
		\mu\frac{dg}{d\mu}\Big|_{g_0}
		=
		-\epsilon g+\beta(g) ,
		\label{eq:betadr}
	\end{equation}
with $\beta(g)$ the four-dimensional beta function.
	Therefore, $Z(g,\epsilon) $ satisfies the matrix-differential equation
	\begin{equation}\label{eqZ-dimensional}
		\Bigg(\frac{\partial}{\partial g} + \frac{\gamma(g)}{\beta(g,\epsilon)}\Bigg)Z(g,\epsilon) = 0
	\end{equation}
whose solution is
	\begin{equation} \label{ZZ-dimensional}
		Z(g,\epsilon)  = P\exp \Big(\int_g \frac{\gamma(g')}{\beta(g',\epsilon)} dg'\Big)\,.
	\end{equation}
	Since the bare operators are $\mu$ independent at fixed bare parameters $g_0,\epsilon$
	\begin{equation}
		\mu\frac{d}{d\mu}\,\mathcal O^{(k)}_{B s}=0 \,,
		\label{eq:baremuind}
	\end{equation}
it follows from Eq. \eqref{eq:OpRen} that the renormalized operators satisfy the RG equation
	\begin{equation}
		\mu\frac{\partial}{\partial\mu}\,\mathcal O^{(k)}_{s}+	\beta(g,\epsilon)\frac{\partial}{\partial g}O^{(k)}_{s}
		+
		\sum_{i}^{s}\gamma_{si}(g)\,\mathcal O^{(k+s-i)}_{i}
		=0 ,
		\label{eq:RGOps}
	\end{equation}
in matrix notation 
	\begin{equation}
	\mu\frac{\partial}{\partial\mu}\, O^{(k)}_{s}+	\beta(g,\epsilon)\frac{\partial}{\partial g} O^{(k)}_s
	+ \gamma(g)\, O^{(k)}_s
	=0 \,.
	\label{eq:RGOps2}
\end{equation}
	Accordingly, the Callan--Symanzik equation for the renormalized Euclidean correlators 
	\begin{equation}\label{key2}
		\langle \mathcal{O}_{k_1}(x_1)\ldots \mathcal{O}_{k_n}(x_n) \rangle = G^{(n)}_{k_1 \ldots k_n}( x_1,\ldots,  x_n; \mu,\epsilon, g(\mu))
	\end{equation}
reads in dimensioal regularization
	\begin{align}		\label{eq:CSwithOp}
	& \Big(\sum_{\alpha = 1}^n x_\alpha \cdot \frac{\partial}{\partial x_\alpha} + \beta(g,\epsilon)\frac{\partial}{\partial g} +\sum_{\alpha = 1}^n D_{\mathcal{O}_\alpha}\Big)G^{(n)}_{k_1 \ldots k_n}  \nonumber\\
	&+ \sum_a \Big(\gamma_{k_1a}(g) G^{(n)}_{ak_2 \ldots k_n}+ \gamma_{k_2a}(g) G^{(n)}_{k_1 a k_3 \ldots k_n} \cdots +\gamma_{k_n a}(g) G^{(n)}_{k_1 \ldots a}\Big) = 0\,.
\end{align}

\subsubsection{Conformal scheme}

In general, once the twist-$2$ operators have been renormalized in Eq. \eqref{eq:OpRen2}, we still have the freedom of making finite renormalizations
\begin{equation}
	\mathcal O'_s = S(g)\mathcal O_s\,,
\end{equation}
where $S(g)$ is a finite lower-triangular matrix. \par
A natural way of using this freedom -- often employed in the literature \cite{Braun:2003rp} -- is
 to choose $S(g)$ such that the operators $\mathcal O'_s$ are conformal covariant up to the order $g^2$ with respect to the full conformal group and not only with respect to scale transformations \cite{Braun:2003rp}. This defines the conformal subtraction scheme \cite{Braun:2003rp} that must exist at order $g^2$  because violations of conformal invariance induced by the nonvanishing beta function only affect the solution of the Callan-Symanzik equation at higher orders. \par
\par
To implement conformal covariance for twist-$2$ operators it suffices to require covariance with respect to special conformal transformations, since the conformal group is generated by the Poincaré group, dilatations 
\begin{equation}
	x^\mu
\rightarrow
x'^{\mu}
=\lambda x^\mu
\end{equation}
and special conformal transformations 
 \begin{equation}
 	x^\mu
 	\rightarrow
 	x'^{\mu}
 	=
 	\frac{x^\mu+a^\mu x^2}
 	{1+2a\cdot x+a^2x^2}\,,
 \end{equation}
the infinitesimal generators of the latter being \begin{equation}\label{Kminusdef}
	K_\mu
	=
	i\left(
	2x_\mu x^\nu\partial_\nu
	-x^2\partial_\mu
	+2D\,x_\mu
	+2x^\nu\Sigma_{\mu\nu}
	\right),
\end{equation}
where $D$ is the canonical dimension of the field on which $K_\mu$ acts and $\Sigma_{\mu\nu}$ the corresponding spin generator. \par
Hence, to construct the conformal scheme it suffices to consider the action of special conformal transformations on the operators $\mathcal{O}^{(k)}_s$. Since they are projected along the $p_+$ direction, the relevant special conformal transformation is the one generated by $K_-$ that lowers by one the number of light-cone derivatives. \par
 Thus, the variation of the above operators under $K_-$ reads \cite{Braun:2003rp} 
\begin{equation}
\label{gammacdef}
	\delta_{K_-}\mathcal{O}^{(k)}_s
	=
	i\sum_{i\leq s}
	\left[
	a_s(k)\delta_{si}
	+
	\gamma^c_{0,si}(k) g^2+\cdots
	\right]
	\mathcal{O}^{(k+s-i-1)}_i\,,
\end{equation}
with
\begin{equation}\label{ask}
a_s(k)=-2k(2s+k+3)\,,
\end{equation}
and $\gamma^{c}_{0,si}$ the first coefficient of the matrix of the special conformal anomaly
 \begin{equation}
 	\gamma^c(g,k)
 	=
 	g^2\gamma^{c}_0(k)
 	+\cdots .
 \end{equation}
It was shown in \cite{Braun:2003rp} that the basis of conformal operators in the conformal scheme -- where $\gamma^{c}_{0}(k)$ vanishes in Eq.\eqref{gammacdef} -- reads
\begin{equation}\label{Sconf1}
	\mathcal{O}^{(k)\,\mathrm{conf}}_s
	=
	\mathcal{O}^{(k)}_s
	+
	g^2
	\sum_{i<s}
	\frac{\gamma^{c}_{0,si}}{a(s,i)}
	\mathcal{O}^{(k+s-i)}_i
	+ \cdots \, ,
\end{equation}
with
\begin{equation}\label{asi}
	a(s,i)=2(s-i)(s+i+3)\,.
\end{equation}
In the language of the present paper, we think of the change of basis in Eq. \eqref{Sconf1} as induced by the gauge transformation 
\begin{equation}
	S(g)=\mathbb{I}+g^2S_1+\cdots\,,
\end{equation}
corresponding to
\begin{equation}
	\mathcal{O}^{(k)\,\mathrm{conf}}_s
	=
	\mathcal{O}^{(k)}_s
	+
	g^2\sum_{i<s}\mathcal S_{1,si} O_i^{(k+s-i)}
	+ \cdots
	\,,
\end{equation}
We now determine how $\gamma(g)$ transforms in the conformal scheme. Since $\gamma(g) = -\beta(g)A(g)$, with
 \begin{equation}
	A(g)
	=
	-\frac{\gamma(g)}{\beta(g)}
	=
	\frac{1}{g}
	\left(
	\frac{\gamma_0}{\beta_0}+\sum_{n\geq1}A_{2n}g^{2n}
	\right),
\end{equation}
which transforms as a connection
\begin{equation}
	A'(g)
	=
	S(g)A(g)S^{-1}(g)
	+
	\frac{\partial S(g)}{\partial g}S^{-1}(g)\,,
\end{equation}
it follows
\begin{equation}\label{gammaconftransf}
	\gamma^{\mathrm{conf}}(g)
	=
	S(g)\gamma^{\overline{\rm MS}}(g)S^{-1}(g)
	-
	\beta(g)\,\frac{\partial S(g)}{\partial g}S^{-1}(g)\,.
\end{equation}
Then, Eq. \eqref{gammaconftransf} gives us
\begin{equation}
	\gamma^{\mathrm{conf}}(g)
	=
	g^2\gamma^{\overline{\rm MS}}_0
	+
	g^4
	\left(
	\gamma^{\overline{\rm MS}}_1
	+
	[S_1,\gamma^{\overline{\rm MS}}_0]
	+
	2\beta_0 S_1
	\right)
	+ \cdots \,.
\end{equation}
Therefore,
\begin{equation} \label{cs000}
	\gamma^{\mathrm{conf}}_0=\gamma^{\overline{\rm MS}}_0\,,
\end{equation}
as the gauge rotation $S(g)$ does not affect $\gamma_0$ to order $g^2$, so that the dilatation generator remains diagonal to order $g^2$,
while
\begin{equation} \label{cs100}
	\gamma^{\mathrm{conf}}_1
	=
	\gamma^{\overline{\rm MS}}_1+[S_1,\gamma^{\overline{\rm MS}}_0]+2\beta_0 S_1\,.
\end{equation}

\subsubsection{Nonresonant scheme}

Alternatively, we can choose $S(g)$ order by order in perturbation theory such that the operators $\mathcal O'_s$ belong to the nonresonant diagonal basis. In the nonresonant scheme, constructed in Sec. \ref{app:C2}, the connection in Eq. \eqref{eq:connection} is one-loop exact and the finite renormalization factor is
\begin{equation} \label{zz-nonresonant}
	Z_{\mathcal{O}_i}(\lambda)
	=
	\Bigg(\frac{g(\mu)}{g(\frac{\mu}{\lambda})}\Bigg)^{\frac{\gamma_{0\mathcal{O}_i}}{\beta_0}} \,,
\end{equation}
where $\gamma_{0\mathcal O_i}$ are the eigenvalues of the one-loop anomalous-dimension matrix $\gamma_0$.\par
As opposed to the conformal scheme, the nonresonant one employs crucially the asymptotic freedom due to the nonvanishig beta fuction $\beta(g)$, because it uses the affine part of the gauge transformation to make the connection $A(g)$ one-loop exact. Thus, in the nonresonant scheme the anomalous dimension matrix takes the form
\begin{equation} \label{NRR}
\gamma^{{\rm NR}}(g) = -\frac{\gamma^{\overline{\rm MS}}_0}{\beta_0 g}\beta(g) = \gamma^{\overline{\rm MS}}_0 g^2+ \gamma^{\overline{\rm MS}}_0 \frac{\beta_1}{\beta_0}g^4+\cdots\,.
\end{equation}
Hence, we may now compare the conformal scheme with the nonresonant one.\par
Obviously, the values of $\gamma_0$ coincide in the conformal and nonresonant schemes by Eqs. \eqref{cs000} and \eqref{NRR}, as it should be because $\gamma_0$ is scheme independent.\par
But already at order $g^4$ the two schemes are deeply different: In the conformal scheme $\gamma_1$ has a smooth limit as $\beta_0 \rightarrow 0$ by Eq. \eqref{cs100}, while in the nonresont scheme it becomes singular as $\beta_0\to 0$ by Eq. \eqref{NRR}. Both behaviors persist at higher orders according to \cite{Braun:2003rp} and Eq. \eqref{NRR}, respectively.

	\subsection{UV asymptotics in the resonant case}
	
 In the geometric approach to operator mixing \cite{Bochicchio:2021geometry}, it was introduced the connection
	\begin{align}
		&A(g)\equiv -\frac{\gamma(g)}{\beta(g)}= \frac{A_0}{g}+ \cdots
		\label{eq:Aconnection}
	\end{align}
that has a pole at $g=0$ and regular terms represented by ellipses. Naively, the reader may think that a formal analytic gauge transformation always exists 
	\begin{equation}
		A'(g)=S(g)A(g)S^{-1}(g)+\frac{\partial S(g)}{\partial g}\,S^{-1}(g) 
		\label{eq:Agauge}
	\end{equation}
that removes all the regular terms, in order to set the connection in the one-loop exact canonical form
\begin{equation}
		A'(g)=\frac{A_0}{g} ,
		\label{eq:nonrescanon}
	\end{equation}
However, it turns out that this is not the case, so that the above scheme necessarily exists only in the nonresonant case previously discussed in the present paper.\par
Instead, for example, if all the eigenvalues, arranged in non-increasing order, of the residue of the pole of the connection are resonant 
	\begin{equation}
		\lambda_i-\lambda_j = 2k ,
		\qquad i < j,
		\label{eq:nonrescond}
	\end{equation}
	for some positive integer $k$,
	then a formal analytic gauge transformation exists that sets the connection in a different canonical form \cite{Becchetti:2021for} 
	\begin{equation}
		A'(g)=\frac{1}{g}\Big(\Lambda+N_0+\sum_{k\ge 1} N_{2k}\,g^{2k}\Big) ,
		\label{eq:rescanon}
	\end{equation}
	where $\Lambda$ is diagonal, $N_0=0$ provided that the theory is unitary in the conformal limit \cite{Becchetti:2021for}, and $N_{2k}$ are certain nilpotent upper triangular matrices \cite{Becchetti:2021for}.  \par	
Consequently, the renormalized mixing matrix reads in the coordinate representation \cite{Becchetti:2021for}
	\begin{equation}
		Z(x,\mu)
		=
		\left(\frac{g(\mu)}{g(x)}\right)^{\Lambda}
		\exp\!\left(
		\sum_{k\ge 0} g^{2k}(x)\,N_{2k}\,
		\log\frac{g(\mu)}{g(x)}
		\right)  = 	\exp\!\left(
		\sum_{k\ge 0} g^{2k}(\mu)\,N_{2k}\,
		\log\frac{g(\mu)}{g(x)}
		\right)	\left(\frac{g(\mu)}{g(x)}\right)^{\Lambda}
		\label{eq:Zres}
	\end{equation}
that enters the solution of the Callan-Symanzik equation for the matrix of $2$-point correlators in the case of resonant mixing
\begin{equation}
G^{(2)}(x;\mu,g(\mu)) = 	Z(x,\mu) \mathcal{G}^{(2)}(x;\mu,g(\mu))Z^T(x,\mu) \, ,
\end{equation}
where $\mathcal{G}^{(2)}(x;\mu,g(\mu))$ is the RG-invariant part of the correlator, which satisfies
\begin{equation}
\left(x\cdot\frac{\partial}{\partial x}+\beta(g)\frac{\partial}{\partial g}+2 D_O\right)\mathcal{G}^{(2)}(x;\mu,g(\mu)) = 0 \,.
\end{equation}
Therefore, the resonance of the eigenvalues of the one-loop anomalous dimension matrix produces an intrinsically nondiagonal UV asymptotics in Eq.\eqref{eq:Zres}, where in addition to the usual RG scaling dictated by $\Lambda= \frac{\gamma_0}{\beta_0}$, we get extra contributions involving factors of powers \footnote{They arise from the expansion of the exponentials in Eq.\eqref{eq:Zres} in power series that necessarily terminates to a certain order because of the nilpotency of the matrices involved.} of $\log(g(\mu)/g(x))$ that translate asymptotically into powers of $\log\log(1/x^2\Lambda_{RG}^2)$. \par
In fact, the resonant case is the closest analog \cite{Becchetti:2021for} in a unitary asymptotically free theory of a logarithmic conformal field theory that is nonunitary.

\subsection{Existence of the nonresonant renormalization scheme for twist-2 operators in \texorpdfstring{$\mathcal N=1$}{N=1} SUSY YM theory}

In \cite{S2}, by extending previous work in \cite{S1} for the pure YM case, it was demonstrated that the mixing of all twist-$2$ operators in  $\mathcal N=1$ SUSY $SU(N)$ YM theory is nonresonant. \par
Since $\gamma_0$ is already diagonal in the standard basis of balanced and unbalanced superfield operators \cite{Belitsky:2004sc,Belitsky:2003sh,SS1} and the corresponding eigenvalues are naturally ordered by increasing spin, the nonresonance condition becomes
\begin{equation}
	\lambda_j-\lambda_i\neq 2k, \, \, \, \,  j\geq i
	\label{eq:Nonres}
\end{equation}
with $k$ a positive integer and
\begin{equation}
	\beta_0=\frac{3}{(4\pi)^2}.
	\label{eq:Beta0}
\end{equation}
Given now the increasing order of the eigenvalues, the nilpotent matrices potentially arising in case of resonance would be lower triangular in Eq. \eqref{eq:rescanon}, as it may a-priori happen for the mixing of twist-$2$ operators that involves lower triangular matrices in Eq. \eqref{eq:OpRen}.\par
The balanced supermultiplets read  \cite{SS1}
\begin{equation}
	W_s(x,\theta,\bar\theta)\sim S^{(2)}_{s+1}+\theta \bar M_{s+1}+\bar\theta M_{s+1}
	+\theta\bar\theta S^{(1)}_{s+2},
	\label{eq:Balanced}
\end{equation}
where, for even spin $s\ge 2$,
\begin{equation}
	S_s^{(1)}=\frac{6}{s-1}O_s^A-O_s^\lambda,
	\qquad
	S_s^{(2)}=\frac{6}{s+2}O_s^A+O_s^\lambda,
	\label{eq:Se}
\end{equation}
and for odd spin $s\ge 3$,
\begin{equation}
	\widetilde S_s^{(1)}=-\frac{6}{s-1}\widetilde O_s^A-\widetilde O_s^\lambda,
	\qquad
	\widetilde S_s^{(2)}=\frac{6}{s+2}\widetilde O_s^A+\widetilde O_s^\lambda .
	\label{eq:So}
\end{equation}
The unbalanced supermultiplets read
\begin{equation}
	W_s^+(x,\theta,\bar\theta)\sim T_{s-1}+\theta S_s^A+\bar\theta \bar S_s^\lambda
	+\theta\bar\theta\,\partial_+ T_{s-1},
	\label{eq:UnbalancedPlus}
\end{equation}
and
\begin{equation}
	W_s^-(x,\theta,\bar\theta)\sim \bar T_{s-1}+\theta \bar S_s^A+\bar\theta S_s^\lambda
	+\theta\bar\theta\,\partial_+ \bar T_{s-1}.
	\label{eq:UnbalancedMinus}
\end{equation}
In the balanced sector, we get
\begin{equation}
	\gamma_{0\,O_s}
	=
	\frac{1}{(4\pi)^2}\left(\widetilde\gamma_{0\,O_s}-\frac{3}{2}\right),
	\qquad
	O_s=S_s^{(1)},S_s^{(2)},\widetilde S_s^{(1)},\widetilde S_s^{(2)},M_s,\bar M_s,
	\label{eq:BalancedGamma}
\end{equation}
with
\begin{equation}
	\widetilde\gamma_{0\,S_s^{(1)}}
	=
	\psi(s+2)+\psi(s-1)-2\psi(1)-\frac{2(-1)^s}{(s+1)s(s-1)},
	\label{eq:GammaS1}
\end{equation}
\begin{equation}
	\widetilde\gamma_{0\,S_s^{(2)}}
	=
	\psi(s+3)+\psi(s)-2\psi(1)+\frac{2(-1)^s}{(s+2)(s+1)s},
	\label{eq:GammaS2}
\end{equation}
and
\begin{equation}
	\widetilde\gamma_{0\,\widetilde S_s^{(i)}}=\widetilde\gamma_{0\,S_s^{(i)}},
	\qquad
	\widetilde\gamma_{0\,M_s}=\widetilde\gamma_{0\,\bar M_s}=\widetilde\gamma_{0\,S_s^{(2)}}.
	\label{eq:GammaBalancedEqual}
\end{equation}
In the unbalanced sector, we instead get
\begin{equation}
	\gamma_{0\,O_s}
	=
	\frac{1}{(4\pi)^2}\left(\widetilde\gamma_{0\,O_s}-\frac{3}{2}\right),
	\qquad
	O_s=S_s^A,\widetilde S_s^A,S_s^\lambda,\widetilde S_s^\lambda,T_s,\bar T_s,
	\label{eq:UnbalancedGamma}
\end{equation}
with
\begin{equation}
	\widetilde\gamma_{0\,S_s^A}
	=
	\widetilde\gamma_{0\,S_s^\lambda}
	=
	2\psi(s+1)-2\psi(1),
	\label{eq:GammaSA}
\end{equation}
and
\begin{equation}
	\widetilde\gamma_{0\,T_s}
	=
	\begin{cases}
		\psi(s+1)-\psi(1), & s=2,4,\ldots,\\[2mm]
		\psi(s+2)-\psi(1), & s=1,3,\ldots.
	\end{cases}
	\label{eq:GammaT}
\end{equation}
The proof \cite{S2} of nonresonance of the eigenvalues is based on the identity
\begin{equation}
	\psi(n+1)-\psi(1)=H_n,
	\qquad
	H_n=\sum_{k=1}^n\frac1k,
	\label{eq:DigammaH}
\end{equation}
which rewrites all the differences of eigenvalues in terms of harmonic numbers or harmonic-type sums. 
\subsubsection{Strategy for a number-theoretic proof}\label{sec:s2}

The proof in \cite{S2} involves the concept of $p$-adic order $\nu_p$ together with Bertrand postulate. \par
The $p$-adic order of an integer $n$ is defined as the exponent of the highest power of a prime number $p$ that divides $n$ \cite{ireland1990classical}. 
Specifically, the $p$-adic order of an integer is the function
\begin{equation}
	\nu_p(n)=
	\begin{cases}
		\mathrm{max}\{k \in \mathbb{N} : p^k\,\text{divides}\,n\} & \text{if } n \neq 0\\
		\infty & \text{if } n=0\,.
	\end{cases}
\end{equation}
For example, $\nu_3(24) = \nu_3(3\times2^3) = 1$ and $\nu_2(24) = 3$.\par
This concept can be extended to rational numbers through the property \cite{ireland1990classical}
\begin{equation}
	\label{eq:p1}
	\nu_p\left(\frac{a}{b}\right)  =\nu_p(a)-\nu_p(b)\,.
\end{equation}
Consequently, rational numbers may have a negative $p$-adic order, whereas integers can only have nonnegative values for any prime $p$.
Additional properties include \cite{ireland1990classical}
\begin{align}
	\label{eq:p2}
	&\nu_p(a\cdot b) = \nu_p(a)+\nu_p(b)\nonumber\\
	&\nu_p(a+b)\geq\min\bigl\{ \nu_p(a), \nu_p(b)\bigr\}\,.
\end{align}
and, if $\nu_p(a) \neq \nu_p(b)$,
\begin{equation}
	\label{equal}
	\nu_p(a+b)= \min\bigl\{ \nu_p(a), \nu_p(b)\bigr\}
\end{equation}
\cite{ireland1990classical}, a fact which is crucial for the proof in \cite{S2} .\par 
Bertrand postulate\footnote{It is actually a theorem.} \cite{ramanujan1919proof} asserts that for any integer $n\geq2$, there is at least one prime number $p$ satisfying
\begin{equation}\label{eq:bertrand}
	\frac{n}{2}+1\leq p\leq n\,.
\end{equation} 
This implies that for any prime $p\in[\frac{n}{2}+1,n]$, its double, $2p$, cannot lie within the same interval, since $2p\geq n+2$. \par
We will apply Bertrand postulate to prove the classical result that Harmonic numbers
\begin{align}\label{eq:Hn}
	H_n = \sum_{k = 1}^{n}\frac{1}{k}\,,
\end{align}
are never integers for any $n\geq 2$.
The key point in the proof is a standard argument \cite{S1,S2} in number theory, i.e. that if a prime $p$ occurs in one denominator and in no other denominator of the sum for a fixed $n$, then the total $p$-adic order is negative and the sum cannot be an integer, thus implying the nonresonant condition
for the eigenvalues. \par
Sometimes, it is necessary to extend the above argument -- that we will refer to as the generalized argument -- to sums of the type \cite{S2}
\begin{equation}
	\Omega_n=\sum_{k=1}^n \frac{c_k}{k},
	\qquad
	c_k\in\{\pm1,\pm2,\pm3,0\}
	\label{eq:Omega}
\end{equation}
that occur in the balanced sector, the proof being then organized by sectors.\par
\subsubsection{The actual proof}
For the unbalanced operators $S_s^A$ and $S_s^\lambda$, the eigenvalues are monotonically increasing and
\begin{equation}
	\Delta_m^{S^{A,\lambda}}(x)
	\equiv
	\frac{\gamma_{0,m+x}^{S^{A,\lambda}}-\gamma_{0,m}^{S^{A,\lambda}}}{\beta_0}
	=
	\frac43\sum_{k=m+1}^{m+x}\frac1k
	\equiv \frac43\,\Sigma_m(x).
	\label{eq:DeltaSA}
\end{equation}
For $x\ge m$, by writing $x=m+t$, Bertrand postulate \cite{ramanujan1919proof} gives us a prime in the interval 
\begin{equation}
	m+1+\frac t2\le p\le 2m+t,
	\label{eq:PrimeSA}
\end{equation}
so that $\nu_p(\Sigma_m(m+t))=-1$ and therefore $\Delta_m^{S^{A,\lambda}}(m+t)$ cannot be an integer. For the remaining range $x<m$, 
the bound holds \cite{S2}
\begin{equation}
	\Sigma_m(x)<\log 2,
	\qquad
	\Delta_m^{S^{A,\lambda}}(x)
	<
	\frac43\log 2<0.93,
	\label{eq:BoundSA}
\end{equation}
which excludes the resonance as well.  \par
For the unbalanced fermionic operators $T_s$ and $\bar T_s$, we write 
\begin{equation}
	\gamma_{0n}^{T}
	=
	\frac{2}{(4\pi)^2}\left(H_{2n}-\frac32\right),
	\qquad n=1,2,3,\ldots ,
	\label{eq:GammaTn}
\end{equation}
so that
\begin{equation}
	\Delta_m^T(x)
	\equiv
	\frac{\gamma_{0,m+x}^{T}-\gamma_{0,m}^{T}}{\beta_0}
	=
	\frac23\sum_{k=2m+1}^{2m+2x}\frac1k
	\equiv \frac23\,\Sigma'_m(x).
	\label{eq:DeltaT}
\end{equation}
Exactly the same strategy applies: For $x\ge m$ the standard $p$-adic argument excludes integrality, while for $x<m$ the bound holds \cite{S2} 
\begin{equation}
	\Sigma'_m(x)<\log 2,
	\qquad
	\Delta_m^T(x)
	<
	\frac23\log 2<0.5.
	\label{eq:BoundT}
\end{equation}
For the first family of even-spin balanced operators, we get
\begin{equation}
	\gamma_{0n}^{S^{(1)}}=
	\frac{2}{(4\pi)^2}
	\left(
	2H_{n-2}+\frac{3}{n}-\frac32
	\right),
	\qquad n=2,4,6,\ldots 
	\label{eq:GammaSeven1}
\end{equation}
and
\begin{equation}
	\Delta_m^{S^{(1)}}(x)
	=
	\frac{\gamma_{0,m+x}^{S^{(1)}}-\gamma_{0,m}^{S^{(1)}}}{\beta_0}
	=
	\frac23\,K_m(x),
	\qquad
	K_m(x)=2\sum_{k=m-1}^{m-2+x}\frac1k+\frac{3}{m+x}-\frac{3}{m}.
	\label{eq:DeltaSeven1}
\end{equation}
Here, the generalized argument applies for $x\ge m$ because $K_m(m+t)$ has the form in eq. \eqref{eq:Omega}, with coefficients $\pm2,\pm3$.
Moreover, for $x<m$, the bound holds \cite{S2}
\begin{equation}
	K_m(x)<2\log 2,
	\qquad
	\Delta_m^{S^{(1)}}(x)<\frac43\log 2<0.93.
	\label{eq:BoundSeven1}
\end{equation}
For the second family of even-spin balanced operators, we obtain
\begin{equation}
	\gamma_{0n}^{S^{(2)}}=
	\frac{2}{(4\pi)^2}
	\left(
	2H_n+\frac{2(-1)^n}{(n+2)(n+1)n}-\frac32
	\right),
	\qquad n=2,4,6,\ldots ,
	\label{eq:GammaSeven2}
\end{equation}
and
\begin{equation}
	\Delta_m^{S^{(2)}}(x)
	=
	\frac23\,J_m(x),
	\qquad
	J_m(x)=
	2\sum_{k=m+1}^{m+x}\frac1k
	+\frac{2(-1)^{m+x}}{(m+x+2)(m+x+1)(m+x)}
	-\frac{2(-1)^m}{(m+2)(m+1)m}.
	\label{eq:DeltaSeven2}
\end{equation}
Again, the generalized argument controls the case $x\ge m$, while for $x<m$ the bound holds \cite{S2}
\begin{equation}
	J_m(x)<2\log 2,
	\qquad
	\Delta_m^{S^{(2)}}(x)<\frac43\log 2<0.93.
	\label{eq:BoundSeven2}
\end{equation}
Since $\widetilde\gamma_{0\,M_s}=\widetilde\gamma_{0\,\bar M_s}=\widetilde\gamma_{0\,S_s^{(2)}}$, we arrive at the same conclusion for $M_s$ and $\bar M_s$.\par
For odd-spin balanced operators, we get for the first family
\begin{equation}
	\gamma_{0n}^{S^{(1)}}=
	\frac{2}{(4\pi)^2}
	\left(
	2H_{n-2}+\frac{2}{n-1}-\frac1n+\frac{2}{n+1}-\frac32
	\right),
	\qquad n=3,5,7,\ldots ,
	\label{eq:GammaSodd1}
\end{equation}
with
\begin{equation}
	\Delta_m^{S^{(1)}}(x)
	=
	\frac23\,L_m(x),
	\qquad
	L_m(x)=
	2\sum_{k=m-1}^{m+x-2}\frac1k
	+\frac{2(-1)^{m+x-1}}{(m+x)(m+x-1)}
	-\frac{2(-1)^{m-1}}{m(m-1)}.
	\label{eq:DeltaSodd1}
\end{equation}
For $x\ge m$ the generalized argument applies, while for $x<m$ the bound holds \cite{S2}
\begin{equation}
	L_m(x)<2\log 2+\frac32,
	\qquad
	\Delta_m^{S^{(1)}}(x)
	<
	\frac43\log 2+1<1.93
	\label{eq:BoundSodd1}
\end{equation}
that it is still sufficient, since it excludes that $\Delta_m^{S^{(1)}}(x)$ be an even integer. \par
Finally, for the second odd-spin family we obtain
\begin{equation}
	\gamma_{0n}^{S^{(2)}}=
	\frac{2}{(4\pi)^2}
	\left(
	2H_{n+1}-\frac{2(-1)^n}{(n+2)(n+1)}-\frac32
	\right),
	\qquad n=1,3,5,\ldots ,
	\label{eq:GammaSodd2}
\end{equation}
and
\begin{equation}
	\Delta_m^{S^{(2)}}(x)
	=
	\frac23\,U_m(x),
	\qquad
	U_m(x)=
	2\sum_{k=m+2}^{m+x+1}\frac1k
	-\frac{2(-1)^{m+x}}{(m+x+2)(m+x+1)}
	+\frac{2(-1)^m}{(m+2)(m+1)}.
	\label{eq:DeltaSodd2}
\end{equation}
Again, the generalized argument implies nonintegrality for $x\ge m$, whereas for $x<m$ we get the boud \cite{S2}
\begin{equation}
	U_m(x)<2\log 2,
	\qquad
	\Delta_m^{S^{(2)}}(x)<\frac43\log 2<0.93.
	\label{eq:BoundSodd2}
\end{equation}
Hence, all twist-$2$ operators in $\mathcal N=1$ SUSY YM theory satisfy the nonresonance condition in Eq. \eqref{eq:Nonres}, thus providing the arithmetic foundation of the existence of the nonresonant diagonal scheme in the present paper and in our first installment \cite{BPS41}.
	
	\subsection{Topology of the large-$N$ expansion in SU($N$) versus U($N$) YM and $\mathcal{N}=1$ SUSY  YM theory and spin-statistics theorem}\label{sec:114}
	
	\subsubsection{A sign puzzle involving the spin-statistics theorem}
\begin{figure}[h!]
		\centering
		\includegraphics[width=0.5\linewidth]{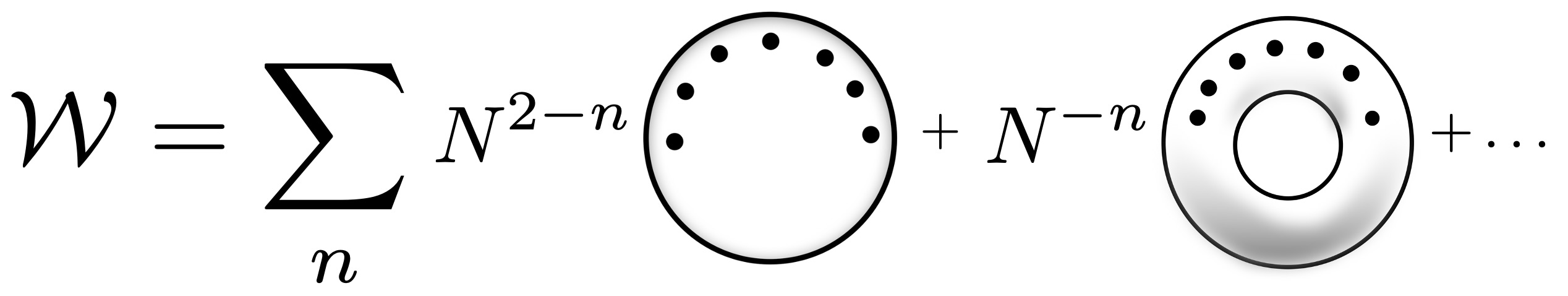}
		\caption{'t Hooft topological expansion of the generating functional, with $n$ the number of operator insertions and the ellipses higher genera.}
		\label{fig:spheretorus}
	\end{figure}

	In \cite{BPSL} we observed that assuming the Euclidean generating functional of connected correlators of single-trace operators ${\mathcal O}$ in pure YM theory,
	\begin{align}
		&\mathcal{Z}^E[J_{\mathcal{O}}]=\frac{1}{\mathcal{Z}^E} \int \mathcal{D}A\, e^{-  S_{YM}+\sum_s\, \int J_{\mathcal{O}_s}\mathcal{O}_s}\,,
	\end{align}
	admits the 't Hooft large-$N$ expansion only consisting of smooth Riemann surfaces (Fig. \ref{fig:spheretorus})
	\begin{equation}
		\label{thooft-pure}
		\mathcal{W}^E[J_{\mathcal O}]=\mathcal{W}^E_{\text{sphere}}[J_{\mathcal O}]+\mathcal{W}^E_{\text{torus}}[J_{\mathcal O}]+ \cdots \,,
	\end{equation}
	leads to the following puzzle about the spin-statistics theorem: After the RG improvement of the perturbative contribution, using the scheme discussed above and in \cite{BPSpaper2,BPSL}, we found the asymptotic generating functional for correlators of balanced twist-$2$ operators
	\begin{align}
		\label{Wasymsign}
		\mathcal{W}^E_{\text{asym}}[J_{\mathbb{O}^E},\lambda]
		=
		-(N^2-1)\log\text{Det}\left(\mathbb{I}+\frac{1}{2}\frac{Z_{\mathbb{O}_s}(\lambda)}{\lambda^{s+2}}\Delta^{-1}\frac{J_{\mathbb{O}^E_{s}}}{N}\otimes\mathcal{Y}_{s-2}^{E\frac{5}{2}}\right) \,,
	\end{align}
	whose leading-order nonplanar contribution reads
	\begin{align}\label{eq:we}
		\mathcal{W}^E_{\text{asym\,torus}}[J_{\mathbb{O}^E},\lambda]
		=
		+\log\text{Det}\left(\mathbb{I}+\frac{1}{2}\frac{Z_{\mathbb{O}_s}(\lambda)}{\lambda^{s+2}}\Delta^{-1} \frac{J_{\mathbb{O}^E_s}}{N}\otimes\mathcal{Y}_{s-2}^{E\frac{5}{2}}\right) \,.
	\end{align}
	The above object has the correct determinant structure to match asymptotically the glueball one-loop effective action, but with the sign opposite to the one required by bosonic statistics for the glueballs, since glueball interpolating fields in large-$N$ YM theory have integer spin and thus glueballs should be bosons by the spin-statistics theorem.\par Nevertheless, the sign in Eq. \eqref{eq:we} is correct, since it is inherited from the perturbative gluon determinant  at one-loop and the fact that in the SU($N$) YM theory there are exactly $N^2-1$ gluons. This is the sign puzzle that involves the spin-statistics theorem for the glueballs in large-$N$ YM theory \cite{BPSL}.

	\subsubsection{Solution to the sign puzzle and refinement of the large-$N$ expansion in the SU($N$) theory}
	
\begin{figure}[h]
		\centering
		\includegraphics[width=0.7\linewidth]{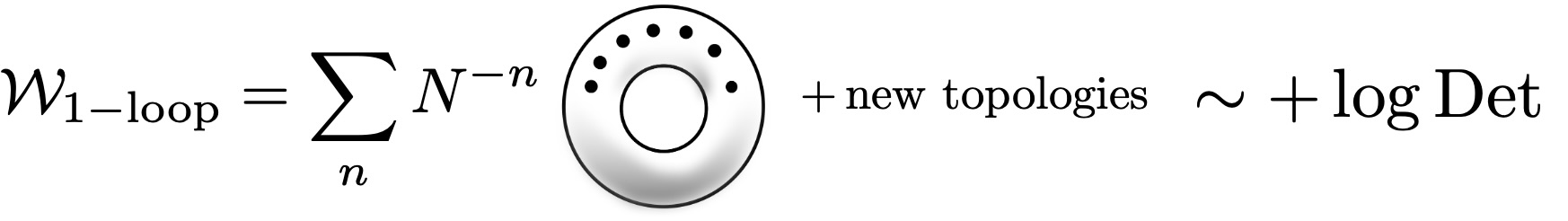}
		\caption{New topologies  solve the sign puzzle compatibly with the spin-statistics theorem.}
		\label{fig:newtopo}
	\end{figure}

\begin{figure}[h]
		\centering
		\includegraphics[width=0.4\linewidth]{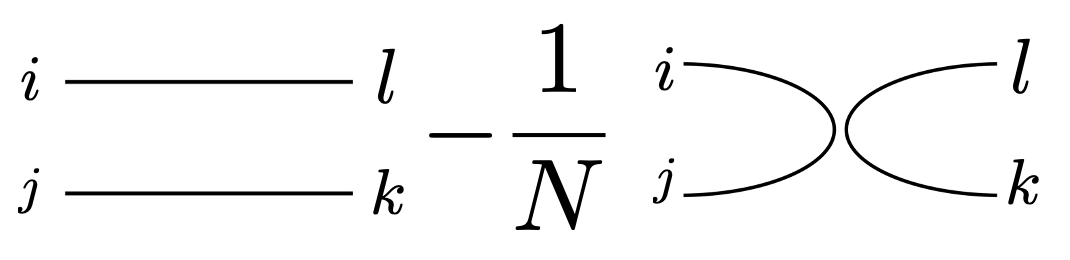}
		\caption{Leading and subleading SU($N$) gluon propagator in double line.}
		\label{fig:propagator}
	\end{figure}

\begin{figure}[h!]
		\centering
		\includegraphics[width=0.2\linewidth]{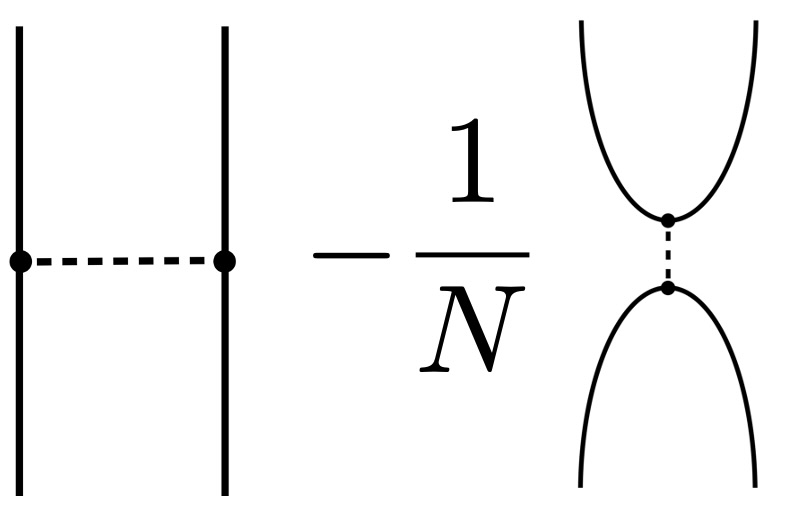}
		\caption{Local vertex for $2$-gluon operators in double-line. The dotted lines identify the punctures.}
		\label{fig:2vertex}
	\end{figure}

	The sign puzzle can be solved by observing that in the SU($N$) YM theory 't Hooft expansion involving only smooth closed Riemann surfaces is actually incomplete \cite{BPSL}  (Fig. \ref{fig:newtopo}).
	To understand why new topologies arise, we reconsidered the proof \cite{BPSL} of the 't Hooft expansion in the SU($N$) theory. Contrary to the original U($N$) case, the SU($N$) gluon propagator has both a leading and subleading term in double-line notation (Fig. \ref{fig:propagator}),
	\begin{equation}
		\langle A^i_{\, j} A^k_{\, l}\rangle \propto I^{i \,  k}_{\, j \, l}-P^{i \, k}_{\, j \, l}\,,
	\end{equation}
	where $I^{i \, k}_{\, j\, l}=\delta^i_{\, l}\delta^k_{\, j}$ and $P^{i \, k}_{\, j \, l}=\frac{1}{N}\delta^i_{\, j}\delta^k_{\, l}$ is the U(1) projector.
	The same statement holds for $2$-gluon operators and, in particular, twist-$2$ operators. Their local vertex involves $\delta^{ab}$ and carries a $\frac{1}{N}$ correction, exactly as the gluon propagator does (Fig. \ref{fig:2vertex}). 
	Interestingly, this is not cured by passing to the U($N$) theory. In fact, we found for the generating functional of RG-improved correlators in the U($N$) theory \cite{BPSL}
	\begin{align} \label{15}
		\mathcal{W}^E_{{\text{U}(N)}\text{asym}}[J_{\mathbb{O}},\lambda]
		=&-(N^2-1)\log\Det\left(\mathbb{I}+\frac{1}{2}\frac{Z_{\mathbb{O}_s}(\lambda)}{\lambda^{s+2}}\Delta^{-1}\frac{J_{\mathbb{O}^E_{s}}}{N}\otimes\mathcal{Y}_{s-2}^{E\frac{5}{2}}\right)\nonumber\\
		&-\log\Det\left(\mathbb{I}+\frac{1}{2}\frac{1}{\lambda^{s+2}}\Delta^{-1}\frac{J_{\mathbb{O}^E_{s}}}{N}\otimes\mathcal{Y}_{s-2}^{E\frac{5}{2}}\right)\,.
	\end{align}
	The first term in the above equation is the SU($N$) contribution that dominates the UV because twist-$2$ operators have positive $\gamma_0$, but for the stress-energy tensor. The second term is the free U($1$) contribution that is conformal and decouples. Thus, the sign problem in the renormalized U($N$) theory is exactly the same as in the SU($N$) theory. \par
\begin{figure}[h!]
		\centering
		\includegraphics[width=0.7\linewidth]{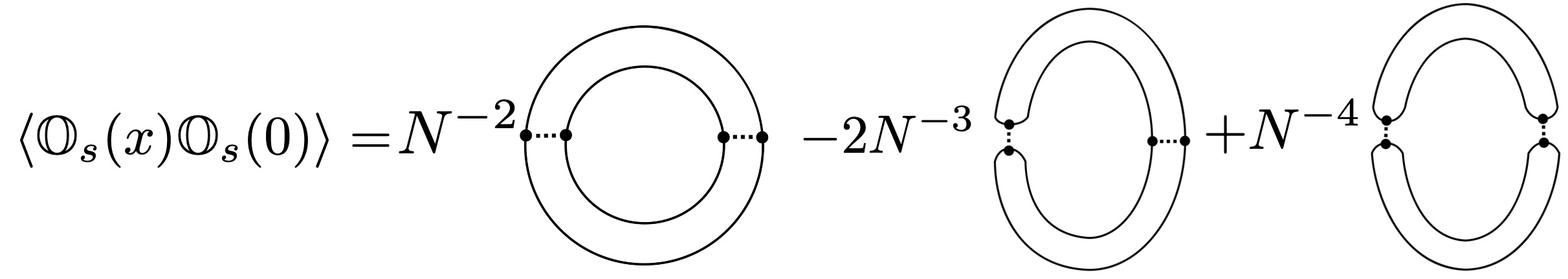}
		\caption{$2$-point correlators of twist-$2$ operators to the leading perturbative order, with signs and weights as in Fig. \ref{fig:nonrenorm1}.}
		\label{fig:2pointcolor}
	\end{figure}
\begin{figure}[h!]
		\centering
		\includegraphics[width=0.7\linewidth]{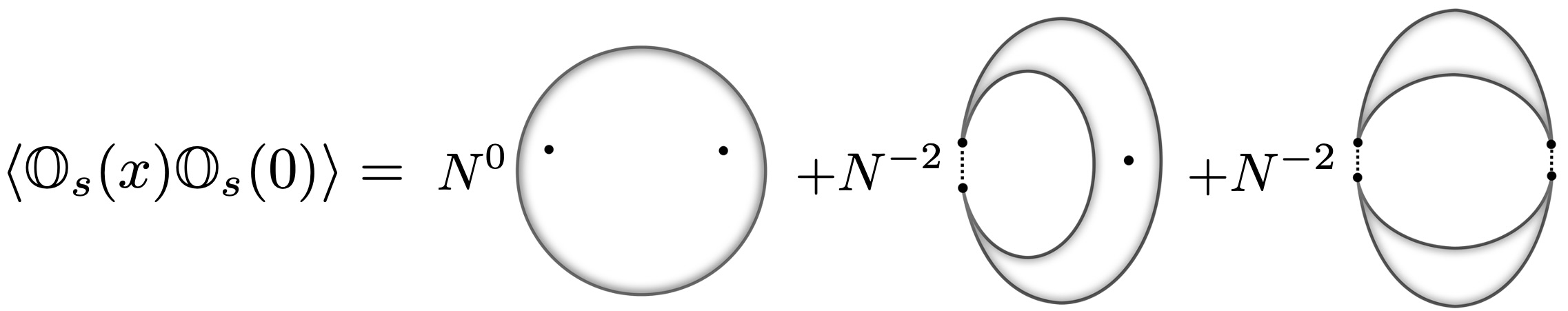}
		\caption{Topology and weights of diagrams in Fig. \ref{fig:2pointcolor}.}
		\label{fig:2pointnewtopos}
	\end{figure}
\begin{figure}[h!]
		\centering
		\includegraphics[width=0.7\linewidth]{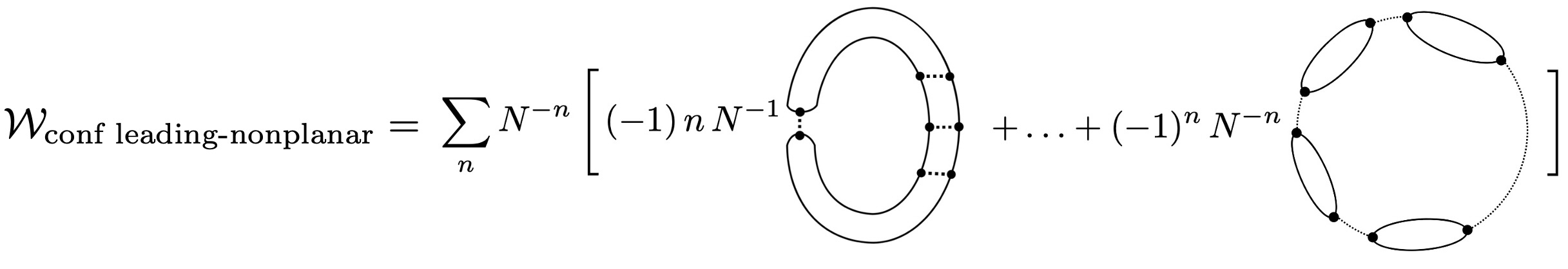}
		\caption{The refined perturbative expansion of the conformal leading-nonplanar generating functional: The square brackets contain the sum over $1 \leq p \leq n$ of possibly pinched annuli involving $n-p$ pairs of punctures and $p$ pinches arising from $p$ insertions of $P$, which give rise to the weights $N^{-p}$ compensated for by the color traces represented by loops. Some relative signs and combinatorics in the expansion of the second line of Eq. \eqref{Wconf1} with respect to the corresponding planar contributions in the first line are also displayed.}
		\label{fig:nonrenorm1}
	\end{figure}
\begin{figure}[h!]
		\centering
		\includegraphics[width=0.7\linewidth]{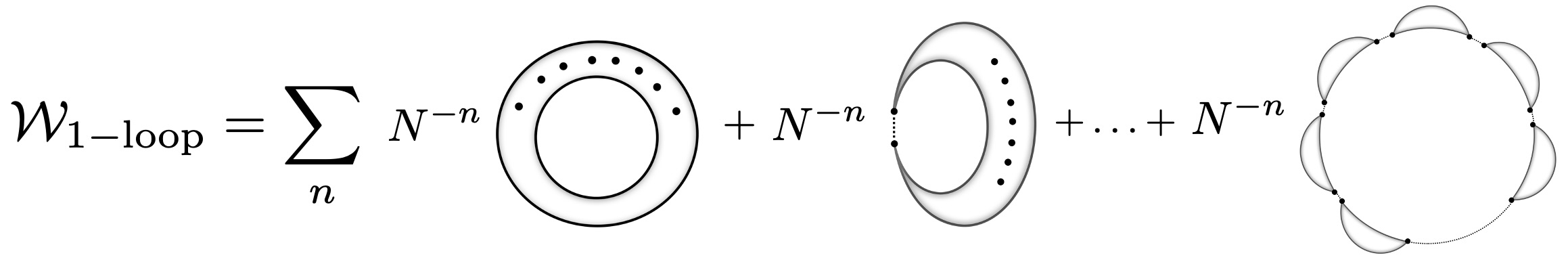}
		\caption{The refined topological expansion of the glueball one-loop generating functional.}
		\label{fig:w1loop}
	\end{figure}
\begin{figure}[h!]
		\centering
		\includegraphics[width=0.7\linewidth]{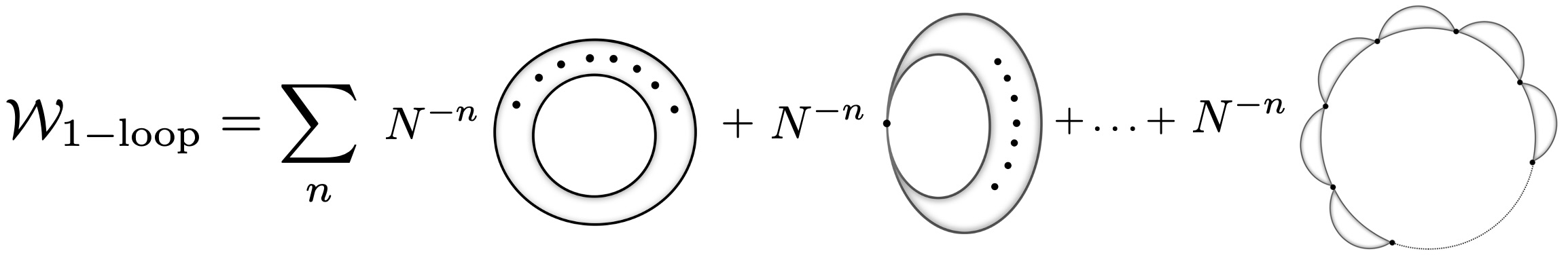}
		\caption{The new topologies in Fig. \ref{fig:w1loop} arise from the components of the normalization of $p$-pinched tori with $n-p$ punctures obtained by cutting the pinches.}
		\label{fig:w2loop}
	\end{figure}

	To solve the sign problem in the SU($N$) theory, we rewrote the conformal generating functional identically
	\begin{align}
		\label{Wconf0}
		\mathcal{W}^E_{\text{conf}}[J_{\mathbb{O}^E}]&=-(N^2-1)\log\Det\left(\mathcal{I}+\frac{1}{2}\Delta^{-1}\frac{J_{\mathbb{O}^E_{s}}}{N}\otimes\mathcal{Y}_{s-2}^{E\frac{5}{2}}\right)\nonumber\\ &=-\log\Det\left(\mathcal{I}+\frac{1}{2}(I-P)\Delta^{-1}\frac{J_{\mathbb{O}^E_{s}}}{N}\otimes\mathcal{Y}_{s-2}^{E\frac{5}{2}}\right) \,,
	\end{align}
	where the term containing $I-P$ involves the SU($N$) propagator (Fig. \ref{fig:propagator}) and $\mathcal{I}$ is the identity in both color and space-time.
	The aforementioned equality follows by noticing that the color trace $\Tr$ in the loop expansion in the second line of Eq. \eqref{Wconf0} containing the insertion of $n$ sources $J_{\mathbb{O}^E_{s}}$ produces the overall factor 
	\begin{align}
		\Tr(I-P)^n= \Tr(I-P)=N^2-1
	\end{align}
	that occurs in the first line of Eq. \eqref{Wconf0}.
	In order to keep the 't Hooft double-line representation that only involves the leading propagator also beyond the planar limit of the SU($N$) theory, we transferred in Eq. \eqref{Wconf0} the $\frac{1}{N}$ dependence from the propagator to the local vertex for $2$-gluon operators (Fig. \ref{fig:2vertex}) by the identity
	\begin{align}
		\label{Wconf01}
		&\mathcal{W}^E_{\text{conf}}[J_{\mathbb{O}^E}] \nonumber\\
		&=-\log\Det\left(\mathcal{I}+\frac{1}{2}(I-P)\Delta^{-1}\frac{J_{\mathbb{O}^E_{s}}}{N}\otimes\mathcal{Y}_{s-2}^{E\frac{5}{2}}\right)\nonumber\\
		&=-\log\Det\left(\mathcal{I}+\frac{1}{2}I\Delta^{-1}\frac{J_{\mathbb{O}^E_{s}}}{N}\otimes\mathcal{Y}_{s-2}^{E\frac{5}{2}}-\frac{1}{2}P\Delta^{-1}\frac{J_{\mathbb{O}^E_{s}}}{N}\otimes\mathcal{Y}_{s-2}^{E\frac{5}{2}}\right) \nonumber \\
		&=-\log\Det\Bigg[\Big(\mathcal{I}+\frac{1}{2}\Delta^{-1} I \frac{J_{\mathbb{O}^E_{s}}}{N}\otimes\mathcal{Y}_{s-2}^{E\frac{5}{2}}\Big) \nonumber \\
		&\,\,\,\,\,\,\,\,\, \left(\mathcal{I}-\frac{1}{2}\Big(\mathcal{I}+\frac{1}{2}\Delta^{-1} I \frac{J_{\mathbb{O}^E_{s}}}{N}\otimes\mathcal{Y}_{s-2}^{E\frac{5}{2}}\Big)^{-1}\Delta^{-1}P\frac{J_{\mathbb{O}^E_{s}}}{N}\otimes\mathcal{Y}_{s-2}^{E\frac{5}{2}}\right)\Bigg] \,,
	\end{align}
	so that
	\begin{align}
		\label{Wconf1}
		\mathcal{W}^E_{\text{conf}}[J_{\mathbb{O}^E}]
		&=-\log\Det\left(\mathcal{I}+\frac{1}{2}\Delta^{-1} I \frac{J_{\mathbb{O}^E_{s}}}{N}\otimes\mathcal{Y}_{s-2}^{E\frac{5}{2}}\right)\nonumber\\
		&\quad-\log\Det\Bigg(\mathcal{I}-\frac{1}{2}\Big(\mathcal{I}+\frac{1}{2}\Delta^{-1} I \frac{J_{\mathbb{O}^E_{s}}}{N}\otimes\mathcal{Y}_{s-2}^{E\frac{5}{2}}\Big)^{-1} \Delta^{-1}P\frac{J_{\mathbb{O}^E_{s}}}{N}\otimes\mathcal{Y}_{s-2}^{E\frac{5}{2}}\Bigg) \,. \nonumber\\
	\end{align}
	The first $\log \Det$ above is the ('t Hooft-)planar contribution -- involving the leading local vertex (Fig. \ref{fig:2vertex}) -- and the second $\log \Det$ the leading-nonplanar one -- involving at least one subleading local vertex (Fig. \ref{fig:2vertex}) carrying the factor of $P$.
	The latter gives rise to new topologies. For example, for $2$-point correlators (Fig. \ref{fig:2pointcolor}),
	by the standard 't Hooft gluing of reversely oriented lines, the first diagram in Fig. \ref{fig:2pointcolor} -- that is the planar contribution -- leads to a $2$-punctured sphere and the remaining ones -- by representing a $2$-punctured disk as an infinite strip and gluing the opposite edges of the strip to get an infinite cylinder, i.e., a $2$-punctured sphere (Fig. \ref{fig:2pointnewtopos}) -- to possibly disconnected punctured spheres with at least two punctures pairwise identified (Fig. \ref{fig:2pointnewtopos}). 
	By generalizing the new diagrams in Fig. \ref{fig:2pointcolor}, the leading-nonplanar contributions due to the second $\log \Det$ in Eq. \eqref{Wconf1} involve the new topological sector (Fig. \ref{fig:nonrenorm1}).
	After the RG improvement, the leading-nonplanar asymptotic generating functional reads
	\begin{align}
		\label{Wasnp}
		&\mathcal{W}^E_{\text{asym nonplanar}}[J_{\mathbb{O}^E},\lambda] \nonumber\\
		&=-\log\Det\Bigg(\mathcal{I}-\frac{1}{2}\Big(\mathcal{I}+ \frac{1}{2} \frac{Z_{\mathbb{O}_s}(\lambda)}{\lambda^{s+2}}  \Delta^{-1} I \frac{J_{\mathbb{O}^E_{s}}}{N}\otimes\mathcal{Y}_{s-2}^{E\frac{5}{2}}\Big)^{-1} \frac{Z_{\mathbb{O}_s}(\lambda)}{\lambda^{s+2}}\Delta^{-1}P\frac{J_{\mathbb{O}^E_{s}}}{N}\otimes\mathcal{Y}_{s-2}^{E\frac{5}{2}}\Bigg)\nonumber \\
		&=+\log\text{Det}\left(\mathbb{I}+\frac{1}{2}\frac{Z_{\mathbb{O}_s}(\lambda)}{\lambda^{s+2}}\Delta^{-1} \frac{J_{\mathbb{O}^E_s}}{N}\otimes\mathcal{Y}_{s-2}^{E\frac{5}{2}}\right) \,.
	\end{align}
	Most remarkably, now the overall sign of the first $\log \Det$ in Eq. \eqref{Wasnp} is consistent with the bosonic statistics for the glueballs, but at the price of 
	introducing a refined topological expansion where, in addition to the $n$-punctured tori, punctured spheres (Fig. \ref{fig:w1loop}) that are the normalization \cite{Chan2021ModuliSO,Witten:2012ga} of $p$-pinched and $n-p$-punctured tori (Fig. \ref{fig:w2loop}) arise, with $1 \leq p \leq n$. 
	Interestingly, in the new sector the weights differ \footnote{Indeed, adding one pinch increases $\chi$ by one with respect to the corresponding smooth surface.} from $N^{\chi}$, unless the pinched tori are thought of as a specific degeneration of the smooth ones, where some cycles of the smooth tori collapse to points and collide with an equal number of punctures.
	
	\subsubsection{Nonperturbative interpretation of the refined topological expansion}
	
	We provided a nonperturbative interpretation of our refined topological expansion for twist-$2$ operators \cite{BPSL}.\par
\begin{figure}[h!]
		\centering
		\includegraphics[width=0.5\linewidth]{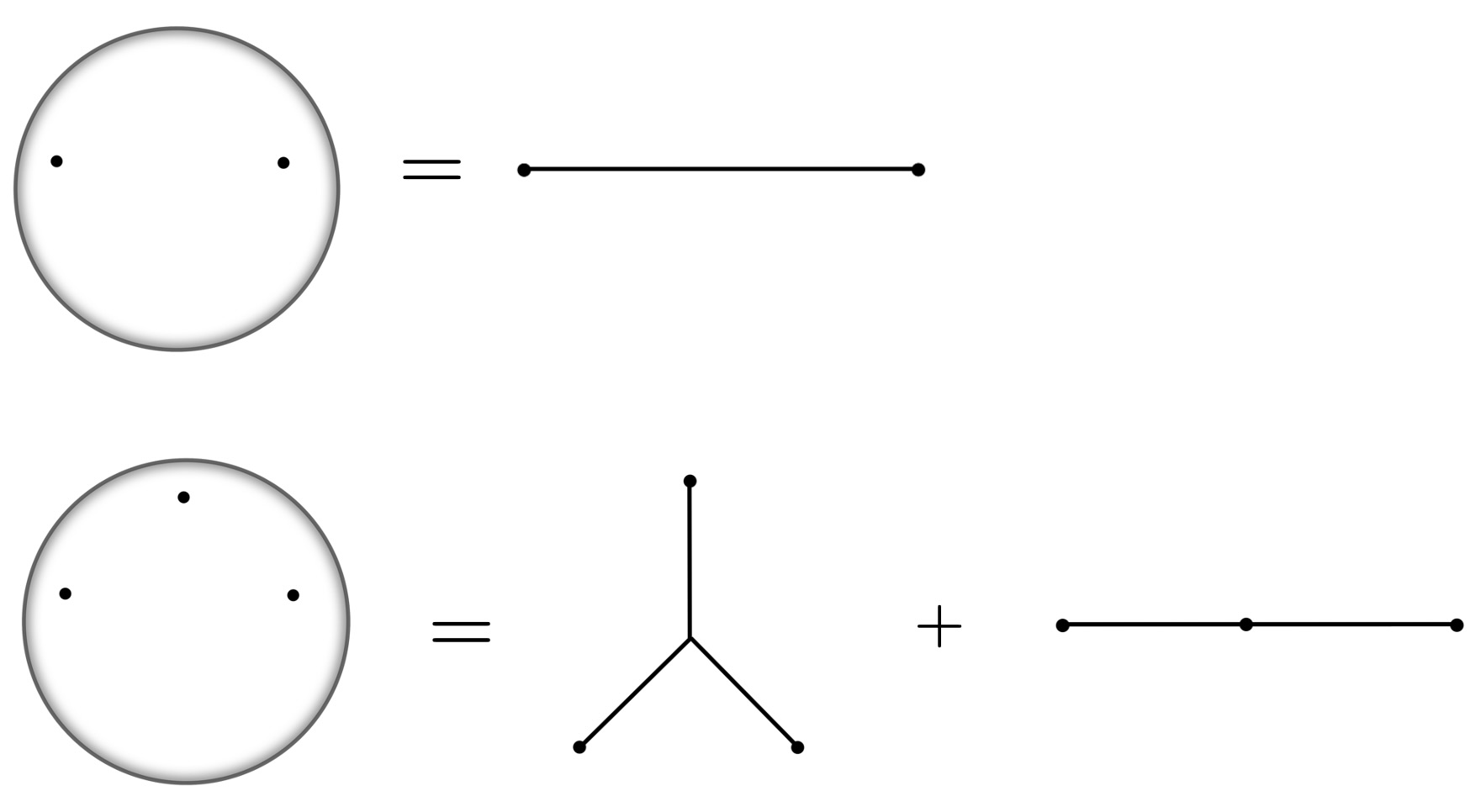}
		\caption{Glueball tree graphs dual to punctured spheres in the effective theory of glueballs.}
		\label{fig:spheres}
	\end{figure}

	It has been known for fifty years that, in the effective theory of glueballs, punctured spheres correspond to glueball tree graphs \cite{Migdal:1977nu,Witten:1979kh}. 
	Specifically, the sphere with two punctures corresponds to an infinite sum of glueball propagators \cite{Migdal:1977nu}, while the sphere with three punctures corresponds in Minkowskian space-time to vertices that are sums of three or two glueball poles \cite{Witten:1979kh} (Fig. \ref{fig:spheres}).\par
	Incidentally, we pointed out that the last graph in Fig. \ref{fig:spheres} contributes zero to the $S$ matrix because of the missing glueball external leg.
		\par
\begin{figure}[h]
		\centering
		\includegraphics[width=0.7\linewidth]{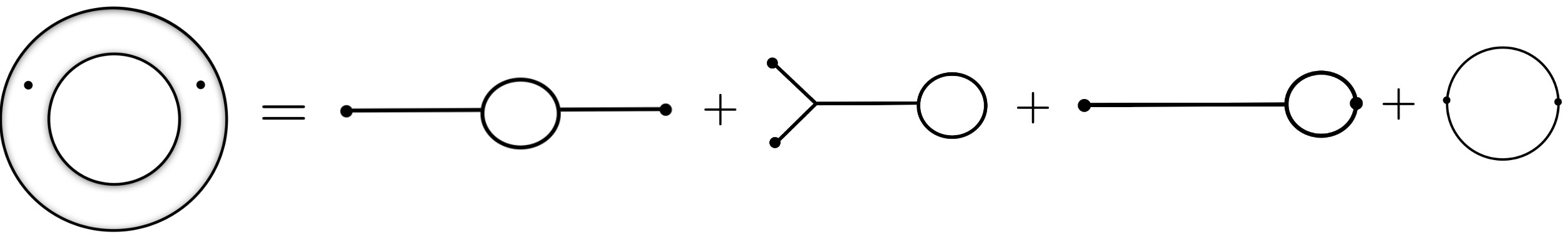}
		\caption{Glueball one-loop graphs dual to the $2$-punctured torus.}
		\label{fig:torus}
	\end{figure}
\begin{figure}[h]
		\centering
		\includegraphics[width=0.5\linewidth]{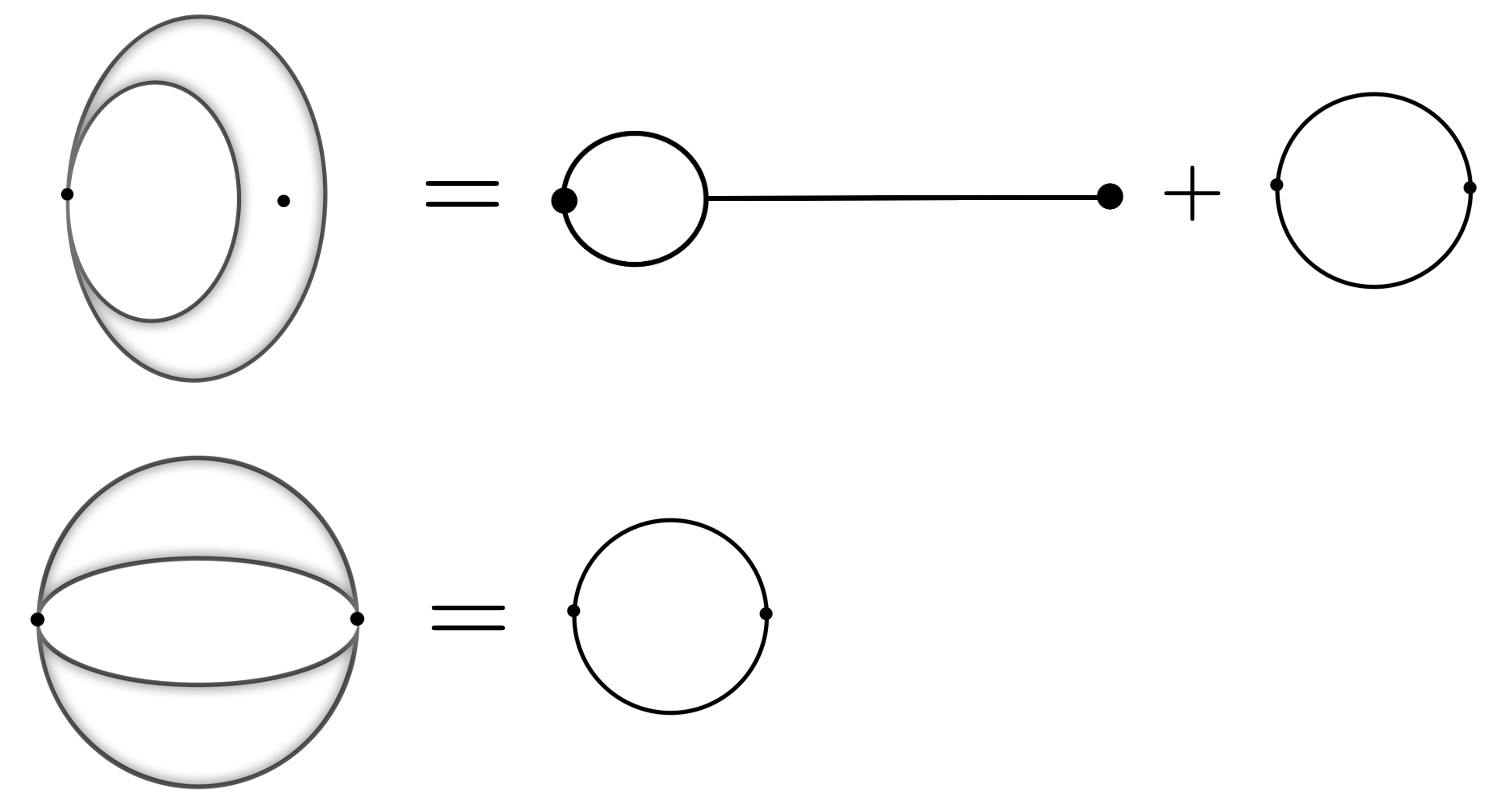}
		\caption{Glueball one-loop graphs dual to pinched tori.}
		\label{fig:maxdisc}
	\end{figure}
\begin{figure}[h!]
		\centering
		\includegraphics[width=0.7\linewidth]{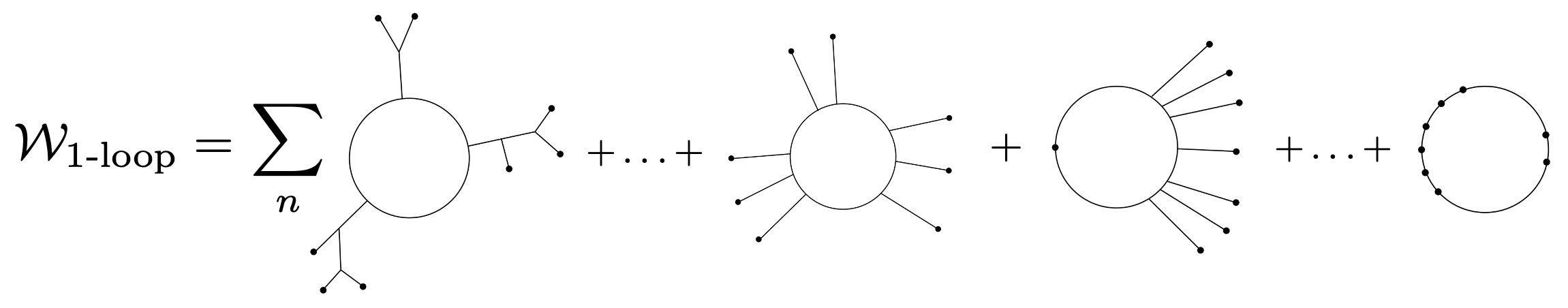}
		\caption{Glueball one-loop generating functional}
		\label{fig:w1loopspectral}
	\end{figure}

	Analogously, the $2$-punctured torus may contribute, in addition to glueball one-loop two-leg graphs, also graphs with only one or zero external glueball legs (Fig. \ref{fig:torus})
	and, similarly, our new topologies may contribute the graphs in Fig. \ref{fig:maxdisc}. Analogous statements hold for the graphs in Figs. \ref{fig:w2loop} and \ref{fig:w1loopspectral}.
	Both graphs in Fig. \ref{fig:maxdisc} contribute zero to the $S$ matrix as well. More generally, all the new topologies contribute zero to the $S$ matrix, since they miss at least one external glueball leg, as for the last two graphs in Fig. \ref{fig:w1loopspectral}. \par
For the graphs that only contain bivalent vertices, i.e., the maximally pinched ones in Fig. \ref{fig:w2loop}, which are in one-to-one correspondence with diagrams only involving the subleading vertex and no leading vertex in Eq. \eqref{Wasnp}, we verified the spin-statistics theorem directly by means of our asymptotic computation thanks to the minus sign in the last line below
	\begin{align}
		\label{Wasnp1}
		\mathcal{W}^E_{\text{asym maximally pinched}}[J_{\mathbb{O}^E}, \lambda] 
		&=-\log\Det\Bigg(\mathcal{I} - \frac{1}{2} \frac{Z_{\mathbb{O}_s}(\lambda)}{\lambda^{s+2}}\Delta^{-1}P\frac{J_{\mathbb{O}^E_{s}}}{N}\otimes\mathcal{Y}_{s-2}^{E\frac{5}{2}}\Bigg)\nonumber \\
		&=-\log\text{Det}\left(\mathbb{I}-\frac{1}{2}\frac{Z_{\mathbb{O}_s}(\lambda)}{\lambda^{s+2}}\Delta^{-1} \frac{J_{\mathbb{O}^E_s}}{N}\otimes\mathcal{Y}_{s-2}^{E\frac{5}{2}}\right) \, .
	\end{align}
	
	\subsubsection{Extension to SU($N$) $\mathcal{N}=1$ YM theory }
	
	Exactly the same sign puzzle and extra topologies occur in the $\mathcal{N}=1$ SUSY extension of SU($N$) YM theory for the corresponding twist-$2$ operators.\par 
	In this case, however, the relevant features occur for the functional superdeterminant that arises from the one-loop glueball/gluinoball effective action, instead of the determinant in the pure YM theory. \par
	Yet, a new issue arises in the SUSY theory: For the comparison of signs to be meaningful, we should first identify the perturbative kernels that are asymptotic in the ultraviolet to the nonperturbative glueball and gluinoball propagators. \par
	Once this is done, the sign puzzle and its resolution are completely analogous to the pure YM case, the only difference being that now bosonic glueballs and fermionic gluinoballs simultaneously appear.\par
	
	For instance, for the fermionic twist-$2$ operators -- that nonperturbatively should interpolate for the gluinoballs -- the leading-nonplanar asymptotic generating functional reads
	\begin{align}\label{eq:asymgluinotorus2}
		\resizebox{0.89\textwidth}{!}{%
			$\begin{aligned}
				&\Gamma^E_{\text{asym torus}}\left[0,0,0,0,\bar{j}_{T^E},j_{\bar{T}^E},\lambda\right]\nonumber\\
				&= 
				+\log\Det\left[I\,\delta_{s_1 s_2} \delta_{k_1 k_2} -{s_1\choose k_1}{s\choose k+1}\partial_{z}^{s_1-k_1+k}\Laplace^{-1}\frac{Z_{T^E}(\lambda)}{\lambda^{s+2}}\frac{1}{N}\bar {j}_{T_{sk}^E}{s+1\choose k+2}{s_2+1\choose k_2}\partial_{z}^{s-k+k_2}\partial_z\Laplace^{-1} \frac{Z_{\bar{T}^E}(\lambda)}{\lambda^{s_2+2}} \frac{1}{N}j_{\bar{T}_{s_2k_2}^E}\right]\nonumber\\
				&= 
				-\log\Det\left[I\,\delta_{s_1 s_2} \delta_{k_1 k_2} -{s_1\choose k_1}{s\choose k+1}\partial_{z}^{s_1-k_1+k} \partial_z\Laplace^{-1} \frac{Z_{\bar{T}^E}(\lambda)}{\lambda^{s+2}} \frac{1}{N}j_{\bar{T}_{sk}^E} {s+1\choose k+2}{s_2+1\choose k_2}\partial_{z}^{s-k+k_2}\Laplace^{-1}\frac{Z_{T^E}(\lambda)}{\lambda^{s_2+2}}\frac{1}{N}\bar {j}_{T_{s_2k_2}^E}\right]\,,
			\end{aligned}$
		}\\
	\end{align}
	where the $-$ sign in the second line arises from the identity between Eqs. \eqref{wgenT1} and 	\eqref{wgenT2}.
	We find it convenient to rewrite the second line of Eq \eqref{eq:asymgluinotorus2} as
	\begin{align}
	&	\mathcal{W}^E_{\text{asym\, torus}}\left[0,0,0,0,\bar{J}_{T},J_{\bar T},\lambda\right]\nonumber\\ &=-\log\Det\Bigg(I+\frac{1}{4}\frac{	Z_{T_{s_1}}(\lambda)}{\lambda^{s_1+2}}\partial_z\Laplace^{-1}\frac{\bar{J}_{T_{s_1}}}{N}\otimes  \mathcal{G}_{s_1-1}^{E\,(1,2)}\frac{	Z_{T_{s_2}}(\lambda)}{\lambda^{s_2+2}}\Laplace^{-1}\frac{J_{\bar{T}_{s_2}}}{N}\otimes\mathcal{G}_{s_2-1}^{E\,(2,1)}\Bigg) \label{eq:257}
	\end{align}
	where the $-$ sign that corresponds to bosonic statistics is opposite to the one arising from the glueball/gluinoball sector shown in  Eq. \eqref{glueballW1loop_gluino}
	\begin{align}
		\resizebox{0.89\textwidth}{!}{%
			$\begin{aligned}
				\mathcal{W}^E_{\text{glueball/gluinoball 1-loop }}[0,J_{\Psi}] =+\frac{1}{2}\log\text{Det}
				\left[\mathcal{I}-\left(\ast'_2(-\Delta+M^2)\right)^{-1}\frac{1}{N}\ast'_3\ast'_3\Psi_J\left(\ast_2(-\Delta+M^2)\right)^{-1}\frac{1}{N}\ast'_3\ast'_3\Psi_J\right] .\nonumber
			\end{aligned}$
		}\\ 	\end{align}
	This is the sign problem in the supersymmetric setting: Since a loop with only external gluinoball lines necessarily requires alternating glueball and gluinoball propagators due to the cubic coupling, the first propagator in Eq. \eqref{eq:257} should be asymptotically identified as a gluinoball propagator
		\begin{equation}\label{eq:gluino}
 Z_{T_s}(\lambda)\partial_z\Laplace^{-1}  \sim \left(\ast'_2(-\Delta+M^2)\right)^{-1}
	\end{equation}
	and the second one as a glueball propagator
	\begin{equation}\label{eq:glue}
  Z_{T_s}(\lambda)\Laplace^{-1}	\sim	\left(\ast_2(-\Delta+M^2)\right)^{-1} \, ,
	\end{equation}
	according to Eqs. \eqref{eq:action} and \eqref{glueballW}.\par
	Then, the resolution of the puzzle in the SU($N$) theory is that we must keep track of the projector $1-P$ in the gluon and gluino propagators and rewrite the conformal generating functional as
	\begin{align}
		\mathcal{W}^E_{\text{conf}}\left[0,0,0,0,\bar{J}_{T},J_{\bar T}\right] =\log\Det\Bigg(\mathcal{I}+\frac{1}{4}(1-P)\partial_z\Laplace^{-1}\frac{\bar{J}_{T_{s_1}}}{N}\otimes  \mathcal{G}_{s_1-1}^{E\,(1,2)}(1-P)\Laplace^{-1}\frac{J_{\bar{T}_{s_2}}}{N}\otimes\mathcal{G}_{s_2-1}^{E\,(2,1)}\Bigg)\,,
		\label{eq:SUprojector}
	\end{align}
	where we denote by $\mathcal{I}$ the identity in both color and space-time.
	Separating the planar and leading-nonplanar contributions, we obtain identically
	\begin{equation}
		\begin{aligned}
		&	W_{\mathrm{conf}}^{E}\left[0,0,0,0,\bar{J}_{T},J_{\bar T}\right]\\
			&=\log\Det\Bigg(\mathcal{I}+\frac{1}{4}\partial_z\Laplace^{-1}\frac{\bar{J}_{T_{s_1}}}{N}\otimes  \mathcal{G}_{s_1-1}^{E\,(1,2)}\Laplace^{-1}\frac{J_{\bar{T}_{s_2}}}{N}\otimes\mathcal{G}_{s_2-1}^{E\,(2,1)}\Bigg)
			\\
			&\quad
			+\log \operatorname{Det}\!\Bigg[
			\mathcal{I}-\frac{1}{4}\Bigg(\mathcal{I}+\frac{1}{4}\partial_z\Laplace^{-1}\frac{\bar{J}_{T_{s_1}}}{N}\otimes  \mathcal{G}_{s_1-1}^{E\,(1,2)}\Laplace^{-1}\frac{J_{\bar{T}_{s_2}}}{N}\otimes\mathcal{G}_{s_2-1}^{E\,(2,1)}\Bigg)^{-1}\\
			&\qquad\Bigl(\partial_z\Laplace^{-1}P\frac{\bar{J}_{T_{s_3}}}{N}\otimes  \mathcal{G}_{s_3-1}^{E\,(1,2)}\Laplace^{-1}\frac{J_{\bar{T}_{s_4}}}{N}\otimes\mathcal{G}_{s_4-1}^{E\,(2,1)}+\partial_z\Laplace^{-1}\frac{\bar{J}_{T_{s_3}}}{N}\otimes  \mathcal{G}_{s_3-1}^{E\,(1,2)}\Laplace^{-1}P\frac{J_{\bar{T}_{s_4}}}{N}\otimes\mathcal{G}_{s_4-1}^{E\,(2,1)}&\\
			&\qquad-\partial_z\Laplace^{-1}P\frac{\bar{J}_{T_{s_3}}}{N}\otimes  \mathcal{G}_{s_3-1}^{E\,(1,2)}\Laplace^{-1}P\frac{J_{\bar{T}_{s_4}}}{N}\otimes\mathcal{G}_{s_4-1}^{E\,(2,1)}\Bigr)
			\Bigg]\,,
		\end{aligned}
		\label{eq:SUdecomposition}
	\end{equation}
	the second term above being the conformal leading-nonplanar generating functional
	\begin{align}
		&\mathcal{W}^E_{\text{conf\, leading-nonplanar}}\left[0,0,0,0,\bar{J}_{T},J_{\bar T}\right]\nonumber\\ &=-\log\Det\Bigg(I+\frac{1}{4}\partial_z\Laplace^{-1}\frac{\bar{J}_{T_{s_1}}}{N}\otimes  \mathcal{G}_{s_1-1}^{E\,(1,2)}\Laplace^{-1}\frac{J_{\bar{T}_{s_2}}}{N}\otimes\mathcal{G}_{s_2-1}^{E\,(2,1)}\Bigg)\nonumber\\
		&=	+\log \operatorname{Det}\!\Bigg[
		\mathcal{I}-\frac{1}{4}\Bigg(\mathcal{I}+\frac{1}{4}\partial_z\Laplace^{-1}\frac{\bar{J}_{T_{s_1}}}{N}\otimes  \mathcal{G}_{s_1-1}^{E\,(1,2)}\Laplace^{-1}\frac{J_{\bar{T}_{s_2}}}{N}\otimes\mathcal{G}_{s_2-1}^{E\,(2,1)}\Bigg)^{-1}\nonumber\\
		&\qquad\Bigl(\partial_z\Laplace^{-1}P\frac{\bar{J}_{T_{s_3}}}{N}\otimes  \mathcal{G}_{s_3-1}^{E\,(1,2)}\Laplace^{-1}\frac{J_{\bar{T}_{s_4}}}{N}\otimes\mathcal{G}_{s_4-1}^{E\,(2,1)}+\partial_z\Laplace^{-1}\frac{\bar{J}_{T_{s_3}}}{N}\otimes  \mathcal{G}_{s_3-1}^{E\,(1,2)}\Laplace^{-1}P\frac{J_{\bar{T}_{s_4}}}{N}\otimes\mathcal{G}_{s_4-1}^{E\,(2,1)}&\nonumber\\
		&\qquad-\partial_z\Laplace^{-1}P\frac{\bar{J}_{T_{s_3}}}{N}\otimes  \mathcal{G}_{s_3-1}^{E\,(1,2)}\Laplace^{-1}P\frac{J_{\bar{T}_{s_4}}}{N}\otimes\mathcal{G}_{s_4-1}^{E\,(2,1)}\Bigr)
		\Bigg]\,.
	\end{align}
	Therefore, the introduction of the new sector consisting of pinched tori allows us to reconcile the sign of the second $\log \Det $ above with the spin-statistics theorem for the gluinoballs, in complete analogy with the pure YM case for the glueballs \cite{BPSL}. In particular, in the maximally pinched sector we find
	\begin{align}
		&W_{\mathrm{asym\,maximally \,pinched}}^{E}\left[0,0,0,0,\bar{J}_{T},J_{\bar T},\lambda\right]
		\nonumber\\
		&=+\log \operatorname{Det}\!\Bigg[
		\mathcal{I}+\frac{1}{4}\frac{	Z_{T_{s_1}}(\lambda)}{\lambda^{s_1+2}}\partial_z\Laplace^{-1}P\frac{\bar{J}_{T_{s_1}}}{N}\otimes  \mathcal{G}_{s_1-1}^{E\,(1,2)} \frac{	Z_{T_{s_2}}(\lambda)}{\lambda^{s_2+2}}      \Laplace^{-1}P\frac{J_{\bar{T}_{s_2}}}{N}\otimes\mathcal{G}_{s_2-1}^{E\,(2,1)}\Bigr)
		\Bigg]\,,
		\label{eq:maxpinched}
	\end{align}
	where we have reintroduced the finite renormalization factors $Z_T(\lambda)$.\par
	Analogous considerations apply to Eq. \eqref{glueballW1loop_glue}
	\begin{align}\label{eq:nonpe}
		\mathcal{W}^E_{\text{glueball/gluinoball 1-loop }}[J_{\Phi},0] 
		=&+\frac{1}{2}\log\text{Det}
		\left(\mathcal{I}+(\ast'_2(-\Delta+M^2))^{-1}\frac{1}{N}\ast'_3\Phi_J\ast'_3\right)\nonumber\\
		&-\frac{1}{2}\log\text{Det}
		\left(\mathcal{I}+(\ast_2(-\Delta+M^2))^{-1}\frac{1}{N}\ast_3\Phi_J\ast_3\right)
	\end{align}
	where the $+$ sign in the first line corresponds to gluinoballs propagating in the loop with glueballs in the external lines and the $-$ sign in the second line corresponds to glueballs propagating in the loop with glueballs in the external lines. \par
	Besides, Eq. \eqref{eq:asymgluetorus}
	\begin{align}
		\resizebox{0.89\textwidth}{!}{%
			$\begin{aligned}
				&\Gamma^E_{\text{asym torus}} \left[ \overline{j}_{\VarPrimeSup{S}{A}{E}},\; j_{\VarPrimeSup{\overline{S}}{A}{E}},\; \overline{j}_{\VarPrimeSup{S}{\lambda}{E}},\; j_{\VarPrimeSup{\overline{S}}{\lambda}{E}},\; 0,\; 0,\; \lambda \right] \nonumber \\
				&= +\frac{1}{2} \log \Det \Bigg[I\begin{pmatrix} \delta_{s_1 s_2} \delta_{k_1 k_2} &0 \nonumber\\ 
					0 & \delta_{s_1 s_2} \delta_{k_1 k_2}  \end{pmatrix} +\partial_z^{s_1 - k_1 + k_2} \Laplace^{-1}\begin{pmatrix}0&\sqrt{2} \binom{s_1}{k_1} \binom{s_2}{k_2+2}
					\frac{Z_{\VarPrimeSup{\overline{S}}{A}{E}}(\lambda)}{\lambda^{s_2+2}}\frac{1}{N} j_{\VarPrimeSup{\overline{S}}{A}{E}_{s_2k_2}}\nonumber \\ 
					\sqrt{2}\binom{s_1}{k_1} \binom{s_2}{k_2 + 2} 
					\frac{Z_{\VarPrimeSup{\overline{S}}{A}{E}_{s_2}}(\lambda)}{\lambda^{s_2 + 2}} 
					\frac{1}{N}\overline{j}_{\VarPrimeSup{S}{E}{A}_{s_2 k_2}} &0 \end{pmatrix} \Bigg]\nonumber \\
				&\quad -\frac{1}{2} \log \Det \Bigg[I\begin{pmatrix} \delta_{s_1 s_2} \delta_{k_1 k_2} &0\nonumber \\ 
					0 &\delta_{s_1 s_2} \delta_{k_1 k_2}  \end{pmatrix} +\partial_z^{s_1 - k_1 + k_2-1}\partial_z \Laplace^{-1}\begin{pmatrix}0&\sqrt{2}\binom{s_1}{k_1} \binom{s_2}{k_2+1}
					\frac{Z_{\VarPrimeSup{\overline{S}}{\lambda}{E}}(\lambda)}{\lambda^{s_2+2}}\frac{1}{N} j_{\VarPrimeSup{\overline{S}}{\lambda}{E}_{s_2k_2}}\nonumber \\ 
					\sqrt{2}	\binom{s_1}{k_1} \binom{s_2}{k_2 + 1} \frac{Z_{\VarPrimeSup{\overline{S}}{\lambda}{E}_{s_2}}(\lambda)}{\lambda^{s_2 + 2}} 
					\frac{1}{N}	\overline{j}_{\VarPrimeSup{S}{E}{\lambda}_{s_2 k_2}} &0 \end{pmatrix} \Bigg]
			\end{aligned}$
		}\\
	\end{align}
	also displays the sign problem, since the nonperturbative interpretation of the asymptotic propagators above exactly inverts the roles of glueballs and gluinoballs compared to the nonperturbative propagators in Eq. \eqref{eq:nonpe}. Obviously, in this case as well, the sign problem is solved by introducing the projector $(1-P)$ and the corresponding topologies.

	\subsection{Matching between the nonperturbative and RG-improved perturbative structure of one-loop effective action revisited}
	
        Hence, we summarize as follows the status of the matching between the structure of the one-loop glueball/gluinoball effective action and the 
        corresponding RG-improved perturbative object.\par
	The structure of the logarithm of a (super-)determinant for the nonperturbative one-loop effective action is in a sense a-specific, since it is a general 
	feature of the (supersymmetric) one-loop effective action.\par
	It is nontrivial, however, that the above structure is matched by the perturbative computation because, on the perturbative side, it involves composite 
	operators.\par
	In fact, a simple matching is obtained only for twist-$2$ operators that are bilinear in the fundamental fields: Computationally, it would be 
	considerably more difficult to verify it for composite operators that are multilinear in the fundamental fields.\par
	Moreover, the structure of the perturbative result for twist-$2$ operators survives the RG improvement thanks to the existence of the 
	nonresonant renormalization scheme for their mixing, another nontrivial feature of our computation. \par
	Finally, even for twist-$2$ operators, the matching of the sign of the nonperturbative effective action with the corresponding perturbative and RG-
	improved objects involves a number of nontrivial facts, including the existence of extra topologies in the large-$N$ SU($N$) theory, with respect to the 
	U($N$) one, and their nonperturbative interpretation outlined in the previous section.

	\subsection{Nonperturbative large-$N$ coupling from the planar and the maximally pinched contributions}

	Nonperturbatively, according to Section \ref{sec:114} the planar contribution to $2$-point correlators must correspond to either a sum of glueball or gluinoball tree propagators depicted in the first graph of Fig. \ref{fig:spheres}, while the maximally pinched contribution must correspond to just one self-energy loop depicted in the last graph of Fig. \ref{fig:maxdisc}. 
	Therefore, we may evaluate the nonperturbative large-$N$ coupling in the maximally pinched sector from the relative normalization of the aforementioned contributions, provided that we take into account the different nonperturbative space-time structure. \par
For example, we choose the unbalanced glueball $2$-point correlators in  $\mathcal{N}=1$ SUSY $SU(N)$ YM theory, but the result for the nonperturbative coupling is actually universal, i.e. it is the very same in the maximally pinched sector for all the twist-$2$ operators both in large-$N$ pure YM and $\mathcal{N}=1$ SUSY YM theory. \par The lowest-order contribution for unbalanced glueball $2$-point correlators in $\mathcal{N}=1$ SUSY YM theory reads in Euclidean space-time
	\begin{align} \label{CE}
		\nonumber
		\mathcal{C}^E_{s}(x,y)=\langle S^{A\,E}_s(x)S^{A\,E}_s(y)\rangle = &\frac{1}{(4\pi^2)^2} \frac{N^2-1}{4} \frac{2^{2s+2}}{(4!)^2}(s+1)^2(s+2)^2
		\frac{(x-y)_{z}^{2s}}{((x-y)^2)^{2s+2}}\\*\nonumber
		&\sum_{k_1 = 0}^{s-2}\sum_{k_2 = 0}^{s-2}{s\choose k_1}{s\choose k_1+2}{s\choose k_2}{s\choose k_2+2}(-1)^{s-k_2+k_1}\\*
		&(s-k_1+k_2)!(s+k_1-k_2)!\,.
	\end{align}
Restricting to the maximally pinched sector is equivalent in the equation above to changing the sign of the leading nonplanar contribution according to Eq. \eqref{Wasnp}.
We thus get from Eq. \eqref{CE} 
	\begin{align}
		& \left(\frac{ 4 \pi^2 \, 4!}{N (s+1)(s+2)}\right)^2   \mathcal{C}^{'E}_{s}(x,y) = \text{Planar}+\text{Maximally Pinched}
	\end{align}
	where now we rewrite \cite{BPS1} the space-time structure of the planar contribution in order to match asymptotically the nonperturbative structure in Eq. \eqref{eq:kallen} 
	\begin{align}
		\text{Planar}=&\frac{1}{2s+1}\frac{s(s-1)}{(s+1)(s+2)}\,\partial^{2s}_{x^z}\frac{1}{(x-y)^4}\,,
	\end{align}
	while we write the space-time structure of the maximally pinched contribution in order to match the structure of a single self-energy glueball loop
	\begin{align}
	\text{Maximally Pinched} =&\frac{1}{N^2}\sum_{k_1 = 0}^{s-2}\sum_{k_2 = 0}^{s-2}{s\choose k_1}{s\choose k_1+2}{s\choose k_2}{s\choose k_2+2}\nonumber\\*
		&\partial^{s-k_1+k_2}_{x^z}\frac{1}{(x-y)^2}\partial^{s+k_1-k_2}_{y^z}\frac{1}{(y-x)^2}\,.
	\end{align}
The corresponding objects in the momentum representation read
	\begin{align}\label{eq:4pi}
		\nonumber
		&   \left(\frac{ 4 \pi \, 4!}{N (s+1)(s+2)}\right)^2     \mathcal{C}^{'E}_{s}(p)  \\*\nonumber
		&=-\frac{1}{2s+1}\frac{s(s-1)}{(s+1)(s+2)}
		p_{z}^{2s}\log(\frac{p^2}{\mu^2})\\*
		&+\left(\frac{4 \pi}{N}\right)^2
		\sum_{k_1 = 0}^{s-2}\sum_{k_2 = 0}^{s-2}{s\choose k_1}{s\choose k_1+2}{s\choose k_2}{s\choose k_2+2}\int \frac{d^4q}{(2 \pi)^4}\,\frac{(p-q)_{z}^{s-k_1+k_2}}{(p-q)^2}\frac{q_{z}^{s-k_2+k_1}}{q^2}\,.
	\end{align}
	After RG-improvement, we obtain
		\begin{align}\label{eq:4pirg}
		\nonumber
		&   \left(\frac{ 4 \pi \, 4!}{N (s+1)(s+2)}\right)^2      \mathcal{C}^{'E}_{s}(p)  \\*\nonumber
		&=-\frac{1}{2s+1}\frac{s(s-1)}{(s+1)(s+2)}
		Z_{\mathcal{O}_s}^2(p) p_{z}^{2s}\log(\frac{p^2}{\mu^2})\\*
		&+\left(\frac{4 \pi}{N}\right)^2
		\sum_{k_1 = 0}^{s-2}\sum_{k_2 = 0}^{s-2}{s\choose k_1}{s\choose k_1+2}{s\choose k_2}{s\choose k_2+2}\int \frac{d^4q}{(2 \pi)^4}\,\frac{Z_{\mathcal{O}_s}(p-q)(p-q)_{z}^{s-k_1+k_2}}{(p-q)^2}\frac{Z_{\mathcal{O}_s}(q)q_{z}^{s-k_2+k_1}}{q^2}\,.
	\end{align}
	 Nonperturbatively, the first term corresponds to Eq. \eqref{eq:2asym} up to possibly divergent contact terms in the ellipses, with $P_s^E(p)$ restricted to the Euclidean continuation of the light-cone projector along the $p_+$ direction
	 	\begin{align}
	 	\sum_{n=1}^{\infty}  P_s^{E} (p) \frac{ R^{(s)}_n}{p^2+m^{(s)2}_n  }  
	 	&\sim - (-p^2)^{s} P_s^{E} (p) \frac{g^{-2}(p)}{\beta_0(\gamma'+1)} \biggl( \frac{g(\mu)}{g(p)} \biggr)^{2 \gamma' } 
	 	+ \cdots \nonumber \\
	 	&= - (-p^2)^{s} P_s^{E}(p) \frac{g^{-2}(\mu)}{\beta_0(\gamma'+1)} \biggl( \frac{g(\mu)}{g(p)} \biggr)^{2 \gamma' +2} 
	 	+ \cdots
	 \end{align}
	Eqs. \eqref{eq:4pi} and \eqref{eq:4pirg} determine the previously unknown large-$N$ nonperturbative coupling, $\frac{4\pi}{N}$, in the maximally pinched sector. The calculation of the nonperturbative coupling has been possible because the maximally pinched sector contains only a one-loop diagram, as opposed to the remaining graphs in Fig. \ref{fig:w1loopspectral}. 
	Therefore, this computation crucially depends on the theory outlined in Sec. \ref{sec:114}.\par 
	Remarkably, the occurrence of the coupling $\frac{4\pi}{N}$ -- that coincides with the Chern-Simons  coupling -- matches the prediction in \cite{Bochicchio:2016toi}  that the topological part of the nonperturbative glueball effective action arises from a Chern-Simons theory.

	\appendix

	\section{Jacobi and Gegenbauer polynomials \label{appB}}

	We derive several formulas for the Jacobi and Gegenbauer polynomials used in this paper. 
	For a real variable $x$, the Jacobi polynomials $P^{(\alpha,\beta)}_l(x)$ have the representation \cite{szego1959orthogonal}
	\begin{align}
		\label{jaco1}
		P^{(\alpha,\beta)}_l(x) = \sum_{k = 0}^{l}{l+\alpha\choose k}{l+\beta\choose k+\beta}\left(\frac{x-1}{2}\right)^k\left(\frac{x+1}{2}\right)^{l-k}\,,
	\end{align}
	where $\alpha, \beta$ are real and $l$ is a natural number. Additionally, they satisfy the symmetry relation
	\begin{align}
		\label{jacox}
		P^{(\alpha,\beta)}_l(-x) = (-1)^{l}P^{(\beta,\alpha)}_l(x)\,.
	\end{align} 
	The Gegenbauer polynomials $C^{\alpha'}_l(x)$ are a special instance of the Jacobi polynomials
	\begin{equation} \label{GPol}
		C^{\alpha'}_l(x) = \frac{\Gamma(l+2\alpha')\Gamma(\alpha'+\frac{1}{2})}{\Gamma(2\alpha')\Gamma(l+\alpha'+\frac{1}{2})}P_l^{(\alpha'-\frac{1}{2},\alpha'-\frac{1}{2})}(x)\,,
	\end{equation}
	with the symmetry property
	\begin{align}
		\label{symmgegen}
		C^{\alpha'}_l(-x) = (-1)^l C^{\alpha'}_l(x) \,.
	\end{align}
	We define
	\begin{equation}
		x = \frac{b-a}{a+b}\,,
	\end{equation}
	so that
	\begin{equation}
		\left(\frac{x-1}{2}\right)^k\left(\frac{x+1}{2}\right)^{l-k} = (-1)^{l-k} \frac{a^{l-k} b^k}{(a+b)^l}\,.
	\end{equation}
	Eq. (\ref{jaco1}) then becomes
	\begin{equation}
		\label{jaco2}
		P^{(\alpha,\beta)}_l(x) = \sum_{k = 0}^{l}{l+\alpha\choose k}{l+\beta\choose k+\beta}(-1)^{l-k} \frac{a^{l-k} b^k}{(a+b)^l}\,.
	\end{equation}
	Moreover, by substituting $l = J-\alpha'+\frac{1}{2}$ into Eq. \eqref{GPol} and $\alpha=\beta=\alpha'-\frac{1}{2}$ into Eq. \eqref{jaco1}, we find
	\begin{align}
		C^{\alpha'}_{J-\alpha'+\frac{1}{2}}\left(x\right) =\frac{\Gamma(J+\frac{1}{2}+\alpha')\Gamma(\alpha'+\frac{1}{2})}{\Gamma(2\alpha')\Gamma(J+1)}
		\sum_{k=0}^{J-\alpha'+\frac{1}{2}} {J\choose k}{J\choose k+\alpha'-\frac{1}{2}}
		(-1)^{J-\alpha'+\frac{1}{2}-k} \frac{a^{J-\alpha'+\frac{1}{2}-k} b^k}{(a+b)^{J-\alpha'+\frac{1}{2}}}\,.
	\end{align}
	Specializing the above formula to $J = s$ and $\alpha' = \frac{5}{2}$ gives
	\begin{align}
		\label{physicalgegen}
		C^{\frac{5}{2}}_{s-2}\left(x\right) = \frac{\Gamma(s+3)\Gamma(3)}{\Gamma(5)\Gamma(s+1)}
		\sum_{k=0}^{s-2} {s\choose k}{s\choose k+2}(-1)^{s-k} \frac{a^{s-k-2} b^k}{(a+b)^{s-2}}\,.
	\end{align}
	Furthermore, for $J = s$ and $\alpha' = \frac{3}{2}$, we find 
	\begin{align}
		\label{physicalgegen2}
		C^{\frac{3}{2}}_{s-1}\left(x\right)=  \frac{\Gamma(s+2)\Gamma(1)}{\Gamma(3)\Gamma(s+1)} 
		\sum_{k=0}^{s-1} {s\choose k}{s\choose k+1}(-1)^{s-k-1} \frac{a^{s-k-1} b^k}{(a+b)^{s-1}}\,.
	\end{align}
	We now restrict $\alpha, \beta$ to the natural numbers, and consequently, $\alpha'$ to positive half-integers and $J$ to natural numbers. 
	Using the identity
	\begin{align}
		{l+\alpha\choose k}{l+\beta\choose k+\beta}	=\frac{(l+\beta)!(l+\alpha)!}{l!(l+\alpha+\beta)!}{l\choose k}{l+\beta+\alpha\choose k+\beta}\,,
	\end{align}
	it follows from Eq. \eqref{jaco2} that
	\begin{align}
		\label{jaco3}
		P^{(\alpha,\beta)}_l(x) = \frac{(l+\beta)!(l+\alpha)!}{l!(l+\alpha+\beta)!}
		\sum_{k = 0}^{l} {l\choose k}{l+\beta+\alpha\choose k+\beta}(-1)^{l-k} \frac{a^{l-k} b^k}{(a+b)^l}\,.
	\end{align}
	Correspondingly, Eq. \eqref{GPol} is
	\begin{align}
		C^{\alpha'}_l\left(x\right)  =\frac{\Gamma(l+2\alpha')\Gamma(\alpha'+\frac{1}{2})}{\Gamma(2\alpha')\Gamma(l+\alpha'+\frac{1}{2})}\frac{(l+\alpha'-\frac{1}{2})!(l+\alpha'-\frac{1}{2})!}{l!(l+2\alpha'-1)!}\sum_{k = 0}^{l} {l\choose k}{l+2\alpha'-1\choose k+\alpha'-\frac{1}{2}}(-1)^{l-k} \frac{a^{l-k} b^k}{(a+b)^l}
	\end{align}
	which simplifies to
	\begin{align} \label{bin}
		C^{\alpha'}_l\left(x\right) 
		= \frac{\Gamma(\alpha'+\frac{1}{2})\Gamma(l+\alpha'+\frac{1}{2})}{\Gamma(2\alpha')\Gamma(l+1)}
		\sum_{k = 0}^{l} {l\choose k}{l+2\alpha'-1\choose k+\alpha'-\frac{1}{2}}
		(-1)^{l-k} \frac{a^{l-k} b^k}{(a+b)^l}\,.
	\end{align}

	\section{Conformal properties of the standard basis} \label{B}

	The gauge-invariant collinear twist-2 operators in the light-cone gauge that are components of the unbalanced superfields, expressed in the standard basis, are given by~\cite{BPS1,Belitsky:2004sc, Belitsky:2003sh}
	\begin{equation}\label{1000}
		\resizebox{0.47\textwidth}{!}{%
			$\begin{aligned}
				S^A_s &= \frac{1}{2\sqrt{2}} \partial_+ \bar{A}^a(i\overrightarrow{\partial}_++ i\overleftarrow{\partial}_+)^{s-2}C^{\frac{5}{2}}_{s-2}\Bigg(\frac{\overrightarrow{\partial}_+- \overleftarrow{\partial}_+}{\overrightarrow{\partial_+}+\overleftarrow{\partial}_+}\Bigg)\partial_+ \bar{A}^a \\
				\bar{S}^A_s &= \frac{1}{2\sqrt{2}} \partial_+ A^a(i\overrightarrow{\partial}_++ i\overleftarrow{\partial}_+)^{s-2}C^{\frac{5}{2}}_{s-2}\Bigg(\frac{\overrightarrow{\partial}_+- \overleftarrow{\partial}_+}{\overrightarrow{\partial_+}+\overleftarrow{\partial}_+}\Bigg)\partial_+ A^a\\
				S^\lambda_s &=  \frac{1}{2\sqrt{2}} \bar{\lambda}^a(i\overrightarrow{\partial}_+ + i\overleftarrow{\partial}_+)^{s-1}C^{\frac{3}{2}}_{s-1}\Bigg(\frac{\overrightarrow{\partial}_+ - \overleftarrow{\partial}_+}{\overrightarrow{\partial}_++\overleftarrow{\partial}_+}\Bigg) \bar{\lambda}^a \\
				\bar{S}^\lambda_s &=  \frac{1}{2\sqrt{2}} \lambda^a(i\overrightarrow{\partial}_+ + i\overleftarrow{\partial}_+)^{s-1}C^{\frac{3}{2}}_{s-1}\Bigg(\frac{\overrightarrow{\partial}_+ - \overleftarrow{\partial}_+}{\overrightarrow{\partial}_++\overleftarrow{\partial}_+}\Bigg) \lambda^a \\
				T_s& =  \frac{1}{2}\lambda^a(i\overrightarrow{\partial}_+ + i\overleftarrow{\partial}_+)^{s-1}P^{(1,2)}_{s-1}\Bigg(\frac{\overrightarrow{\partial}_+ - \overleftarrow{\partial}_+}{\overrightarrow{\partial}_++\overleftarrow{\partial}_+}\Bigg) \partial_+\bar{A}^a \\
				\bar{T}_s &=
				\frac{1}{2} \partial_+A^a(i\overrightarrow{\partial}_+ + i\overleftarrow{\partial}_+)^{s-1}P^{(2,1)}_{s-1}\Bigg(\frac{\overrightarrow{\partial}_+ - \overleftarrow{\partial}_+}{\overrightarrow{\partial}_++\overleftarrow{\partial}_+}\Bigg) \bar{\lambda}^a\,,
			\end{aligned}$
		}
	\end{equation}
	where $C^{\alpha'}_l(x)$ are the Gegenbauer polynomials (\ref{appB}) with the symmetry relation
	\begin{align}
		C_{l}^{\alpha'}(-x)=(-1)^{l}C_{l}^{\alpha'}(x)\,.
	\end{align}
	These are the restrictions to the components with maximal-spin projection $s$ along the $p_+$ axis of linear combinations of twist-2 operators of the form
	\begin{align} \label{1}
		&\mathbb{S}^{A\,\mathcal{T}=2}_{s}=\quad \Tr\, (F_{\mu(\nu}+i\tilde{F}_{\mu(\nu})\overleftarrow{D}_{\rho_1}\ldots \overrightarrow{D}_{\rho_{s-2}}(F_{\lambda)\sigma}+i\tilde{F}_{\lambda)\sigma})-\,\text{traces}\nonumber\\
		&S^{\lambda\,\mathcal{T}=2}_{s} \quad=\quad \Tr\, \bar{\chi}\sigma_{\mu(\rho_1}\overleftarrow{D}_{\rho_2}\ldots \overrightarrow{D}_{\rho_{s-1})}\chi-\,\text{traces}\qquad\qquad \nonumber\\
		&T^{\mathcal{T}=2}_{s+\frac{1}{2}} \quad=\quad \Tr\, F_{(\rho_1}^{\nu}\overleftarrow{D}_{\rho_2}\ldots \overrightarrow{D}_{\rho_{s-1}}\sigma_{\rho_s)\nu}\chi-\,\text{traces}\,,
	\end{align}
	including all possible combinations of right and left derivatives \cite{makeenko,Belitsky:2007jp}, where the parentheses denote symmetrization of all enclosed indices and the trace subtraction guarantees that the contraction of any two indices is zero. \par
	Specific linear combinations of the above operators are twist-$2$ operators conserved at the leading order of perturbation theory \cite{makeenko,Belitsky:2007jp} and, consequently, they transform as primary operators with respect to the conformal group \cite{makeenko,Belitsky:2007jp,Beisert:2004fv}.
	By projecting onto the maximal-spin component along the $p_+$ direction, they reduce to 
	\begin{align}\label{OO}
		S^A_s &= \frac{1}{2\sqrt{2}}  f_{11}^a(i\overrightarrow{\partial}_++ i\overleftarrow{\partial}_+)^{s-2}C^{\frac{5}{2}}_{s-2}\Bigg(\frac{\overrightarrow{\partial}_+- \overleftarrow{\partial}_+}{\overrightarrow{\partial_+}+\overleftarrow{\partial}_+}\Bigg) f_{11}^a\nonumber \\
		\bar{S}^A_s &= \frac{1}{2\sqrt{2}} f_{\dot{1}\dot{1}}^a (i\overrightarrow{\partial}_++ i\overleftarrow{\partial}_+)^{s-2}C^{\frac{5}{2}}_{s-2}\Bigg(\frac{\overrightarrow{\partial}_+- \overleftarrow{\partial}_+}{\overrightarrow{\partial_+}+\overleftarrow{\partial}_+}\Bigg)f_{\dot{1}\dot{1}}^a\nonumber \\
		S^\lambda_s &=  \frac{1}{2\sqrt{2}} \bar{\lambda}^a(i\overrightarrow{\partial}_+ + i\overleftarrow{\partial}_+)^{s-1}C^{\frac{3}{2}}_{s-1}\Bigg(\frac{\overrightarrow{\partial}_+ - \overleftarrow{\partial}_+}{\overrightarrow{\partial}_++\overleftarrow{\partial}_+}\Bigg) \bar{\lambda}^a\nonumber\\
		\bar{S}^\lambda_s &=  \frac{1}{2\sqrt{2}} \lambda^a(i\overrightarrow{\partial}_+ + i\overleftarrow{\partial}_+)^{s-1}C^{\frac{3}{2}}_{s-1}\Bigg(\frac{\overrightarrow{\partial}_+ - \overleftarrow{\partial}_+}{\overrightarrow{\partial}_++\overleftarrow{\partial}_+}\Bigg) \lambda^a\nonumber \\
		T_s& =  -\frac{1}{2}\lambda^a(i\overrightarrow{\partial}_+ + i\overleftarrow{\partial}_+)^{s-1}P^{(1,2)}_{s-1}\Bigg(\frac{\overrightarrow{\partial}_+ - \overleftarrow{\partial}_+}{\overrightarrow{\partial}_++\overleftarrow{\partial}_+}\Bigg) f_{11}^a\nonumber \\
		\bar{T}_s &=-
		\frac{1}{2} f_{\dot{1}\dot{1}}^a (i\overrightarrow{\partial}_+ + i\overleftarrow{\partial}_+)^{s-1}P^{(2,1)}_{s-1}\Bigg(\frac{\overrightarrow{\partial}_+ - \overleftarrow{\partial}_+}{\overrightarrow{\partial}_++\overleftarrow{\partial}_+}\Bigg) \bar{\lambda}^a\,,
	\end{align}
	which are primaries with respect to the collinear conformal subgroup $SL(2,R)$ \cite{Belitsky:1998gc}. In the light-cone gauge, these simplify to the operators in Eq. \eqref{1000} with $f_{11}=- \partial_+ \bar A$. By permitting operator mixing with derivatives of lower-spin twist-2 operators, they can be extended to primary conformal operators at the next-to-leading order within the conformal renormalization scheme \cite{Braun:2003rp}, which differs from the $\overline{MS}$ scheme by a finite renormalization.\par

	\section{Minkowskian conformal correlators \label{npoint}}

	From the generating functional in Eq. (\ref{wgen1}), we derive the $n$-point conformal correlators in various sectors.  
	
	\subsection{$S^A$ and $\bar{S}^A$ correlators}
	We obtain
	\begin{align}
		&\langle S^A_{s_1}(x_1)\ldots S^A_{s_n}(x_n)\bar{S}^A_{s'_{1}}(y_{1})\ldots \bar{S}^A_{s'_{n}}(y_{n})\rangle \nonumber\\
		& =\frac{\delta}{\delta J_{S^A_{s_1}}(x_1)}\cdots\frac{\delta}{\delta J_{S^A_{s_n}}(x_n)}\frac{\delta}{\delta J_{\bar{S}^A_{s'_{1}}}(y_{1})}\cdots\frac{\delta}{\delta J_{\bar{S}^A_{s'_{n}}}(y_{n})}\mathcal{W}_{\text{conf}}\left[J_{S^A},J_{\bar{S}^A},0,0,0,0\right]\nonumber\\
		&=-\frac{1}{2}\frac{\delta}{\delta J_{S^A_{s_1}}(x_1)}\cdots\frac{\delta}{\delta J_{S^A_{s_n}}(x_n)}\frac{\delta}{\delta J_{\bar{S}^A_{s'_{1}}}(y_{1})}\cdots\frac{\delta}{\delta J_{\bar{S}^A_{s'_{n}}}(y_{n})}
		\log\Det \Bigg(\mathcal{I}-\frac{1}{2} i\square^{-1}\bar{J}_{S^A_{s_1}} \otimes\mathcal{Y}_{s_1-2}^{\frac{5}{2}} i\square^{-1}J_{\bar{S}^A_{s_2}} \otimes\mathcal{Y}_{s_2-2}^{\frac{5}{2}} \Bigg) ,
	\end{align}
	which reproduces the established result \cite{BPS1,BPSpaper2}
	\begin{align} \label{SAcorr}
		\nonumber
		&\langle S^A_{s_1}(x_1)\ldots S^A_{s_n}(x_n)\bar{S}^A_{s'_{1}}(y_{1})\ldots \bar{S}^A_{s'_{n}}(y_{n})\rangle\nonumber\\
		&=\frac{1}{(4\pi^2)^{2n}}\frac{N^2-1}{2^{2n}}2^{\sum_{l=1}^n s_l+{s'}_l}i^{\sum_{l=1}^n\nonumber s_l+{s'}_l}\frac{\Gamma(3)\Gamma(s_1+3)}{\Gamma(5)\Gamma(s_1+1)}\ldots \frac{\Gamma(3)\Gamma(s_n+3)}{\Gamma(5)\Gamma(s_n+1)}\frac{\Gamma(3)\Gamma({s'}_1+3)}{\Gamma(5)\Gamma({s'}_1+1)}\ldots \frac{\Gamma(3)\Gamma({s'}_n+3)}{\Gamma(5)\Gamma({s'}_n+1)}\\\nonumber
		&\quad\sum_{k_1=0}^{s_1-2}\ldots \sum_{k_n = 0}^{s_n-2}{s_1\choose k_1}{s_1\choose k_1+2}\ldots {s_n\choose k_n}{s_n\choose k_n+2}\\\nonumber
		&\quad\sum_{{k'}_1=0}^{{s'}_1-2}\ldots \sum_{{k'}_n = 0}^{{s'}_n-2}{{s'}_1\choose {k'}_1}{{s'}_1\choose {k'}_1+2}\ldots {{s'}_n\choose {k'}_n}{{s'}_n\choose {k'}_n+2}\\\nonumber
		&\quad\frac{2^{n-1}}{n}\sum_{\sigma\in P_n}\sum_{\rho\in P_n}
		(s_{\sigma(1)}-k_{\sigma(1)}+{k'}_{\rho(1)})!({s'}_{\rho(1)}-{k'}_{\rho(1)}+k_{\sigma(2)})!\\\nonumber
		&\quad\ldots(s_{\sigma(n)}-k_{\sigma(n)}+{k'}_{\rho(n)})!({s'}_{\rho(n)}-{k'}_{\rho(n)}+k_{\sigma(1)})!\\\nonumber
		&\quad\frac{(x_{\sigma(1)}-y_{\rho(1)})_+^{s_{\sigma(1)}-k_{\sigma(1)}+{k'}_{\rho(1)}}}{\left(\rvert x_{\sigma(1)}-y_{\rho(1)}\rvert^2\right)^{s_{\sigma(1)}-k_{\sigma(1)}+{k'}_{\rho(1)}+1}}\frac{(y_{\rho(1)}-x_{\sigma(2)})_+^{{s'}_{\rho(1)}-{k'}_{\rho(1)}+k_{\sigma(2)}}}{\left(\rvert y_{\rho(1)}-x_{\sigma(2)}\rvert^2\right)^{{s'}_{\rho(1)}-{k'}_{\rho(1)}+k_{\sigma(2)}+1}}\\
		&\quad\ldots\frac{(x_{\sigma(n)}-y_{\rho(n)})_+^{s_{\sigma(n)}-k_{\sigma(n)}+{k'}_{\rho(n)}}}{\left(\rvert x_{\sigma(n)}-y_{\rho(n)}\rvert^2\right)^{s_{\sigma(n)}-k_{\sigma(n)}+{k'}_{\rho(n)}+1}}
		\frac{(y_{\rho(n)}-x_{\sigma(1)})_+^{{s'}_{\rho(n)}-{k'}_{\rho(n)}+k_{\sigma(1)}}}{\left(\rvert y_{\rho(n)}-x_{\sigma(1)}\rvert^2\right)^{{s'}_{\rho(n)}-{k'}_{\rho(n)}+k_{\sigma(1)}+1}}\,.
	\end{align}
	where the $-i\epsilon$ prescription in the propagators is implied.
	
	\subsection{$S^{\lambda}$  and $\bar{S}^\lambda$ correlators}
	Similarly, we find
	\begin{align}
		&\langle S^{\lambda}_{s_1}(x_1)\ldots S^{\lambda}_{s_n}(x_n)\bar{S}^{\lambda}_{s'_1}(y_1)\ldots \bar{S}^{\lambda}_{s'_n}(y_n)\rangle \nonumber\\
		&  =\frac{\delta}{\delta \bar{J}_{S^\lambda_{s_1}}(x_1)}\cdots\frac{\delta}{\delta \bar{J}_{S^\lambda_{s_n}}(x_n)}\frac{\delta}{\delta J_{\bar{S}^\lambda_{s'_{1}}}(y_{1})}\cdots\frac{\delta}{\delta J_{\bar{S}^\lambda_{s'_{n}}}(y_{n})}\mathcal{W}_{\text{conf}}\left[0,0,J_{S^{\lambda}},J_{\bar{S}^{\lambda}},0,0\right]\nonumber\\
		&=\frac{1}{2} \frac{\delta}{\delta \bar{J}_{S^\lambda_{s_1}}(x_1)}\cdots\frac{\delta}{\delta \bar{J}_{S^\lambda_{s_n}}(x_n)}\frac{\delta}{\delta J_{\bar{S}^\lambda_{s'_{1}}}(y_{1})}\cdots\frac{\delta}{\delta J_{\bar{S}^\lambda_{s'_{n}}}(y_{n})}\nonumber\\
		&\qquad\log\Det\Bigg(\mathcal{I}-\frac{1}{2} i\partial_+i\square^{-1}\bar{J}_{S^\lambda_{s_1}} \otimes\mathcal{Y}_{s_1-1}^{\frac{3}{2}} i\partial_+i\square^{-1}J_{\bar{S}^\lambda_{s_2}} \otimes\mathcal{Y}_{s_2-1}^{\frac{3}{2}} \Bigg)\nonumber\\
		&=\frac{\delta}{\delta \bar{J}_{S^\lambda_{s_1}}(x_1)}\cdots\frac{\delta}{\delta \bar{J}_{S^\lambda_{s_n}}(x_n)}\frac{\delta}{\delta J_{\bar{S}^\lambda_{s'_{1}}}(y_{1})}\cdots\frac{\delta}{\delta J_{\bar{S}^\lambda_{s'_{n}}}(y_{n})}\nonumber\\
		&-\frac{N^2-1}{2}\sum_{l=1}^{\infty}\frac{1}{l}\frac{1}{2^l}\sum_{s_1}\ldots\sum_{s_l}\sum_{s'_1}\ldots\sum_{s'_l}\int d^4u_1\ldots d^4u_ld^4v_1\ldots d^4v_l  \nonumber\\
		&\quad i\partial_{u_1^+}i\Box^{-1}(u_1-v_1)\mathcal{Y}_{s_1-1}^{\frac{3}{2}}(\overleftarrow{\partial}_{v_1^+},\overrightarrow{\partial}_{v_1^+})\otimes \bar{J}_{S^{\lambda}_{s_1}}(v_1)i\partial_{v_1^+}i\Box^{-1}(v_1-u_2)\mathcal{Y}_{s'_2-1}^{\frac{3}{2}}(\overleftarrow{\partial}_{u_2^+},\overrightarrow{\partial}_{u_2^+})\otimes J_{\bar{S}^{\lambda}_{s'_2}}(u_2)\nonumber\\
		&\quad\ldots i\partial_{u_l^+}i\Box^{-1}(u_l-v_l)\mathcal{Y}_{s_l-1}^{\frac{3}{2}}(\overleftarrow{\partial}_{v_l^+},\overrightarrow{\partial}_{v_l^+})\otimes \bar{J}_{S^{\lambda}_{s_l}}(v_l)i\partial_{v_l^+}i\Box^{-1}(v_l-u_1)\mathcal{Y}_{s'_1-1}^{\frac{3}{2}}(\overleftarrow{\partial}_{u_1^+},\overrightarrow{\partial}_{u_1^+})\otimes J_{\bar{S}^{\lambda}_{s'_1}}(u_1)\,.
	\end{align}
	Executing the functional derivatives gives
	\begin{align}
		&\langle S^{\lambda}_{s_1}(x_1)\ldots S^{\lambda}_{s_n}(x_n)\bar{S}^{\lambda}_{s'_1}(y_1)\ldots \bar{S}^{\lambda}_{s'_n}(y_n)\rangle =-\frac{N^2-1}{2}\sum_{\sigma \in P_n}\sum_{\rho \in P_n}\frac{1}{n}\frac{1}{2^n}\nonumber\\
		& i\partial_{y_{\rho(1)}^+}i\Box^{-1}(y_{\rho(1)}-x_{\sigma(1)})\mathcal{Y}_{s_{\sigma(1)}-1}^{\frac{3}{2}}(\overleftarrow{\partial}_{x_{\sigma(1)}^+},\overrightarrow{\partial}_{x_{\sigma(1)}^+})i\partial_{x_{\sigma(1)}^+}i\Box^{-1}(x_{\sigma(1)}-y_{\rho(2)})\mathcal{Y}_{s'_{\rho(2)}-1}^{\frac{3}{2}}(\overleftarrow{\partial}_{y_{\rho(2)}^+},\overrightarrow{\partial}_{y_{\rho(2)}^+})\nonumber\\
		&\ldots i\partial_{y_{\rho(n)}^+}i\Box^{-1}(y_{\rho(n)}-x_{\sigma(n)})\mathcal{Y}_{s_{\sigma(n)}-1}^{\frac{3}{2}}(\overleftarrow{\partial}_{x_{\sigma(n)}^+},\overrightarrow{\partial}_{x_{\sigma(n)}^+})i\partial_{x_{\sigma(n)}^+}i\Box^{-1}(x_{\sigma(n)}-y_{\rho(1)})\mathcal{Y}_{s'_{\rho(1)}-1}^{\frac{3}{2}}(\overleftarrow{\partial}_{y_{\rho(1)}^+},\overrightarrow{\partial}_{y_{\rho(1)}^+})\,,
	\end{align}
	where $P_{n}$ denotes the permutation group of the set $\{1,\ldots,n\}$. 
	From Eq. \eqref{propagator}, it follows that
	\begin{align}
		&\langle S^{\lambda}_{s_1}(x_1)\ldots S^{\lambda}_{s_n}(x_n)\bar{S}^{\lambda}_{s'_1}(y_1)\ldots \bar{S}^{\lambda}_{s'_n}(y_n)\rangle =-\frac{N^2-1}{2}\frac{1}{2^n}\frac{1}{n}\frac{1}{(4\pi^2)^{2n}}\sum_{\sigma \in P_n}\sum_{\rho \in P_n}\nonumber\\
		& i\partial_{y_{\rho(1)}^+}\frac{1}{\rvert y_{\rho(1)}-x_{\sigma(1)}\rvert^2}\mathcal{Y}_{s_{\sigma(1)}-1}^{\frac{3}{2}}(\overleftarrow{\partial}_{x_{\sigma(1)}^+},\overrightarrow{\partial}_{x_{\sigma(1)}^+})i\partial_{x_{\sigma(1)}^+}\frac{1}{\rvert x_{\sigma(1)}-y_{\rho(2)}\rvert^2}\mathcal{Y}_{s'_{\rho(2)}-1}^{\frac{3}{2}}(\overleftarrow{\partial}_{y_{\rho(2)}^+},\overrightarrow{\partial}_{y_{\rho(2)}^+})\nonumber\\
		&\ldots i\partial_{y_{\rho(n)}^+}\frac{1}{\rvert y_{\rho(n)}-x_{\sigma(n)}\rvert^2}\mathcal{Y}_{s_{\sigma(n)}-1}^{\frac{3}{2}}(\overleftarrow{\partial}_{x_{\sigma(n)}^+},\overrightarrow{\partial}_{x_{\sigma(n)}^+})i\partial_{x_{\sigma(n)}^+}i\Box^{-1}(x_{\sigma(n)}-y_{\rho(1)})\mathcal{Y}_{s'_{\rho(1)}-1}^{\frac{3}{2}}(\overleftarrow{\partial}_{y_{\rho(1)}^+},\overrightarrow{\partial}_{y_{\rho(1)}^+})\,,
	\end{align}
	Using the definition of $\mathcal{Y}^{\frac{3}{2}}$ in Eq. (\ref{defY}), we find
	\begin{align}
		&\langle S^{\lambda}_{s_1}(x_1)\ldots S^{\lambda}_{s_n}(x_n)\bar{S}^{\lambda}_{s'_1}(y_1)\ldots \bar{S}^{\lambda}_{s'_n}(y_n)\rangle \nonumber\\
		&=-\frac{N^2-1}{2}\frac{1}{2^n}\frac{1}{(4\pi^2)^{2n}}i^{\sum_{l=1}^n s_l+s'_l-2n}\frac{(s_1+1)}{2}\ldots \frac{(s_n+1)}{2}\frac{(s'_1+1)}{2}\ldots \frac{(s'_n+1)}{2}\nonumber\\
		&\quad\sum_{k_1 = 0}^{s_1-1}{s_1\choose k_1}{s_1\choose k_1+1}(-1)^{s_1-k_1-1}\ldots \sum_{k_n = 0}^{s_n-1}{s_n\choose k_n}{s_n\choose k_n+1}(-1)^{s_n-k_n-1}\nonumber\\
		&\quad\sum_{k'_1 = 0}^{s'_1-1}{s'_1\choose k'_1}{s'_1\choose k'_1+1}(-1)^{s'_1-k'_1-1}\ldots \sum_{k'_n = 0}^{s'_n-1}{s'_n\choose k'_n}{s'_n\choose k'_n+1}(-1)^{s'_n-k'_n-1}\nonumber\\
		&\quad\frac{1}{n}\sum_{\sigma \in P_n}\sum_{\rho \in P_n}i\partial_{y_{\rho(1)}^+}\frac{1}{\rvert y_{\rho(1)}-x_{\sigma(1)}\rvert^2}\overleftarrow{\partial}_{x^+_{\sigma(1)}}^{s_{\sigma(1)}-k_{\sigma(1)}-1}\overrightarrow{\partial}_{x^+_{\sigma(1)}}^{k_{\sigma(1)}}i\partial_{x_{\sigma(1)}^+}\frac{1}{\rvert x_{\sigma(1)}-y_{\rho(2)}\rvert^2}\overleftarrow{\partial}_{y^+_{\rho(2)}}^{s'_{\rho(2)}-k'_{\rho(2)}-1}\overrightarrow{\partial}_{y^+_{\rho(2)}}^{k'_{\rho(2)}}\nonumber\\
		&\quad\ldots i\partial_{y_{\rho(n)}^+}\frac{1}{\rvert y_{\rho(n)}-x_{\sigma(n)}\rvert^2}\overleftarrow{\partial}_{x^+_{\sigma(n)}}^{s_{\sigma(n)}-k_{\sigma(n)}-1}\overrightarrow{\partial}_{x^+_{\sigma(n)}}^{k_{\sigma(n)}}i\partial_{x_{\sigma(n)}^+}\frac{1}{\rvert x_{\sigma(n)}-y_{\rho(1)}\rvert^2}\overleftarrow{\partial}_{y^+_{\rho(1)}}^{s'_{\rho(1)}-k'_{\rho(1)}-1}\overrightarrow{\partial}_{y^+_{\rho(1)}}^{k'_{\rho(1)}}\,.
	\end{align}
	We move all derivatives to the left using Eq. \eqref{arrowkey}
	\begin{align}
		&\langle S^{\lambda}_{s_1}(x_1)\ldots S^{\lambda}_{s_n}(x_n)\bar{S}^{\lambda}_{s'_1}(y_1)\ldots \bar{S}^{\lambda}_{s'_n}(y_n)\rangle \nonumber\\
		&=-\frac{N^2-1}{2}\frac{1}{2^n}\frac{1}{(4\pi^2)^{2n}}i^{\sum_{l=1}^n s_l+s'_l}\frac{(s_1+1)}{2}\ldots \frac{(s_n+1)}{2}\frac{(s'_1+1)}{2}\ldots \frac{(s'_n+1)}{2}\nonumber\\
		&\quad\sum_{k_1 = 0}^{s_1-1}{s_1\choose k_1}{s_1\choose k_1+1}\ldots \sum_{k_n = 0}^{s_n-1}{s_n\choose k_n}{s_n\choose k_n+1}\nonumber\\
		&\quad\sum_{k'_1 = 0}^{s'_1-1}{s'_1\choose k'_1}{s'_1\choose k'_1+1}\ldots \sum_{k'_n = 0}^{s'_n-1}{s'_n\choose k'_n}{s'_n\choose k'_n+1}\nonumber\\
		&\quad\frac{1}{n}\sum_{\sigma \in P_n}\sum_{\rho \in P_n}\partial_{y_{\rho(1)}^+}^{s_{\sigma(1)}-k_{\sigma(1)}+k'_{\rho(1)}}\frac{1}{\rvert y_{\rho(1)}-x_{\sigma(1)}\rvert^2}\partial_{x_{\sigma(1)}^+}^{s'_{\rho(2)}-k'_{\rho(2)}+k_{\sigma(1)}}\frac{1}{\rvert x_{\sigma(1)}-y_{\rho(2)}\rvert^2}\nonumber\\
		&\quad\ldots \partial_{y_{\rho(n)}^+}^{s_{\sigma(n)}-k_{\sigma(n)}+k'_{\rho(n)}}\frac{1}{\rvert y_{\rho(n)}-x_{\sigma(n)}\rvert^2}\partial_{x_{\sigma(n)}^+}^{s'_{\rho(1)}-k'_{\rho(1)}+k_{\sigma(n)}}\frac{1}{\rvert x_{\sigma(n)}-y_{\rho(1)}\rvert^2}\,.
	\end{align}
	We then employ the relation from \cite{BPS1}
	\begin{align}
		\label{doubleder}
		\partial_{x^+}^{a}\frac{1}{\rvert x-y\rvert^2} =\partial_{x^+}^{a}\frac{1}{2 (x-y)_+(x-y)_--(x-y)^2_\perp} =(-1)^{a} a!\,2^{a} \frac{(x-y)_+^{a}}{(\rvert x-y\rvert^2)^{a+1}}\,,
	\end{align}
	so that
	\begin{align}
		&\langle S^{\lambda}_{s_1}(x_1)\ldots S^{\lambda}_{s_n}(x_n)\bar{S}^{\lambda}_{s'_1}(y_1)\ldots \bar{S}^{\lambda}_{s'_n}(y_n)\rangle \nonumber\\
		&=-\frac{N^2-1}{2}\frac{1}{2^n}\frac{1}{(4\pi^2)^{2n}}i^{\sum_{l=1}^n s_l+s'_l}2^{\sum_{l=1}^n s_l+s'_l}\frac{(s_1+1)}{2}\ldots \frac{(s_n+1)}{2}\frac{(s'_1+1)}{2}\ldots \frac{(s'_n+1)}{2}\nonumber\\
		&\quad\sum_{k_1 = 0}^{s_1-1}{s_1\choose k_1}{s_1\choose k_1+1}\ldots \sum_{k_n = 0}^{s_n-1}{s_n\choose k_n}{s_n\choose k_n+1}\sum_{k'_1 = 0}^{s'_1-1}{s'_1\choose k'_1}{s'_1\choose k'_1+1}\ldots \sum_{k'_n = 0}^{s'_n-1}{s'_n\choose k'_n}{s'_n\choose k'_n+1}\nonumber\\
		&\quad\frac{1}{n}\sum_{\sigma \in P_n}\sum_{\rho \in P_n}(s_{\sigma(1)}-k_{\sigma(1)}+k'_{\rho(1)})!\frac{(x_{\sigma(1)}-y_{\rho(1)})_+^{s_{\sigma(1)}-k_{\sigma(1)}+k'_{\rho(1)}}}{(\rvert x_{\sigma(1)}-y_{\rho(1)}\rvert^2)^{s_{\sigma(1)}-k_{\sigma(1)}+k'_{\rho(1)}+1}}\nonumber\\
		&\quad(s'_{\rho(2)}-k'_{\rho(2)}+k_{\sigma(1)})!\frac{(y_{\rho(2)}-x_{\sigma(1)})_+^{s'_{\rho(2)}-k'_{\rho(2)}+k_{\sigma(1)}}}{(\rvert y_{\rho(2)}-x_{\sigma(1)}\rvert^2)^{s'_{\rho(2)}-k'_{\rho(2)}+k_{\sigma(1)}+1}}\nonumber\\
		&\quad\ldots(s_{\sigma(n)}-k_{\sigma(n)}+k'_{\rho(n)})!\frac{(x_{\sigma(n)}-y_{\rho(n)})_+^{s_{\sigma(n)}-k_{\sigma(n)}+k'_{\rho(n)}}}{(\rvert x_{\sigma(n)}-y_{\rho(n)}\rvert^2)^{s_{\sigma(n)}-k_{\sigma(n)}+k'_{\rho(n)}+1}}\nonumber\\
		&\quad(s'_{\rho(1)}-k'_{\rho(1)}+k_{\sigma(n)})!\frac{(y_{\rho(1)}-x_{\sigma(n)})_+^{s'_{\rho(1)}-k'_{\rho(1)}+k_{\sigma(n)}}}{(\rvert y_{\rho(1)}-x_{\sigma(n)}\rvert^2)^{s'_{\rho(1)}-k'_{\rho(1)}+k_{\sigma(n)}+1}}\,.
	\end{align}
	After relabeling the permutations as $\sigma(n)\rightarrow\sigma(2)$, $\sigma(n-1)\rightarrow\sigma(3)$, and so on, while keeping $\sigma(1)$ unchanged, and similarly relabeling $\rho(n)\rightarrow\rho(2)$, $\rho(n-1)\rightarrow\rho(3)$, etc., while fixing $\rho(1)$, we find
	\begin{align}
		\label{Sncorr}
		&\langle S^{\lambda}_{s_1}(x_1)\ldots S^{\lambda}_{s_n}(x_n)\bar{S}^{\lambda}_{s'_1}(y_1)\ldots \bar{S}^{\lambda}_{s'_n}(y_n)\rangle \nonumber\\
		&=-\frac{N^2-1}{2}\frac{1}{2^n}\frac{1}{(4\pi^2)^{2n}}i^{\sum_{l=1}^n s_l+s'_l}2^{\sum_{l=1}^n s_l+s'_l}\frac{(s_1+1)}{2}\ldots \frac{(s_n+1)}{2}\frac{(s'_1+1)}{2}\ldots \frac{(s'_n+1)}{2}\nonumber\\
		&\quad\sum_{k_1 = 0}^{s_1-1}{s_1\choose k_1}{s_1\choose k_1+1}\ldots \sum_{k_n = 0}^{s_n-1}{s_n\choose k_n}{s_n\choose k_n+1}\sum_{k'_1 = 0}^{s'_1-1}{s'_1\choose k'_1}{s'_1\choose k'_1+1}\ldots \sum_{k'_n = 0}^{s'_n-1}{s'_n\choose k'_n}{s'_n\choose k'_n+1}\nonumber\\
		&\quad\frac{1}{n}\sum_{\sigma \in P_n}\sum_{\rho \in P_n}(s_{\sigma(1)}-k_{\sigma(1)}+k'_{\rho(1)})!(s'_{\rho(1)}-k'_{\rho(1)}+k_{\sigma(2)})!\nonumber\\
		&\quad\ldots (s_{\sigma(n)}-k_{\sigma(n)}+k'_{\rho(n)})!(s'_{\rho(n)}-k'_{\rho(n)}+k_{\sigma(1)})!\nonumber\\
		&\quad\frac{(x_{\sigma(1)}-y_{\rho(1)})_+^{s_{\sigma(1)}-k_{\sigma(1)}+k'_{\rho(1)}}}{(\rvert x_{\sigma(1)}-y_{\rho(1)}\rvert^2)^{s_{\sigma(1)}-k_{\sigma(1)}+k'_{\rho(1)}+1}}\frac{(y_{\rho(1)}-x_{\sigma(2)})_+^{s'_{\rho(1)}-k'_{\rho(1)}+k_{\sigma(2)}}}{(\rvert y_{\rho(1)}-x_{\sigma(n)}\rvert^2)^{s'_{\rho(1)}-k'_{\rho(1)}+k_{\sigma(2}+1}}\nonumber\\
		&\quad\ldots\frac{(x_{\sigma(n)}-y_{\rho(n)})_+^{s_{\sigma(n)}-k_{\sigma(n)}+k'_{\rho(n)}}}{(\rvert x_{\sigma(n)}-y_{\rho(n)}\rvert^2)^{s_{\sigma(n)}-k_{\sigma(n)}+k'_{\rho(n)}+1}}\frac{(y_{\rho(n)}-x_{\sigma(1)})_+^{s'_{\rho(n)}-k'_{\rho(n)}+k_{\sigma(1)}}}{(\rvert y_{\rho(2)}-x_{\sigma(1)}\rvert^2)^{s'_{\rho(n)}-k'_{\rho(n)}+k_{\sigma(1)}+1}}\,.
	\end{align}

	\subsection{$T$ and $\bar{T}$ correlators}
	
	The gluon-gluino correlators must contain an equal number of $T_s$ and $\bar{T}_s$ operators; otherwise, they vanish. This can be easily seen from the generating functional in Eq. \eqref{wgenT2}
	\begin{align}
		&\langle T_{s_1}(x_1)\bar{T}_{s'_1}(y_1) T_{s_2}(x_2)\bar{T}_{s'_2}(y_2)\ldots T_{s_n}(x_n) \bar{T}_{s'_n}(y_n)\rangle \nonumber\\
		&= \frac{\delta}{\delta \bar{J}_{T_{s_1}}(x_1)}\left(-\frac{\delta}{\delta J_{\bar T_{s'_1}}(y_1)}\right)\cdots\frac{\delta}{\delta \bar{J}_{T_{s_n}}(x_n)}\left(-\frac{\delta}{\delta J_{\bar T_{s'_n}}(y_n)}\right)\mathcal{W}_{\text{conf}}\left[0,0,0,0,J_{\bar T},\bar{J}_{T}\right]\nonumber\\
		&= \frac{\delta}{\delta \bar{J}_{T_{s_1}}(x_1)}\left(-\frac{\delta}{\delta J_{\bar T_{s'_1}}(y_1)}\right)\cdots\frac{\delta}{\delta \bar{J}_{T_{s_n}}(x_n)}\left(-\frac{\delta}{\delta J_{\bar T_{s'_n}}(y_n)}\right)\nonumber\\
		&\quad\log\Det\Bigg(\mathcal{I}+\frac{1}{4}i\partial_+i\square^{-1}\bar{J}_{T_{s_1}}\otimes  \mathcal{G}_{s_1-1}^{(1,2)} i\square^{-1}J_{\bar{T}_{s_2}}\otimes\mathcal{G}_{s_2-1}^{(2,1)}\Bigg)\,.\nonumber\\\
	\end{align}
	Expanding the above $\log\Det$ gives
	\begin{align}
		&\langle T_{s_1}(x_1)\bar{T}_{s'_1}(y_1) T_{s_2}(x_2)\bar{T}_{s'_2}(y_2)\ldots T_{s_n}(x_n) \bar{T}_{s'_n}(y_n)\rangle \nonumber\\
		&=\frac{\delta}{\delta \bar{J}_{T_{s_1}}(x_1)}\left(-\frac{\delta}{\delta J_{\bar T_{s'_1}}(y_1)}\right)\cdots\frac{\delta}{\delta \bar{J}_{T_{s_n}}(x_n)}\left(-\frac{\delta}{\delta J_{\bar T_{s'_n}}(y_n)}\right)\nonumber\\
		&\quad(N^2-1)\sum_{l=1}^{\infty}\frac{(-1)^{l+1}}{l}\frac{1}{2^{2l}}\int d^4u_1\ldots d^4u_ld^4v_1\ldots d^4v_l \sum_{s_1}\ldots \sum_{s_l}\sum_{s'_1}\ldots\sum_{s'_l} \nonumber\\
		&\quad i\partial_{u_1^+}i\Box^{-1}(u_1-v_1)
		\mathcal{G}_{s_1-1}^{(1,2)}(\overleftarrow{\partial}_{v_1^+},\overrightarrow{\partial}_{v_1^+})\otimes \bar J_{ T_{s_1}}(v_1)
		i\Box^{-1}(v_1-u_2) \mathcal{G}_{s'_2-1}^{(2,1)}(\overleftarrow{\partial}_{u_2^+},\overrightarrow{\partial}_{u_2^+})\otimes J_{\bar T_{s'_2}}(u_2)\nonumber\\
		&\quad\ldots i\partial_{u_l^+}i\Box^{-1}(u_l-v_l)\mathcal{G}_{s_l-1}^{(1,2)}(\overleftarrow{\partial}_{v_l^+},\overrightarrow{\partial}_{v_l^+})\otimes \bar J_{ T_{s_l}}(v_l) i\Box^{-1}(v_l-u_1) \mathcal{G}_{s'_1-1}^{(2,1)}(\overleftarrow{\partial}_{u_1^+},\overrightarrow{\partial}_{u_1^+})\otimes J_{\bar T_{s'_1}}(u_1)
	\end{align}
	and after performing the functional derivatives, we obtain
	\begin{align}
		&\langle T_{s_1}(x_1)\bar{T}_{s'_1}(y_1) T_{s_2}(x_2)\bar{T}_{s'_2}(y_2)\ldots T_{s_n}(x_n) \bar{T}_{s'_n}(y_n)\rangle \nonumber\\
		&= (N^2-1)\sum_{\rho\in P_n}\sum_{\sigma\in P_n}\frac{(-1)^{n+1}}{n}\frac{1}{2^{2n}}\text{sgn}(\rho)\text{sgn}(\sigma)\nonumber\\
		&\quad i\partial_{y_{\rho(1)}^+}i\Box^{-1}(y_{\rho(1)}-x_{\sigma(1)})
		\mathcal{G}_{s_{\sigma(1)}-1}^{(1,2)}(\overleftarrow{\partial}_{x_{\sigma(1)}^+},\overrightarrow{\partial}_{x_{\sigma(1)}^+})
		i\Box^{-1}(x_{\sigma(1)}-y_{\rho(2)}) \mathcal{G}_{s'_{\rho(2)}-1}^{(2,1)}(\overleftarrow{\partial}_{y_{\rho(2)}^+},\overrightarrow{\partial}_{y_{\rho(2)}^+})\nonumber\\
		&\quad\ldots i\partial_{y_{\rho(n)}^+}i\Box^{-1}(y_{\rho(n)}-x_{\sigma(n)})\mathcal{G}_{s_{\sigma(n)}-1}^{(1,2)}(\overleftarrow{\partial}_{x_{\sigma(n)}^+},\overrightarrow{\partial}_{x_{\sigma(n)}^+}) i\Box^{-1}(x_{\sigma(n)}-y_{\rho(1)}) \mathcal{G}_{s'_{\rho(1)}-1}^{(2,1)}(\overleftarrow{\partial}_{y_{\rho(1)}^+},\overrightarrow{\partial}_{y_{\rho(1)}^+})\,.
	\end{align}
	Using Eq. \eqref{propagator} yields
	\begin{align}
		&\langle T_{s_1}(x_1)\bar{T}_{s'_1}(y_1) T_{s_2}(x_2)\bar{T}_{s'_2}(y_2)\ldots T_{s_n}(x_n) \bar{T}_{s'_n}(y_n)\rangle \nonumber\\
		&= (N^2-1)\frac{i^n}{(4\pi^2)^{2n}}\sum_{\rho\in P_n}\sum_{\sigma\in P_n}\frac{(-1)^{n+1}}{n}\frac{1}{2^{2n}}\text{sgn}(\rho)\text{sgn}(\sigma)\nonumber\\
		&\quad\partial_{y_{\rho(1)}^+}\frac{1}{\rvert y_{\rho(1)}-x_{\sigma(1)}\rvert^2}	\mathcal{G}_{s_{\sigma(1)}-1}^{(1,2)}(\overleftarrow{\partial}_{x_{\sigma(1)}^+},\overrightarrow{\partial}_{x_{\sigma(1)}^+})\frac{1}{\rvert x_{\sigma(1)}-y_{\rho(2)}\rvert^2}\mathcal{G}_{s'_{\rho(1)}-1}^{(2,1)}(\overleftarrow{\partial}_{y_{\rho(2)}^+},\overrightarrow{\partial}_{y_{\rho(2)}^+})\nonumber\\
		&\quad\ldots \partial_{y_{\rho(n)}^+}\frac{1}{\rvert y_{\rho(n)}-x_{\sigma(n)}\rvert^2}\mathcal{G}_{s_{\sigma(n)}-1}^{(1,2)}(\overleftarrow{\partial}_{x_{\sigma(n)}^+},\overrightarrow{\partial}_{x_{\sigma(n)}^+})\frac{1}{\rvert x_{\sigma(n)}-y_{\rho(1)}\rvert^2} \mathcal{G}_{s'_{\rho(1)}-1}^{(2,1)}(\overleftarrow{\partial}_{y_{\rho(1)}^+},\overrightarrow{\partial}_{y_{\rho(1)}^+})
	\end{align}
	and Eq. \eqref{gdef} gives
	\begin{align}
		&\langle T_{s_1}(x_1)\bar{T}_{s'_1}(y_1) T_{s_2}(x_2)\bar{T}_{s'_2}(y_2)\ldots T_{s_n}(x_n) \bar{T}_{s'_n}(y_n)\rangle= (N^2-1)\frac{i^n}{(4\pi^2)^{2n}}\nonumber\\
		& i^{s_1-1} \sum_{k_1 = 0}^{s_1-1}{s_1\choose k_1}{s_1+1\choose k_1+2} (-1)^{s_1-k_1-1}\ldots i^{s_n-1} \sum_{k_n = 0}^{s_n-1}{s_n\choose k_n}{s_n+1\choose k_n+2} (-1)^{s_n-k_n-1}\nonumber\\
		& i^{s'_1-1}\sum_{k_1 = 0}^{s'_1-1}{s_1+1\choose k_1}{s'_1\choose k_1+1} (-1)^{s'_1-k_1-1}\ldots i^{s'_n-1}\sum_{k_n = 0}^{s'_n-1}{s'_n+1\choose k_n}{s'_n\choose k_n+1} (-1)^{s'_n-k_n-1}\nonumber\\
		&\sum_{\rho\in P_n}\sum_{\sigma\in P_n}\frac{(-1)^{n+1}}{n}\frac{1}{2^{2n}}\text{sgn}(\rho)\text{sgn}(\sigma)\nonumber\\
		&\partial_{y_{\rho(1)}^+}\frac{1}{\rvert y_{\rho(1)}-x_{\sigma(1)}\rvert^2}\overleftarrow{\partial}_{x^+_{\sigma(1)}}^{s_{\sigma(1)}-k_{\sigma(1)}-1}\overrightarrow{\partial}_{x^+_{\sigma(1)}}^{k_{\sigma(1)}+1}\frac{1}{\rvert x_{\sigma(1)}-y_{\rho(2)}\rvert^2}\overleftarrow{\partial}_{y^+_{\rho(2)}}^{s'_{\rho(2)-}k'_{\rho(2)}}\overrightarrow{\partial}_{y^+_{\rho(2)}}^{k'_{\rho(2)}} \nonumber\\
		&\ldots\partial_{y_{\rho(n)}^+}\frac{1}{\rvert y_{\rho(n)}-x_{\sigma(n)}\rvert^2}\overleftarrow{\partial}_{x^+_{\sigma(n)}}^{s_{\sigma(n)}-k_{\sigma(n)}-1}\overrightarrow{\partial}_{x^+_{\sigma(n)}}^{k_{\sigma(n)}+1}\frac{1}{\rvert x_{\sigma(n)}-y_{\rho(1)}\rvert^2}\overleftarrow{\partial}_{y^+_{\rho(1)}}^{s'_{\rho(1)-}k'_{\rho(1)}}\overrightarrow{\partial}_{y^+_{\rho(1)}}^{k'_{\rho(1)}}\,.
	\end{align}
	Finally, using Eq. (\ref{doubleder}), we get
	\begin{align}
		&\langle T_{s_1}(x_1)\bar{T}_{s'_1}(y_1) T_{s_2}(x_2)\bar{T}_{s'_2}(y_2)\ldots T_{s_n}(x_n) \bar{T}_{s'_n}(y_n)\rangle= (N^2-1)\frac{i^n}{(4\pi^2)^{2n}}\nonumber\\
		& i^{s_1-1} \sum_{k_1 = 0}^{s_1-1}{s_1\choose k_1}{s_1+1\choose k_1+2} (-1)^{s_1-k_1-1}\ldots i^{s_n-1} \sum_{k_n = 0}^{s_n-1}{s_n\choose k_n}{s_n+1\choose k_n+2} (-1)^{s_n-k_n-1}\nonumber\\
		& i^{s'_1-1}\sum_{k'_1 = 0}^{s'_1-1}{s'_1+1\choose k'_1}{s'_1\choose k'_1+1} (-1)^{s'_1-k'_1-1}\ldots i^{s'_n-1}\sum_{k'_n = 0}^{s'_n-1}{s'_n+1\choose k'_n}{s'_n\choose k'_n+1} (-1)^{s'_n-k'_n-1}\nonumber\\
		& \sum_{\rho\in P_n}\sum_{\sigma\in P_n}\frac{(-1)^{n+1}}{n}\frac{1}{2^{2n}}\text{sgn}(\rho)\text{sgn}(\sigma)\nonumber\\
		& (-1)^{k'_{\rho(1)}+1}2^{s_{\sigma(1)}-k_{\sigma(1)}+k'_{\rho(1)}}(s_{\sigma(1)}-k_{\sigma(1)}+k'_{\rho(1)})! \frac{(y_{\rho(1)}-x_{\sigma(1)})_+^{s_{\sigma(1)}-k_{\sigma(1)}+k'_{\rho(1)}}}{(\rvert y_{\rho(1)}-x_{\sigma(1)}\rvert^2)^{s_{\sigma(1)}-k_{\sigma(1)}+k'_{\rho(1)}+1}}\nonumber\\
		& (-1)^{k_{\sigma(1)}+1}2^{s'_{\rho(2)}-k'_{\rho(2)}+k_{\sigma(1)}+1}(s'_{\rho(2)}-k'_{\rho(2)}+k_{\sigma(1)}+1)! \frac{(x_{\sigma(1)}-y_{\rho(2)})_+^{s'_{\rho(2)}-k'_{\rho(2)}+k_{\sigma(1)}+1}}{(\rvert x_{\sigma(1)}-y_{\rho(2)}\rvert^2)^{s'_{\rho(2)}-k'_{\rho(2)}+k_{\sigma(1)}+2}}\nonumber\\
		& \ldots\nonumber\\
		& (-1)^{k'_{\rho(n)+1}}2^{s_{\sigma(n)}-k_{\sigma(n)}+k'_{\rho(n)}}(s_{\sigma(n)}-k_{\sigma(n)}+k'_{\rho(n)})! \frac{(y_{\rho(n)}-x_{\sigma(n)})_+^{s_{\sigma(n)}-k_{\sigma(n)}+k'_{\rho(n)}}}{(\rvert y_{\rho(n)}-x_{\sigma(n)}\rvert^2)^{s_{\sigma(n)}-k_{\sigma(n)}+k'_{\rho(n)}+1}}\nonumber\\
		& (-1)^{k_{\sigma(n)}+1}2^{s'_{\rho(1)}-k'_{\rho(1)}+k_{\sigma(n)}+1}(s'_{\rho(1)}-k'_{\rho(1)}+k_{\sigma(n)}+1)! \frac{(x_{\sigma(n)}-y_{\rho(1)})_+^{s'_{\rho(1)}-k'_{\rho(1)}+k_{\sigma(n)}+1}}{(\rvert x_{\sigma(n)}-y_{\rho(1)}\rvert^2)^{s'_{\rho(1)}-k'_{\rho(1)}+k_{\sigma(n)}+2}} ,
	\end{align}
	which simplifies to
	\begin{align}
		&\langle T_{s_1}(x_1)\bar{T}_{s'_1}(y_1) T_{s_2}(x_2)\bar{T}_{s'_2}(y_2)\ldots T_{s_n}(x_n) \bar{T}_{s'_n}(y_n)\rangle \nonumber\\
		&= (N^2-1)\frac{1}{(4\pi^2)^{2n}}2^{\sum_{l=1}^n s'_l+\sum_{l=1}^n s_{l}-n}i^{\sum_{l=1}^n s'_l+\sum_{l=1}^n s_{l}-n}(-1)^{\sum_{l=1}^n s'_l+\sum_{l=1}^n s_{l}}\nonumber\\
		&\quad \sum_{k_1 = 0}^{s_1-1}{s_1\choose k_1}{s_1+1\choose k_1+2} \ldots  \sum_{k_n = 0}^{s_n-1}{s_n\choose k_n}{s_n+1\choose k_n+2} \sum_{k'_1 = 0}^{s'_1-1}{s'_1+1\choose k'_1}{s'_1\choose k'_1+1} \ldots \sum_{k'_n = 0}^{s'_n-1}{s'_n+1\choose k'_n}{s'_n\choose k'_n+1} \nonumber\\
		&\quad \sum_{\rho\in P_n}\sum_{\sigma\in P_n}\frac{(-1)^{n+1}}{n}\text{sgn}(\rho)\text{sgn}(\sigma)\nonumber\\
		&\quad (s_{\sigma(1)}-k_{\sigma(1)}+k'_{\rho(1)})! \frac{(y_{\rho(1)}-x_{\sigma(1)})_+^{s_{\sigma(1)}-k_{\sigma(1)}+k'_{\rho(1)}}}{(\rvert y_{\rho(1)}-x_{\sigma(1)}\rvert^2)^{s_{\sigma(1)}-k_{\sigma(1)}+k'_{\rho(1)}+1}}\nonumber\\
		&\quad (s'_{\rho(2)}-k'_{\rho(2)}+k_{\sigma(1)}+1)! \frac{(x_{\sigma(1)}-y_{\rho(2)})_+^{s'_{\rho(2)}-k'_{\rho(2)}+k_{\sigma(1)}+1}}{(\rvert x_{\sigma(1)}-y_{\rho(2)}\rvert^2)^{s'_{\rho(2)}-k'_{\rho(2)}+k_{\sigma(1)}+2}}\nonumber\\
		&\quad \ldots\nonumber\\
		&\quad (s_{\sigma(n)}-k_{\sigma(n)}+k'_{\rho(n)})! \frac{(y_{\rho(n)}-x_{\sigma(n)})_+^{s_{\sigma(n)}-k_{\sigma(n)}+k'_{\rho(n)}}}{(\rvert y_{\rho(n)}-x_{\sigma(n)}\rvert^2)^{s_{\sigma(n)}-k_{\sigma(n)}+k'_{\rho(n)}+1}}\nonumber\\
		&\quad (s'_{\rho(1)}-k'_{\rho(1)}+k_{\sigma(n)}+1)! \frac{(x_{\sigma(n)}-y_{\rho(1)})_+^{s'_{\rho(1)}-k'_{\rho(1)}+k_{\sigma(n)}+1}}{(\rvert x_{\sigma(n)}-y_{\rho(1)}\rvert^2)^{s'_{\rho(1)}-k'_{\rho(1)}+k_{\sigma(n)}+2}}\,.
	\end{align}
	By exchanging the coordinates $x$ and $y$ and relabeling the permutations as $\sigma(n)\rightarrow\sigma(2)$, $\sigma(n-1)\rightarrow\sigma(3)$, etc., while keeping $\sigma(1)$ fixed, and similarly for $\rho$, with $\rho(n)\rightarrow\rho(2)$, $\rho(n-1)\rightarrow\rho(3)$, etc., while keeping $\rho(1)$ fixed, we find
	\begin{align}
		\label{Tncorr}
		&\langle T_{s_1}(x_1)\bar{T}_{s'_1}(y_1) T_{s_2}(x_2)\bar{T}_{s'_2}(y_2)\ldots T_{s_n}(x_n) \bar{T}_{s'_n}(y_n)\rangle \nonumber\\
		& =(N^2-1)\frac{1}{(4\pi^2)^{2n}}2^{\sum_{l=1}^n s'_l+\sum_{l=1}^n s_{l}-n}i^{\sum_{l=1}^n s'_l+\sum_{l=1}^n s_{l}-n}\nonumber\\
		&\quad \sum_{k_1 = 0}^{s_1-1}{s_1\choose k_1}{s_1+1\choose k_1+2} \ldots  \sum_{k_n = 0}^{s_n-1}{s_n\choose k_n}{s_n+1\choose k_n+2}\sum_{k'_1 = 0}^{s'_1-1}{s'_1+1\choose k'_1}{s'_1\choose k'_1+1} \ldots \sum_{k'_n = 0}^{s'_n-1}{s'_n+1\choose k'_n}{s'_n\choose k'_n+1} \nonumber\\
		&\quad \sum_{\rho\in P_n}\sum_{\sigma\in P_n}\frac{(-1)^{2n+1}}{n}\text{sgn}(\rho)\text{sgn}(\sigma)\nonumber\\
		&\quad (s_{\sigma(1)}-k_{\sigma(1)}+k'_{\rho(1)})! \frac{(x_{\sigma(1)}-y_{\rho(1)})_+^{s_{\sigma(1)}-k_{\sigma(1)}+k'_{\rho(1)}}}{(\rvert x_{\sigma(1)}-y_{\rho(1)}\rvert^2)^{s_{\sigma(1)}-k_{\sigma(1)}+k'_{\rho(1)}+1}}\nonumber\\
		&\quad (s'_{\rho(1)}-k'_{\rho(1)}+k_{\sigma(2)}+1)! \frac{(y_{\rho(1)}-x_{\sigma(2)})_+^{s'_{\rho(1)}-k'_{\rho(1)}+k_{\sigma(2)}+1}}{(\rvert y_{\rho(1)}-x_{\sigma(2)}\rvert^2)^{s'_{\rho(1)}-k'_{\rho(1)}+k_{\sigma(2)}+2}}\nonumber\\
		&\quad \ldots\nonumber\\
		&\quad (s_{\sigma(n)}-k_{\sigma(n)}+k'_{\rho(n)})! \frac{(x_{\sigma(n)}-y_{\rho(n)})_+^{s_{\sigma(n)}-k_{\sigma(n)}+k'_{\rho(n)}}}{(\rvert x_{\sigma(n)}-y_{\rho(n)}\rvert^2)^{s_{\sigma(n)}-k_{\sigma(n)}+k'_{\rho(n)}+1}}\nonumber\\
		&\quad (s'_{\rho(n)}-k'_{\rho(n)}+k_{\sigma(1)}+1)! \frac{(y_{\rho(n)}-x_{\sigma(1)})_+^{s'_{\rho(n)}-k'_{\rho(n)}+k_{\sigma(1)}+1}}{(\rvert y_{\rho(2)}-x_{\sigma(1)}\rvert^2)^{s'_{\rho(2)}-k'_{\rho(2)}+k_{\sigma(1)}+2}}\,.
	\end{align}

	\subsection{Vanishing correlators}
	
	An inspection of Eq. \eqref{wgen1} reveals that all mixed correlators between $S^A,\bar{S}^A$ and $S^\lambda,\bar{S}^\lambda$ are zero.

	\section{Euclidean conformal correlators} \label{9}
	
	Here, we compute the corresponding Euclidean $n$-point correlators. 
	
	\subsection{$S^{A\,E}$ and $\bar{S}^{A\,E}$ correlators}
	
	The $2n$-point correlators for $S^{A\,E}$ and $\bar{S}^{A\,E}$ are
	\begin{align} 
		\nonumber
		&\langle S^{A\,E}_{s_1}(x_1)\ldots S^{A\,E}_{s_n}(x_n)\bar{S}^{A\,E}_{s'_{1}}(y_{1})\ldots \bar{S}^{A\,E}_{s'_{n}}(y_{n})\rangle\nonumber\\\nonumber
		&=\frac{1}{(4\pi^2)^{2n}}\frac{N^2-1}{2^{2n}}2^{\sum_{l=1}^n s_l+{s'}_l}(-1)^{\sum_{l=1}^n s_l+{s'}_l}\\\nonumber
		&\quad \frac{\Gamma(3)\Gamma(s_1+3)}{\Gamma(5)\Gamma(s_1+1)}\ldots \frac{\Gamma(3)\Gamma(s_n+3)}{\Gamma(5)\Gamma(s_n+1)}\frac{\Gamma(3)\Gamma({s'}_1+3)}{\Gamma(5)\Gamma({s'}_1+1)}\ldots \frac{\Gamma(3)\Gamma({s'}_n+3)}{\Gamma(5)\Gamma({s'}_n+1)}\\\nonumber
		&\quad \sum_{k_1=0}^{s_1-2}\ldots \sum_{k_n = 0}^{s_n-2}{s_1\choose k_1}{s_1\choose k_1+2}\ldots {s_n\choose k_n}{s_n\choose k_n+2}\sum_{{k'}_1=0}^{{s'}_1-2}\ldots \sum_{{k'}_n = 0}^{{s'}_n-2}{{s'}_1\choose {k'}_1}{{s'}_1\choose {k'}_1+2}\ldots {{s'}_n\choose {k'}_n}{{s'}_n\choose {k'}_n+2}\\\nonumber
		&\quad \frac{2^{n-1}}{n}\sum_{\sigma\in P_n}\sum_{\rho\in P_n}
		(s_{\sigma(1)}-k_{\sigma(1)}+{k'}_{\rho(1)})!({s'}_{\rho(1)}-{k'}_{\rho(1)}+k_{\sigma(2)})!\\\nonumber
		&\quad \ldots(s_{\sigma(n)}-k_{\sigma(n)}+{k'}_{\rho(n)})!({s'}_{\rho(n)}-{k'}_{\rho(n)}+k_{\sigma(1)})!\\\nonumber
		&\quad \frac{(x_{\sigma(1)}-y_{\rho(1)})_z^{s_{\sigma(1)}-k_{\sigma(1)}+{k'}_{\rho(1)}}}{\left(( x_{\sigma(1)}-y_{\rho(1)})^2\right)^{s_{\sigma(1)}-k_{\sigma(1)}+{k'}_{\rho(1)}+1}}\frac{(y_{\rho(1)}-x_{\sigma(2)})_z^{{s'}_{\rho(1)}-{k'}_{\rho(1)}+k_{\sigma(2)}}}{\left(( y_{\rho(1)}-x_{\sigma(2)})^2\right)^{{s'}_{\rho(1)}-{k'}_{\rho(1)}+k_{\sigma(2)}+1}}\\
		&\quad \ldots\frac{(x_{\sigma(n)}-y_{\rho(n)})_z^{s_{\sigma(n)}-k_{\sigma(n)}+{k'}_{\rho(n)}}}{\left(( x_{\sigma(n)}-y_{\rho(n)})^2\right)^{s_{\sigma(n)}-k_{\sigma(n)}+{k'}_{\rho(n)}+1}}
		\frac{(y_{\rho(n)}-x_{\sigma(1)})_z^{{s'}_{\rho(n)}-{k'}_{\rho(n)}+k_{\sigma(1)}}}{\left(( y_{\rho(n)}-x_{\sigma(1)})^2\right)^{{s'}_{\rho(n)}-{k'}_{\rho(n)}+k_{\sigma(1)}+1}}\,.
	\end{align}
	
	\subsection{$S^{\lambda\,E}$ and $\bar{S}^{\lambda\,E}$ correlators}
	
	The $2n$-point correlators for $S^{\lambda\,E}$ and $\bar{S}^{\lambda\,E}$ are
	\begin{align}
		&\langle S^{\lambda\,E}_{s_1}(x_1)\ldots S^{\lambda\,E}_{s_n}(x_n)\bar{S}^{\lambda\,E}_{s'_1}(y_1)\ldots \bar{S}^{\lambda\,E}_{s'_n}(y_n)\rangle \nonumber\\
		&=-\frac{N^2-1}{2}\frac{1}{2^n}\frac{1}{(4\pi^2)^{2n}}(-1)^{\sum_{l=1}^n s_l+s'_l}2^{\sum_{l=1}^n s_l+s'_l}\frac{(s_1+1)}{2}\ldots \frac{(s_n+1)}{2}\frac{(s'_1+1)}{2}\ldots \frac{(s'_n+1)}{2}\nonumber\\
		&\quad \sum_{k_1 = 0}^{s_1-1}{s_1\choose k_1}{s_1\choose k_1+1}\ldots \sum_{k_n = 0}^{s_n-1}{s_n\choose k_n}{s_n\choose k_n+1}\sum_{k'_1 = 0}^{s'_1-1}{s'_1\choose k'_1}{s'_1\choose k'_1+1}\ldots \sum_{k'_n = 0}^{s'_n-1}{s'_n\choose k'_n}{s'_n\choose k'_n+1}\nonumber\\
		&\quad \frac{1}{n}\sum_{\sigma \in P_n}\sum_{\rho \in P_n}(s_{\sigma(1)}-k_{\sigma(1)}+k'_{\rho(1)})!(s'_{\rho(1)}-k'_{\rho(1)}+k_{\sigma(2)})!\nonumber\\
		&\quad \ldots (s_{\sigma(n)}-k_{\sigma(n)}+k'_{\rho(n)})!(s'_{\rho(n)}-k'_{\rho(n)}+k_{\sigma(1)})!\nonumber\\
		&\quad \frac{(x_{\sigma(1)}-y_{\rho(1)})_z^{s_{\sigma(1)}-k_{\sigma(1)}+k'_{\rho(1)}}}{(( x_{\sigma(1)}-y_{\rho(1)})^2)^{s_{\sigma(1)}-k_{\sigma(1)}+k'_{\rho(1)}+1}}\frac{(y_{\rho(1)}-x_{\sigma(2)})_z^{s'_{\rho(1)}-k'_{\rho(1)}+k_{\sigma(2)}}}{(( y_{\rho(1)}-x_{\sigma(n)})^2)^{s'_{\rho(1)}-k'_{\rho(1)}+k_{\sigma(2}+1}}\nonumber\\
		&\quad \ldots\frac{(x_{\sigma(n)}-y_{\rho(n)})_z^{s_{\sigma(n)}-k_{\sigma(n)}+k'_{\rho(n)}}}{(( x_{\sigma(n)}-y_{\rho(n)})^2)^{s_{\sigma(n)}-k_{\sigma(n)}+k'_{\rho(n)}+1}}\frac{(y_{\rho(n)}-x_{\sigma(1)})_z^{s'_{\rho(n)}-k'_{\rho(n)}+k_{\sigma(1)}}}{(( y_{\rho(2)}-x_{\sigma(1)})^2)^{s'_{\rho(n)}-k'_{\rho(n)}+k_{\sigma(1)}+1}}\,.
	\end{align}
	
	\subsection{$T^E$ and ${\bar T}^E$ correlators}
	
	The $2n$-point correlators for $T^E$ and ${\bar T}^E$ are
	\begin{align}
		&\langle T^E_{s_1}(x_1)\bar{T}^E_{s'_1}(y_1) T_{s_2}(x_2)\bar{T}^E_{s'_2}(y_2)\ldots T^E_{s_n}(x_n) \bar{T}^E_{s'_n}(y_n)\rangle \nonumber\\
		&= (N^2-1)\frac{1}{(4\pi^2)^{2n}}2^{\sum_{l=1}^n s'_l+\sum_{l=1}^n s_{l}-n}(-1)^{\sum_{l=1}^n s'_l+\sum_{l=1}^n s_{l}}\nonumber\\
		&\quad \sum_{k_1 = 0}^{s_1-1}{s_1\choose k_1}{s_1+1\choose k_1+2} \ldots  \sum_{k_n = 0}^{s_n-1}{s_n\choose k_n}{s_n+1\choose k_n+2} \nonumber\\
		&\quad \sum_{k'_1 = 0}^{s'_1-1}{s'_1+1\choose k'_1}{s'_1\choose k'_1+1} \ldots \sum_{k'_n = 0}^{s'_n-1}{s'_n+1\choose k'_n}{s'_n\choose k'_n+1} \sum_{\rho\in P_n}\sum_{\sigma\in P_n}\frac{(-1)^{2n+1}}{n}\text{sgn}(\rho)\text{sgn}(\sigma)\nonumber\\
		&\quad (s_{\sigma(1)}-k_{\sigma(1)}+k'_{\rho(1)})! \frac{(x_{\sigma(1)}-y_{\rho(1)})_z^{s_{\sigma(1)}-k_{\sigma(1)}+k'_{\rho(1)}}}{(( x_{\sigma(1)}-y_{\rho(1)})^2)^{s_{\sigma(1)}-k_{\sigma(1)}+k'_{\rho(1)}+1}}\nonumber\\
		&\quad (s'_{\rho(1)}-k'_{\rho(1)}+k_{\sigma(2)}+1)! \frac{(y_{\rho(1)}-x_{\sigma(2)})_z^{s'_{\rho(1)}-k'_{\rho(1)}+k_{\sigma(2)}+1}}{(( y_{\rho(1)}-x_{\sigma(2)})^2)^{s'_{\rho(1)}-k'_{\rho(1)}+k_{\sigma(2)}+2}}\nonumber\\
		&\quad \ldots\nonumber\\
		&\quad (s_{\sigma(n)}-k_{\sigma(n)}+k'_{\rho(n)})! \frac{(x_{\sigma(n)}-y_{\rho(n)})_z^{s_{\sigma(n)}-k_{\sigma(n)}+k'_{\rho(n)}}}{(( x_{\sigma(n)}-y_{\rho(n)})^2)^{s_{\sigma(n)}-k_{\sigma(n)}+k'_{\rho(n)}+1}}\nonumber\\
		&\quad (s'_{\rho(n)}-k'_{\rho(n)}+k_{\sigma(1)}+1)! \frac{(y_{\rho(n)}-x_{\sigma(1)})_z^{s'_{\rho(n)}-k'_{\rho(n)}+k_{\sigma(1)}+1}}{(( y_{\rho(2)}-x_{\sigma(1)})^2)^{s'_{\rho(2)}-k'_{\rho(2)}+k_{\sigma(1)}+2}}\,.
	\end{align}

	\section{Normalization of $2$-point correlators} \label{app:2point}
	
	\subsection{$2$-point correlators of gluino-gluino operators}

	The $2$-point correlators of $S_s^{\lambda}$ and $\bar{S}_s^{\lambda}$ follow directly from Eq. \eqref{Sncorr}
	\begin{align}
		\langle S^{\lambda}_{s_1}(x)\bar{S}^{\lambda}_{s'_1}(y)\rangle 
		=&-\frac{N^2-1}{4}\frac{1}{(4\pi^2)^{2}}i^{s_1+s'_1}2^{s_1+s'_1}\frac{(s_1+1)(s'_1+1)}{4}
		\sum_{k_1 = 0}^{s_1-1}\sum_{k'_1 = 0}^{s'_1-1}{s_1\choose k_1}{s_1\choose k_1+1}{s'_1\choose k'_1}{s'_1\choose k'_1+1}	\nonumber\\
		&(s_{1}-k_{1}+k'_{1})!(s'_{1}-k'_{1}+k_{1})!(-1)^{s'_1-k'_1+k_1}\frac{(x-y)_+^{s_1+s'_1}}{(\rvert x-y\rvert^2)^{s_1+s'_1+2}} ,
	\end{align}
	which are identical to the $2$-point correlators of the operators $O^{\lambda}_s$ \cite{BPS41}. Since (from Appendix C in \cite{BPS41})
	\begin{align}
		\delta_{s_1s'_1}\frac{s_1}{s_1+1}
		=\sum_{k_1 = 0}^{s_1-1}\sum_{k'_1 = 0}^{s'_1-1}{s_1\choose k_1}{s_1\choose k_1+1}{s'_1\choose k'_1}{s'_1\choose k'_1+1}(-1)^{k_1+k'_1}\frac{1}{{s_1+s'_1\choose k_1+k'_1+1}}\,,
	\end{align}
	we obtain
	\begin{align}
		\langle S^{\lambda}_{s_1}(x)\bar{S}^{\lambda}_{s'_1}(y)\rangle=\delta_{s_1s'_1}\frac{N^2-1}{4}\frac{1}{(4\pi^2)^2}(-1)^{s_1}2^{2s_1}(2s_1)!\frac{s_1(s_1+1)}{4}
		\frac{(x-y)_+^{2s_1}}{(\rvert x-y\rvert^2)^{2s_1+2}}\,.
	\end{align}
	
	\subsection{$2$-point correlators of gluon-gluino operators}
	
	The $2$-point correlators of $T_s$ and $\bar{T}_s$ arise from Eq. \eqref{Tncorr} and are, in fact, identical to the $2$-point correlators of $M_s$ and $\bar{M}_s$ \cite{BPS41}
	\begin{align}
		&\langle T_{s_1}(x)\bar{T}_{s_2}(y)\rangle\nonumber\\
		&= i(N^2-1)\frac{1}{(4\pi^2)^{2}}2^{s_1+s_2-1}i^{s_1+s_2}\nonumber\\
		&\quad\sum_{k_1 = 0}^{s_1-1}\sum_{k_2 = 0}^{s_2-1}{s_1\choose k_1}{s_1+1\choose k_1+2}{s_2+1\choose k_2}{s_2\choose k_2+1}\nonumber\\
		&\quad(s_{1}-k_{1}+k_2)!(s_2-k_2+k_{1}+1)! \frac{(x-y)_+^{s_{1}-k_{1}+k_2}}{(\rvert x-y\rvert^2)^{s_{1}-k_{1}+k_2+1}}\frac{(y-x)_+^{s_2-k_2+k_{1}+1}}{(\rvert y-x\rvert^2)^{s_2-k_2+k_{1}+2}}\,,
	\end{align}
	which become
	\begin{align}
		\label{twopointmix}
		&\langle T_{s_1}(x)\bar{T}_{s_2}(y)\rangle\nonumber\\
		&= -i(N^2-1)\frac{1}{(4\pi^2)^{2}}2^{s_1+s_2-1}i^{s_1+s_2}\nonumber\\
		&\quad\sum_{k_1 = 0}^{s_1-1}\sum_{k_2 = 0}^{s_2-1}{s_1\choose k_1}{s_1+1\choose k_1+2}{s_2+1\choose k_2}{s_2\choose k_2+1}\nonumber\\
		&\quad(s_{1}-k_{1}+k_2)!(s_2-k_2+k_{1}+1)!(-1)^{s_1-k_2+k_{1}}\frac{(x-y)_+^{s_{1}+s_2+1}}{(\rvert x-y\rvert^2)^{s_{1}+s_2+3}}\,.
	\end{align}
	Finally, using the relation from \cite{BPS41} (Appendix C in \cite{BPS41}),
	\begin{align}
		&-\delta_{s_1s_2}\frac{s_1}{s_1+2}=\sum_{k_1 = 0}^{s_1-1}\sum_{k_2 = 0}^{s_2-1}{s_1\choose k_1}{s_1+1\choose k_1+2}{s_2+1\choose k_2}{s_2\choose k_2+1}(-1)^{s_1-k_2+k_1}\frac{1}{{s_1+s_2+1\choose s_1-k_1+k_2}}\,,
	\end{align}
	we get
	\begin{align}
		\label{twopointmix2}
		\langle 	T_{s_1}(x)\bar{T}_{s_2}(y)\rangle
		= \delta_{s_1s_2}i(N^2-1)\frac{1}{(4\pi^2)^{2}}2^{2s_1-1}(-1)^{s_1}\frac{s_1}{s_1+2}(2s_1+1)!\frac{(x-y)^{2s_1+1}}{(\rvert x-y\rvert^2)^{2s_1+3}}\,.
	\end{align} 
	\hspace{1cm}

	\section{Connected generating functional in the momentum representation \label{momgen}}
	
	The conformal generating functional in momentum representation is defined by the functional integral
	\begin{align}
		\label{Zmom}
		\mathcal{Z}_{\text{conf}}[J_{\mathcal{O}}]= \frac{1}{Z}\int \mathcal{D} A \mathcal{D} \bar{A} \mathcal{D} \lambda \mathcal{D} \bar{\lambda}\,  e^{\int \,-i\bar{A}^a \square A^a+\bar{\lambda}^a\Box \partial_+^{-1} \lambda^a\,d^4x}\exp\Bigg(\int \frac{d^4p}{(2\pi)^4}\sum_i\, J_{\mathcal{O}_i}(-p)\mathcal{O}_i(p)\Bigg)\,.
	\end{align}
	The connected correlators in momentum representation then follow \cite{BPSpaper2}
	\begin{align}
		\langle \mathcal{O}_{s_1}(p_1)\ldots\mathcal{O}_{s_n}(p_n)\rangle 
		= (2\pi)^4\frac{\delta }{\delta J_{\mathcal{O}_{s_1}}(-p_1)}\cdots (2\pi)^4\frac{\delta }{\delta J_{\mathcal{O}_{s_n}}(-p_n)}\mathcal{W}[J_{\mathcal{O}}] \,.
	\end{align} 
	The generating functional is found using the kernels in momentum representation
	\begin{equation} \label{AB}
		i\square^{-1} \rightarrow \frac{-i}{\rvert q\rvert^2+i\epsilon} 
	\end{equation}
	and
	\begin{equation} \label{BA}
		(i\partial_+)^{k} i\square^{-1} \rightarrow (-q_+)^{k} \frac{-i}{\rvert q\rvert^2+i\epsilon} \,,
	\end{equation}
	where we utilize the momentum-space measure \cite{BPSpaper2}
	\begin{equation}
		\int\frac{d^4q}{(2\pi)^4}\,.
	\end{equation}
	For the identity in space-time, we also get
	\begin{align}
		&I \to(2\pi)^4\delta^{(4)}(q_1-q_2)\,,
	\end{align}
	for the sources,
	\begin{align}
		&J_{\mathcal{O}}\to J_{\mathcal{O}}(q_1-q_2)
	\end{align}
	and for the differential operators
	\begin{align}
		&\mathcal{Y}^{\frac{5}{2}}_{s-2}(\overrightarrow{\partial}_+,\overleftarrow{\partial}_+)\to\mathcal{Y}_{s-2}^{\frac{5}{2}}(q_{2+},q_{1+}) = q_{1+} (q_{2+}-q_{1+})^{s-2}C^{\frac{5}{2}}_{s-2}\left(\frac{q_{2+}+q_{1+}}{q_{2+}-q_{1+}}\right)q_{2+}\nonumber\\
		&\mathcal{Y}^{\frac{3}{2}}_{s-1}(\overrightarrow{\partial}_+,\overleftarrow{\partial}_+)\to\mathcal{Y}_{s-1}^{\frac{3}{2}}(q_{2+},q_{1+})  =(q_{2+}-q_{1+})^{s-2}C^{\frac{3}{2}}_{s-1}\left(\frac{q_{2+}+q_{1+}}{q_{2+}-q_{1+}}\right)\nonumber\\
		&\mathcal{H}^{\frac{5}{2}}_{s-2}(\overrightarrow{\partial}_+,\overleftarrow{\partial}_+)\to\mathcal{H}_{s-2}^{\frac{5}{2}}(q_{2+},q_{1+}) = q_{1+} (q_{2+}-q_{1+})^{s-2}C^{\frac{5}{2}}_{s-2}\left(\frac{q_{2+}+q_{1+}}{q_{2+}-q_{1+}}\right)q_{2+}\nonumber\\
		&\mathcal{H}^{\frac{3}{2}}_{s-1}(\overrightarrow{\partial}_+,\overleftarrow{\partial}_+)\to\mathcal{H}_{s-1}^{\frac{3}{2}}(q_{2+},q_{1+})  =(q_{2+}-q_{1+})^{s-2}C^{\frac{3}{2}}_{s-1}\left(\frac{q_{2+}+q_{1+}}{q_{2+}-q_{1+}}\right)\nonumber\\
		&	\mathcal{G}_{s-1}^{(1,2)} (\overrightarrow{\partial}_+,\overleftarrow{\partial}_+)\to	\mathcal{G}_{s-1}^{(1,2)} (q_{2+},q_{1+}) =i(q_{2+}-q_{1+})^{s-1}P^{(1,2)}_{s-1}\left(\frac{q_{2+}+q_{1+}}{q_{2+}-q_{1+}}\right) q_{2+}\nonumber\\
		&	\mathcal{G}_{s-1}^{(2,1)} (\overrightarrow{\partial}_+,\overleftarrow{\partial}_+)\to	\mathcal{G}_{s-1}^{(2,1)} (q_{2+},q_{1+}) =-iq_{1+}(q_{2+}-q_{1+})^{s-1}P^{(2,1)}_{s-1}\left(\frac{q_{2+}+q_{1+}}{q_{2+}-q_{1+}}\right) \,.
	\end{align}
	The explicit expression in momentum representation \cite{BPSpaper2} is derived from Eq. \eqref{jaco2}.
	We provide the generating functional restricted to several sectors
	\begin{align}
		&\mathcal{W}_{\text{conf}}\left[\bar{J}_{S^A},J_{\bar{S}^A},0,0,0,0\right] =-\frac{N^2-1}{2}\log\Det \Big((2\pi)^4\delta^{(4)}(q_1-q_2) \nonumber\\
		& - \frac{1}{2}\int\frac{d^4q}{(2\pi)^4} i\square^{-1}(q_1)\bar{J}_{S^A_{s_1}}(q_1-q)\otimes\mathcal{Y}_{s_1-2}^{\frac{5}{2}}(q_{+},q_{1+}) i\square^{-1}(q)J_{\bar{S}^A_{s_2}}(q-q_2)\otimes\mathcal{Y}_{s_2-2}^{\frac{5}{2}}(q_{2+},q_{+})\Big)\nonumber\\
		&+\frac{N^2-1}{2}\log\Det \Big((2\pi)^4\delta^{(4)}(q_1-q_2) \nonumber\\
		& - \frac{1}{2}\int\frac{d^4q}{(2\pi)^4} q_{1+}i\square^{-1}(q_1)\bar{J}_{S^\lambda_{s_1}}(q_1-q)\otimes\mathcal{Y}_{s_1-1}^{\frac{3}{2}}(q_{+},q_{1+}) q_+i\square^{-1}(q)J_{\bar{S}^\lambda_{s_2}}(q-q_2)\otimes\mathcal{Y}_{s_2-1}^{\frac{3}{2}}(q_{2+},q_{+})\Big)\,,
	\end{align}
	where we have shown the corresponding kernel in momentum representation as the argument of the functional determinant and have calculated the color trace \cite{BPSpaper2}.\par
	We calculate the generating functional of correlators of the fermionic operators, $T_s$ and $\bar{T}_s$, using Eq. \eqref{wgenT2}
	\begin{align}
		&\mathcal{W}_{\text{conf}}\left[0,0,0,0,\bar{J}_{T},J_{\bar T}\right] =(N^2-1)\log\Det \Big((2\pi)^4\delta^{(4)}(q_1-q_2) \nonumber\\
		& - \frac{1}{4}\int\frac{d^4q}{(2\pi)^4} q_{1+}i\square^{-1}(q_1)J_{\bar T_{s_1}}(q_1-q)\otimes\mathcal{G}_{s_1-1}^{(1,2)}(q_{+},q_{1+}) i\square^{-1}(q)\bar{J}_{T_{s_2}}(q-q_2)\otimes\mathcal{G}_{s_2-1}^{(2,1)}(q_{2+},q_{+})\Big)\,.
	\end{align}

	\section{RG-improved correlators}\label{rgcorr}

	We calculate the UV asymptotics of the RG-improved correlators by using Eqs. \eqref{zz}, \eqref{eqrg} and the conformal correlators from Eqs. \eqref{SAcorr}, \eqref{Sncorr}, and \eqref{Tncorr}, respectively
	\begin{align} 
		\nonumber
		&\langle S^{A\,E}_{s_1}(x_1)\ldots S^{A\,E}_{s_n}(x_n)\bar{S}^{A\,E}_{s'_{1}}(y_{1})\ldots \bar{S}^{A\,E}_{s'_{n}}(y_{n})\rangle_{\text{asym}}\nonumber\\
		&\sim\frac{1}{(4\pi^2)^{2n}}\frac{N^2-1}{2^{2n}}2^{\sum_{l=1}^n s_l+{s'}_l}i^{\sum_{l=1}^n s_l+{s'}_l}\frac{\Gamma(3)\Gamma(s_1+3)}{\Gamma(5)\Gamma(s_1+1)}\ldots\nonumber \frac{\Gamma(3)\Gamma(s_n+3)}{\Gamma(5)\Gamma(s_n+1)}\frac{\Gamma(3)\Gamma({s'}_1+3)}{\Gamma(5)\Gamma({s'}_1+1)}\ldots \frac{\Gamma(3)\Gamma({s'}_n+3)}{\Gamma(5)\Gamma({s'}_n+1)}\\\nonumber
		&\quad\frac{1}{\lambda^{\sum_{l=1}^{n}(2+s_l)+(2+s'_l)}}\Bigg(\frac{g(\mu)}{g(\frac{\mu}{\lambda})}\Bigg)^{\frac{\gamma_{0 S^A_{s_1}}}{\beta_0}}\ldots\nonumber \Bigg(\frac{g(\mu)}{g(\frac{\mu}{\lambda})}\Bigg)^{\frac{\gamma_{0 S^A_{s_n}}}{\beta_0}}\Bigg(\frac{g(\mu)}{g(\frac{\mu}{\lambda})}\Bigg)^{\frac{\gamma_{0 S^A_{s'_1}}}{\beta_0}}\ldots \Bigg(\frac{g(\mu)}{g(\frac{\mu}{\lambda})}\Bigg)^{\frac{\gamma_{0 S^A_{s'_n}}}{\beta_0}}\nonumber\\\nonumber
		&\quad \sum_{k_1=0}^{s_1-2}\ldots \sum_{k_n = 0}^{s_n-2}{s_1\choose k_1}{s_1\choose k_1+2}\ldots {s_n\choose k_n}{s_n\choose k_n+2}\sum_{{k'}_1=0}^{{s'}_1-2}\ldots \sum_{{k'}_n = 0}^{{s'}_n-2}{{s'}_1\choose {k'}_1}{{s'}_1\choose {k'}_1+2}\ldots {{s'}_n\choose {k'}_n}{{s'}_n\choose {k'}_n+2}\\\nonumber
		&\quad \frac{2^{n-1}}{n}\sum_{\sigma\in P_n}\sum_{\rho\in P_n}
		(s_{\sigma(1)}-k_{\sigma(1)}+{k'}_{\rho(1)})!({s'}_{\rho(1)}-{k'}_{\rho(1)}+k_{\sigma(2)})!\\\nonumber
		&\quad \ldots(s_{\sigma(n)}-k_{\sigma(n)}+{k'}_{\rho(n)})!({s'}_{\rho(n)}-{k'}_{\rho(n)}+k_{\sigma(1)})!\\\nonumber
		&\quad \frac{(x_{\sigma(1)}-y_{\rho(1)})_z^{s_{\sigma(1)}-k_{\sigma(1)}+{k'}_{\rho(1)}}}{\left(( x_{\sigma(1)}-y_{\rho(1)})^2\right)^{s_{\sigma(1)}-k_{\sigma(1)}+{k'}_{\rho(1)}+1}}\frac{(y_{\rho(1)}-x_{\sigma(2)})_z^{{s'}_{\rho(1)}-{k'}_{\rho(1)}+k_{\sigma(2)}}}{\left(( y_{\rho(1)}-x_{\sigma(2)})^2\right)^{{s'}_{\rho(1)}-{k'}_{\rho(1)}+k_{\sigma(2)}+1}}\\
		&\quad \ldots\frac{(x_{\sigma(n)}-y_{\rho(n)})_z^{s_{\sigma(n)}-k_{\sigma(n)}+{k'}_{\rho(n)}}}{\left(( x_{\sigma(n)}-y_{\rho(n)})^2\right)^{s_{\sigma(n)}-k_{\sigma(n)}+{k'}_{\rho(n)}+1}}
		\frac{(y_{\rho(n)}-x_{\sigma(1)})_z^{{s'}_{\rho(n)}-{k'}_{\rho(n)}+k_{\sigma(1)}}}{\left(( y_{\rho(n)}-x_{\sigma(1)})^2\right)^{{s'}_{\rho(n)}-{k'}_{\rho(n)}+k_{\sigma(1)}+1}}\,,
	\end{align}
	\begin{align}
		&\langle S^{\lambda\,E}_{s_1}(x_1)\ldots S^{\lambda\,E}_{s_n}(x_n)\bar{S}^{\lambda\,E}_{s'_1}(y_1)\ldots \bar{S}^{\lambda\,E}_{s'_n}(y_n)\rangle_{\text{asym}} \nonumber\\
		&\sim-\frac{N^2-1}{2}\frac{1}{2^n}\frac{1}{(4\pi^2)^{2n}}i^{\sum_{l=1}^n s_l+s'_l}2^{\sum_{l=1}^n s_l+s'_l}\frac{(s_1+1)}{2}\ldots \frac{(s_n+1)}{2}\frac{(s'_1+1)}{2}\ldots \frac{(s'_n+1)}{2}\nonumber\\
		&\quad\frac{1}{\lambda^{\sum_{l=1}^{n}(2+s_l)+(2+s'_l)}}\Bigg(\frac{g(\mu)}{g(\frac{\mu}{\lambda})}\Bigg)^{\frac{\gamma_{0S^{\lambda}_{s_1}}}{\beta_0}}\ldots \Bigg(\frac{g(\mu)}{g(\frac{\mu}{\lambda})}\Bigg)^{\frac{\gamma_{0S^{\lambda}_{s_n}}}{\beta_0}}\Bigg(\frac{g(\mu)}{g(\frac{\mu}{\lambda})}\Bigg)^{\frac{\gamma_{0S^{\lambda}_{s'_1}}}{\beta_0}}\ldots \Bigg(\frac{g(\mu)}{g(\frac{\mu}{\lambda})}\Bigg)^{\frac{\gamma_{0S^{\lambda}_{s'_n}}}{\beta_0}}\nonumber\\
		&\quad \sum_{k_1 = 0}^{s_1-1}{s_1\choose k_1}{s_1\choose k_1+1}\ldots \sum_{k_n = 0}^{s_n-1}{s_n\choose k_n}{s_n\choose k_n+1}\sum_{k'_1 = 0}^{s'_1-1}{s'_1\choose k'_1}{s'_1\choose k'_1+1}\ldots \sum_{k'_n = 0}^{s'_n-1}{s'_n\choose k'_n}{s'_n\choose k'_n+1}\nonumber\\
		&\quad \frac{1}{n}\sum_{\sigma \in P_n}\sum_{\rho \in P_n}(s_{\sigma(1)}-k_{\sigma(1)}+k'_{\rho(1)})!(s'_{\rho(1)}-k'_{\rho(1)}+k_{\sigma(2)})!\nonumber\\
		&\quad \ldots (s_{\sigma(n)}-k_{\sigma(n)}+k'_{\rho(n)})!(s'_{\rho(n)}-k'_{\rho(n)}+k_{\sigma(1)})!\nonumber\\
		&\quad \frac{(x_{\sigma(1)}-y_{\rho(1)})_z^{s_{\sigma(1)}-k_{\sigma(1)}+k'_{\rho(1)}}}{(( x_{\sigma(1)}-y_{\rho(1)})^2)^{s_{\sigma(1)}-k_{\sigma(1)}+k'_{\rho(1)}+1}}\frac{(y_{\rho(1)}-x_{\sigma(2)})_z^{s'_{\rho(1)}-k'_{\rho(1)}+k_{\sigma(2)}}}{(( y_{\rho(1)}-x_{\sigma(n)})^2)^{s'_{\rho(1)}-k'_{\rho(1)}+k_{\sigma(2}+1}}\nonumber\\
		&\quad \ldots\frac{(x_{\sigma(n)}-y_{\rho(n)})_z^{s_{\sigma(n)}-k_{\sigma(n)}+k'_{\rho(n)}}}{(( x_{\sigma(n)}-y_{\rho(n)})^2)^{s_{\sigma(n)}-k_{\sigma(n)}+k'_{\rho(n)}+1}}\frac{(y_{\rho(n)}-x_{\sigma(1)})_z^{s'_{\rho(n)}-k'_{\rho(n)}+k_{\sigma(1)}}}{(( y_{\rho(2)}-x_{\sigma(1)})^2)^{s'_{\rho(n)}-k'_{\rho(n)}+k_{\sigma(1)}+1}}\,
	\end{align}
	and
	\begin{align}
		&\langle T^E_{s_1}(x_1)\bar{T}^E_{s'_1}(y_1) T_{s_2}(x_2)\bar{T}^E_{s'_2}(y_2)\ldots T^E_{s_n}(x_n) \bar{T}^E_{s'_n}(y_n)\rangle_{\text{asym}} \nonumber\\
		&\sim(N^2-1)\frac{1}{(4\pi^2)^{2n}}2^{\sum_{l=1}^n s'_l+\sum_{l=1}^n s_{l}-n}i^{\sum_{l=1}^n s'_l+\sum_{l=1}^n s_{l}-n}\nonumber\\ 	&\quad\frac{1}{\lambda^{\sum_{l=1}^{n}(2+s_l)+(2+s'_l)}}\Bigg(\frac{g(\mu)}{g(\frac{\mu}{\lambda})}\Bigg)^{\frac{\gamma_{0T_{s_1}}}{\beta_0}}\ldots \Bigg(\frac{g(\mu)}{g(\frac{\mu}{\lambda})}\Bigg)^{\frac{\gamma_{0T_{s_n}}}{\beta_0}}\Bigg(\frac{g(\mu)}{g(\frac{\mu}{\lambda})}\Bigg)^{\frac{\gamma_{0T_{s'_1}}}{\beta_0}}\ldots \Bigg(\frac{g(\mu)}{g(\frac{\mu}{\lambda})}\Bigg)^{\frac{\gamma_{0T_{s'_n}}}{\beta_0}}\nonumber\\
		&\quad \sum_{k'_1 = 0}^{s'_1-1}{s'_1+1\choose k'_1}{s'_1\choose k'_1+1} \ldots \sum_{k'_n = 0}^{s'_n-1}{s'_n+1\choose k'_n}{s'_n\choose k'_n+1} \sum_{\rho\in P_n}\sum_{\sigma\in P_n}\frac{(-1)^{2n+1}}{n}\text{sgn}(\rho)\text{sgn}(\sigma)\nonumber\\
		&\quad (s_{\sigma(1)}-k_{\sigma(1)}+k'_{\rho(1)})! \frac{(x_{\sigma(1)}-y_{\rho(1)})_z^{s_{\sigma(1)}-k_{\sigma(1)}+k'_{\rho(1)}}}{(( x_{\sigma(1)}-y_{\rho(1)})^2)^{s_{\sigma(1)}-k_{\sigma(1)}+k'_{\rho(1)}+1}}\nonumber\\
		&\quad (s'_{\rho(1)}-k'_{\rho(1)}+k_{\sigma(2)}+1)! \frac{(y_{\rho(1)}-x_{\sigma(2)})_z^{s'_{\rho(1)}-k'_{\rho(1)}+k_{\sigma(2)}+1}}{(( y_{\rho(1)}-x_{\sigma(2)})^2)^{s'_{\rho(1)}-k'_{\rho(1)}+k_{\sigma(2)}+2}}\nonumber\\
		&\quad \ldots\nonumber\\
		&\quad (s_{\sigma(n)}-k_{\sigma(n)}+k'_{\rho(n)})! \frac{(x_{\sigma(n)}-y_{\rho(n)})_z^{s_{\sigma(n)}-k_{\sigma(n)}+k'_{\rho(n)}}}{(( x_{\sigma(n)}-y_{\rho(n)})^2)^{s_{\sigma(n)}-k_{\sigma(n)}+k'_{\rho(n)}+1}}\nonumber\\
		&\quad (s'_{\rho(n)}-k'_{\rho(n)}+k_{\sigma(1)}+1)! \frac{(y_{\rho(n)}-x_{\sigma(1)})_z^{s'_{\rho(n)}-k'_{\rho(n)}+k_{\sigma(1)}+1}}{(( y_{\rho(2)}-x_{\sigma(1)})^2)^{s'_{\rho(2)}-k'_{\rho(2)}+k_{\sigma(1)}+2}}\,.
	\end{align}
	
\bibliographystyle{JHEP}
\bibliography{mybib}
	
\end{document}